# Title: Human Sexual Cycles are Driven by Culture and Match Collective Moods

**Authors:** Ian B. Wood[1], Pedro Leal Varela[2], Johan Bollen[1,3], Luis M. Rocha[1,2]*, and Joana Gonçalves-Sá†[2]

**Affiliations:**
[1] School of Informatics & Computing, Indiana University, Bloomington, IN, USA.
[2] Instituto Gulbenkian de Ciência, Oeiras, Portugal.
[3]Wageningen University, The Netherlands.
Correspondence to: † mjsa@igc.gulbenkian.pt and *rocha@indiana.edu

## Abstract

Human reproduction does not happen uniformly throughout the year and what drives human sexual cycles is a long-standing question. The literature is mixed with respect to whether biological or cultural factors best explain these cycles. The biological hypothesis proposes that human reproductive cycles are an adaptation to the seasonal (hemisphere-dependent) cycles, while the cultural hypothesis proposes that conception dates vary mostly due to cultural factors, such as holidays. However, for many countries, common records used to investigate these hypotheses are incomplete or unavailable, biasing existing analysis towards Northern Hemisphere Christian countries. Here we show that interest in sex peaks sharply online during major cultural and religious celebrations, regardless of hemisphere location. This online interest, when shifted by nine months, corresponds to documented human births, even after adjusting for numerous factors such as language and amount of free time due to holidays. We further show that mood, measured independently on Twitter, contains distinct collective emotions associated with those cultural celebrations.

Our results provide converging evidence that the cyclic sexual and reproductive behavior of human populations is mostly driven by culture and that this interest in sex is associated with specific emotions, characteristic of major cultural and religious celebrations.

## Introduction

Human reproduction shows a yearly cyclical pattern and whether this periodicity is driven primarily by cultural or by biological factors has been an open question for several decades. In Western, Northern Hemisphere countries, births tend to peak in September, corresponding to early winter conceptions (1). These conception dates are aligned with the December solstice which has been taken as evidence for the existence of an environment-induced biological clock that drives human reproduction cycles (2, 3). Proposed evolutionary explanations include temperature (4), libido, or the availability of food (1, 5). However, this conception peak also coincides with religious celebrations, like Christmas, suggesting that culture drives the observed birth cycles.

Culture and biology certainly influence each other, and it is very likely that both influence sexual drive. However, whether biological or cultural factors best explain the reproduction cycle has long been debated in the literature, with biological explanations dominating the argument (1).

The biological hypothesis, proposes that human reproductive cycles are an adaptation to the seasonal cycles caused by hemisphere positioning in the yearly orbit of the Earth around the Sun. If true, reproductive periodicity should be similar among Northern Hemisphere countries, less pronounced closer to the equator, and reversed in Southern Hemisphere countries (6). On the other hand, the cultural hypothesis proposes that conception dates vary mostly due to cultural factors, such as holidays or seasonal marriage patterns (3). If true, we should see similar sexual cycles in similar cultures independent of hemisphere. To study these hypotheses we need to measure sexual activity on a planetary scale. Common proxies for such measurements include birth records, incidence of sexually transmitted diseases, or condom sales (7). However, for many countries these records are inaccurate with respect to the timing of sexual activity (8, 9) and a focus on hospital records (for births or sexually transmitted diseases) would largely restrict analysis to "Western" countries, where such data tends to be most commonly available. Thus, previous indicators do not offer sufficiently accurate data from across the globe to help distinguish between the two hypotheses.

The recent availability of large-scale population data from web searches and social media now allows us to study collective social behavior on a global scale. In this work, we gauge interest in sex directly from Google searches and characterize seasonal population sentiment from the analysis of Twitter feeds. We show that analysis of this large-scale online activity can be used as proxies for real-life actions and help answer longstanding scientific questions about human behavior.

## Results

### Worldwide Variations in Sexual Interest

To measure interest in sex, for each country, we retrieved the frequency by which people searched for the word "sex" using Google Trends$^{tm}$ (GT) (10) (Methods 1-3); henceforth referred to as "sex-searches." Interestingly, even in countries where English is not an official language, the English term "sex" is either more searched for than the corresponding word in the local languages or they are strongly correlated (Supplementary Table S1). Moreover, the terms most associated with searches for "sex" in GT refer to direct interest in sex and pornography (Supplementary Table S1). Therefore, GT searches for the term "sex" are a good proxy for interest in sexual behavior in the countries analyzed in this study.

Fig. 1 depicts GT weekly sex-search data for 10 years from January 2004 to February 2014 for a set of Northern countries, which celebrate Christmas on December 25th. Yearly maximum peaks occur during Christmas week (red vertical lines), as previously observed for the USA (11). While one may think that this increased interest in sex results simply from more free time during the holiday season, GT data is normalized by overall search volume (10); even in a situation of increased general online activity, the increase in sexual interest is higher. Conversely, we could expect the holiday season to lead to a decrease in overall searches, led by school vacations for instance, originating an artificial peak for sex-related interest. However, we do not observe similar increases in weekly sex-searches for other widely observed holidays, such as Thanksgiving in the USA or Easter in France (Fig. S1A and S1B). Furthermore, a putative decrease in overall searches is unlikely, as a decrease in searches for school-related material can be compensated by a strong increase in searches for "presents" or "recipes". In fact, when we control for search-volume of very common words, such as "on", "and", or "the", there is some variation around the holiday period but it is in different directions for different search terms (Fig S2A and S2B), probably resulting in an overall neutral change. Therefore, and although other dates lead to an increase in sex-searches (Fig. S1A and B), the Christmas holiday is uniquely associated with the highest peaks in sex-searches observed in these Northern countries. It is also known that, in Western Northern countries, conceptions peak around Christmas, in what some refer to as the "holiday effect" (12). Indeed, the observed sex-search peaks match birth rate increases for this set of countries when shifted by nine months (Fig. S3A), which further confirms GT sex-searches as a good proxy for sexual activity.

Compared to the observation of sex-search peaks in Northern countries that celebrate Christmas on December 25th (and corresponding increase in September birth rates where such data is available), the two hypotheses outlined above would predict quite distinct observations for other cultures and hemisphere locations. If the biological hypothesis is correct, all Northern countries should have similar sex-search peaks around the same time, and these peaks should occur in a counter-phase pattern (six months later) in all Southern countries—irrespective of culture. On the other hand, if the cultural hypothesis is true, these peaks should appear anywhere Christmas is celebrated—irrespective of hemisphere—and other similar celebrations in different cultures should lead to sex-search peaks in other times of the year.

To test these predictions, we extracted GT sex-search time-series data for all 129 countries for which GT offered consistent data. Countries were categorized according to hemisphere (North or South) and their predominant religion (13, 14). Countries where at least 50% of the population self-identifies as Christian were considered culturally Christian countries, and similarly for Muslim countries. Other countries, where neither of these religions is dominant, were grouped separately; Supplementary Table S2 shows the complete list of countries and categorization.

Both Northern and Southern countries show a prominent peak in sex-searches around Christmas and we observe no counter-phase pattern corresponding to the southern hemisphere winter solstice of June 21st (see Fig. S4A, Fig. S5C, and Fig. S5D). In fact, there is a strong significant correlation ($R2= 0.54$, $p\text{-value}< 0.001$) between the mean sex-search time series of Northern and Southern countries (Supplementary Table S3). Since most Northern and Southern countries for which we have data identify as Christian (80 of 129), the observed correlation suggests that a cultural effect, rather than hemisphere location, drives the Christmas sex-search peak. Indeed, the birth data available for Christian, Southern countries peaks with Christmas sex-searches when shifted by nine months in much the same way as for Christian, Northern Countries, even though it is summer in the former and winter in the latter (Fig. S3). Notice further that there is neither a sex-searches increase in December nor a birth peak in September for Northern countries that do not celebrate Christmas on December 25th (Fig. S7). As reliable birth data is not generally available, particularly for Southern and Muslim countries, and is only available for four Southern countries, all of them predominantly Christian, (Methods 6, Supplementary Table S9 and Figs. S3 and S6), we use GT sex-search data instead to observe many more countries and address the two hypotheses.

Parsing all countries by religion (Fig. 2A&B, Fig. S4 and Supplementary Table S3), it is clear that the mean sex-search time-series are periodic but uncorrelated between Christian and Muslim countries (R2 = 0.19, p-value<0.001). The difference in sex-search behavior between these two sets of countries is further revealed in Fig. 2C&D, where we averaged the sex-search yearly time-series across all ten years centered on Christmas week (for Christian countries) or centered on Eid-al-Fitr, the major family holiday that ends Ramadan (for Muslim countries). In Christian countries, the only clear peak occurs during the Christmas week. In contrast, in Muslim countries there is a peak during the week of Eid-al-Fitr and a second peak during the week of Eid-al-Adha, the other major religious and family celebration in Muslim culture; also noteworthy is a steep decrease during Ramadan, consistent with that period of general abstinence (as further discussed below). Both of these groups of countries clearly show sex-search peaks associated with distinct cultural celebrations, rather than with hemisphere. Indeed, it is worth noting that the Muslim calendar does not follow the solar calendar: every year Ramadan shifts by 10 days relative to its date during the previous Gregorian calendar year. Nevertheless, sex-searches peak during the moving week of Eid-al-Fitr (and Eid-al-Adha) in Muslim countries. The moving sex-search peaks associated with major religious events in Muslim countries further emphasizes the cultural driver behind such collective behavior.

To resolve the incompatible predictions of the biological and cultural hypotheses we made country-specific comparisons between hemisphere and culture, beyond the group-average behavior described above. We averaged the yearly sex-search time-series for each of the 129 individual countries across all years in four different ways: centered on Christmas week (fixed relative to the solar calendar), centered on Eid-al-Fitr week (moving relative to the solar calendar), and centered on each of the solstices, fixed on June 21st and December 21st (Methods 4, Supplementary Tables S4-6 and Fig. S5). We then measured the response of countries to a holiday as the sex-search z-score deviation above the mean at Christmas, Eid-al-Fitr and the two solstice weeks (Methods 5 and Supplementary Table S7). Fig. 3 shows a world map with color-coded countries: shades of red indicate countries whose highest sex-search deviation from mean occurs during the Christmas week, and shades of green indicate countries whose highest sex-search deviation from mean occurs during Eid-al-Fitr week (Methods 7). It is clear that this response yields a map organized according to culture rather than hemisphere.

We then compared this new country classification (according to the individual countries' sex-search profile, Supplementary Table S7 and Supplementary Methods S1) with our previous identification based only on the proportion of the population that self-identified as Christian, Muslim or Other (Supplementary Table S2) (13,14). Out of the 30 countries originally identified as Muslim (14), 77% show a significant increase (z>1) in sex-searches during the week of Eid-al-Fitr, and out of the 80 countries originally identified as Christian (13), 80% show a significant increase (z>1) during the Christmas week, regardless of the hemisphere. It is important to note that this correspondence is even higher (91%) when we identify as "Other" the ten Christian countries that do not celebrate Christmas on December 25th. In fact, we do not see an increase in sex searches around December 25th in any of these Northern Russian and Serbian Orthodox Christian countries, which celebrate Christmas in early January, and this further supports the cultural hypothesis (Methods 2, Supplementary Methods S1, Supplementary Figure S7). Moreover, only 14% of Southern countries showed a significant increase in sex-searches during the June solstice (Supplementary Tables S7 and S8B), demonstrating that there is no significant counter-phase sex-search peak in the southern hemisphere, contradicting the biological hypothesis.

**Trends in Holiday Moods**

The Christmas and Eid-al-Fitr holidays carry significant cultural and religious meaning, but they are not directly associated with sex. It is, in fact, very counter-intuitive to think of Christmas and Eid as the times of the year with the most online searches for sex. However, these events may trigger specific and collective moods, leading to a striking correspondence between these holidays and sexual interest. To investigate the emotional factors involved we measured changes in public sentiment on Twitter (21-23). The analysis was performed before, during, and after Christmas and Eid-al-Fitr in a set of seven countries with sufficient Twitter traffic in our data: Australia, Argentina, Brazil, Chile, Indonesia, Turkey, and the USA (Methods 9 and Fig. S8). Although it is not possible to know whether the Google and Twitter populations are the same per country, given the large volume of Google searches and tweets, it is very likely that they provide a significant sample of the same populations.

Twitter sentiment was quantified by rating a random 10% sample of all tweets posted between September 2010 to February 2014 using the Affective Norms for English Words (ANEW) lexicon (18) (Methods 8 and 9). The ANEW lexicon consists of 1,034 English words that carry a sentiment score along three dimensions: Arousal (a), Dominance (d), and Valence (v), corresponding respectively to whether the word makes human raters feel calm vs. excited, controlled vs. in-control, and sad vs. happy. The sentiment value of a single tweet is defined as the mean ANEW score of its words. We translated the lexicon to Spanish and Portuguese to capture public sentiment in those languages as well, but did not have the ability to translate into additional languages. To avoid bias from holiday-related language, we ignored all words used in traditional greetings for all known holidays in the World (Supplementary Table S13); we also removed the word "Christmas" and "valentine" from the lexicon, which does not include other holiday names.

We first observed that the weekly volume of sex-searches significantly correlates with the mean weekly sentiment derived from the three ANEW dimensions in a multiple linear regression (Supplementary Methods S2, Supplementary Table S10). In every country, valence yields a positive coefficient, while dominance a negative coefficient; thus the happier but less in-control the population mood is, the more sex-searches tend to increase in every country (Methods 10 and Supplementary Methods S2). Interestingly, while public sentiment displays a strong linear relationship with sex-search volume when all mood dimensions are considered, there is little correlation with each ANEW dimension on its own (Supplementary Table S11).However, the observed linear correlation does not allow us to characterize the population mood in the target cultural celebrations. To investigate if days that are similar in mood to Christmas in Christian Countries or to Eid-al-Fitr in Muslim Countries also tend to observe increased volume of sex-searches, we need a more nuanced characterization of the mood profile each week.

Because collective mood sentiment, as measured here, is derived from many tweets of large and diverse populations, it can contain distinct and informative components. Thus, we employed an eigenvector-based analysis (20) to characterize the distribution of sentiment values, rather than just average sentiment. We thus obtain the components of public sentiment that explain most of the variance in the data not attributable to regular language use, hereafter referred to as “eigenmoods.” Specifically, an eigenmood is a small set of components (eigenvectors) of a matrix. In this matrix, the rows denote sentiment scores in a given range or bin, and the columns denote the weeks (Methods 11 and Supplementary Methods S3), and elements are the number of tweets during a week that fall in that bin. Thus, an eigenmood is not an average sentiment value (per week in our analysis), but rather a change in the distribution of sentiment that explains a significant proportion of the variation in the time-series data (19).

We found that two components were sufficient to describe public sentiment associated with each holiday and country – a characterization that is independent of sex-search volume, and relies only on measurement of sentiment on Twitter (Methods 10-12, Supplementary Methods S3-5, and Supplementary Fig. S10 and Fig. S11). Fig. 4 (Column A), Fig. S9 and Fig. S14 show the sentiment distribution of a selected eigenmood per every week of the year; redder (greener) colors represent increased (decreased) numbers of tweets falling in the respective mood dimension bins – e.g., for valence, upper bins on vertical axis denote increased happiness and lower bins denote increased sadness. The sentiment distributions of rows 1, 2, and 3 in Fig. 4 column A are centered on Christmas for USA (Northern, Christian) and Brazil (Southern, Christian), and Eid-al-Fitr for Indonesia (Southern, Muslim). While the eigenmood that describes Christmas in the USA uses only the valence dimension of ANEW, the best eigenmood for Christmas in Brazil requires valence and arousal, and for Eid-al-Fitr in Indonesia requires valence and dominance. The sentiment distribution of these eigenmoods per week clearly shows that significant and unique changes in sentiment occur during the target holidays. In all these cases, the public mood of the holiday in question generally shifts to “happy” bins (more red in higher valence) and away from "sad" bins (more green in lower valence). In Brazil, the mood also shifts to more "calm" bins during Christmas week (more red in lower arousal), and in Indonesia it also shifts to neither "in-control" nor "controlled" bins during the Eid al-Fitr week (more red in mid dominance). More detailed characterization of eigenmoods for each country is provided in Supplementary Material (Supplementary Methods S3-5, Fig. S12-13).

Fig. 4, column B, shows all weeks in the data projected onto the selected eigenmood space of two components for each country. It is clear that in this space Christmas weeks (red diamonds) cluster together for the USA and Brazil, and Eid-al-Fitr weeks (green circles) cluster together for Indonesia, demonstrating that the eigenmoods are consistent in different years for each holiday in each country. Fig. 4 column C depicts the linear regression between sex-search volume as calculated before (vertical axis), and mood similarity to the target holiday in the respective eigenmood space (horizontal axis) for all weeks in the data set denoted by black circles in the plot (Methods 14 and Supplementary Methods S6). We observe a significant correlation for all countries studied, with $R2 \geq 0.38$ for Christmas in all Christian Countries and $R2 \geq 0.34$ for Eid-al-Fitr in all Muslim Countries, irrespective of hemisphere (Supplementary Table S12). Thus, in Christian countries we can say that the more the public mood of any week resembles the Christmas eigenmood, the larger the volume of observed sex-searches tends to be. Similarly, in Muslim Countries the more public mood is similar to the Eid-Al-Fitr eigenmood, the larger is the volume of sex-searches. In the case of both Muslim Countries studied (Indonesia and Turkey), there is a striking result pertaining to Ramadan: those 4 weeks (4 lowest green triangles in Fig. 4C, bottom right, for Indonesia), have the lowest sex-search volume by far in the data, consistent with the period of abstinence that marks Ramadan (see also Fig. 2B, Fig. 2D). The public mood during these weeks of Ramadan is also quite distinct from the Eid-al-Fitr mood (horizontal axis in Fig. 4C, bottom right), but, becomes more similar the closer the week is to Eid-al-Fitr in time; and as the mood becomes closer to the Eid-al-Fitr mood as Ramadan approaches its end, the sex-search volume also increases. Naturally, due to the low, outlier sex-search volume during Ramadan weeks, the linear regression is much stronger if those weeks are removed, with $R2 \geq 0.64$ (Supplementary Table S12).

Thus, not only there are specific moods associated with Christmas and Eid-al-Fitr, the eigenmoods that best characterize these holidays significantly correlate with increased interest in sex throughout the calendar. This is true in all countries studied, in both hemispheres and cultures. Moreover, and although these moods, occur at different times in different cultures, they seem to be similar in essence and, in general, the "happier" the mood, the more it associates with sex interest.

**Discussion**

Taken together, our analyses provide strong converging evidence for the cultural hypothesis: human reproductive cycles are driven by culture rather than biological adaptation to seasonal cycles. Furthermore, the observed peaks of interest in sex occur around family-oriented religious holidays, across different hemispheres and cultures, and the measured collective mood on these holidays correlates with interest in sex throughout the year, beyond these holidays. This correlation suggests that the cultural driver of reproductive cycles depends on the collective mood of human societies, though establishing such causality warrants further study. It is also worth noticing that while other major holidays in each country lead to increased sex-search volume (e.g. Eid-al-Adha), not all holidays exhibit this effect (e.g. Easter and Thanksgiving), suggesting that certain holidays have unique eigenmoods which lead to increased interest in sex at the population level. Thus, specific mood states−typically happier, calmer, and neither in-control nor controlled−are associated with interest in sex, and this collective emotion is universal and maximized during cultural celebrations such as Christmas and Eid-al-Fitr. The fact that the Muslim holidays do not follow a solar calendar, with the interest in sex varying according to the religious calendar, provides additional support for the cultural hypothesis.

It is clear from this work that culture (particularly the religious calendar) best explains the pattern of sexual interest. Naturally, it is important to stress that if collective mood states drive interest in sex at the individual level, there must ultimately be a common biological response to the cultural, contextual driver. Several hypotheses can be entertained − though not adaptation to seasonal cycles. For instance, some studies show that depressed people lose interest in sex and that “happy moods,” such as those uncovered for Christmas and Eid-al-Fitr, are usually more conducive to sexual arousal (26,27). Increased food consumption has also been shown to have a relationship with sexual maturation and interest (24,25), however, we do not see similar increase in sex-searches during other holidays associated with high food intake, such as Thanksgiving in the USA or Easter in France. And given the children and family focus of both Christmas and Eid-al-Fitr, it is reasonable to consider psychological and symbolic triggers to the observed behavior, but the neurological and biochemical pathways involved in such responses are as yet unknown.

That the culturally motivated surge in sexual interest can be so easily anticipated and measured has implications for public health and policy. Hospitals should be prepared for an increase in STD testing and possibly even abortions in the weeks following such holidays and when the corresponding collective mood is observed at other times of the year.

Overall, this work emphasizes the need for more world-scale studies and the importance of a better understanding of global collective behaviors at the level of individual countries. These will enable better-informed decisions and the more effective fine-tuning of policy towards the distinct needs of countries, cultures, and communities.

## Methods

### 1. Google Trends Data

Google Trends (GT) provides a time series index of the search volume of a given Google query (10). GT allows for searches in a selected region (country, state, city, etc.) and for a selected time range starting in January 2004 for most countries. Google normalizes the resulting query index relative to the total amount of query volume for a search term in the chosen area, per week, so that the maximum query share of the time series is set to be 100. GT queries are also broad matched, meaning that queries such as ”sex videos” are counted in the calculation of the query index for ”sex”.

### 2. Country Selection and Categorization

We considered all countries for which GT is available and for which a search for “sex” had a least two contributing cities and had enough time points to analyze at least four consecutive holiday seasons (Christmas and Ramadan), thus starting at least in the last week of 2009. This was the case for 129 countries in all continents. In the paper these countries are identified either by their name or by the country code, as in Supplementary Table S2.

Countries were categorized according to their major religion and geographical location (continent and Northern or Southern Hemisphere according to Wikipedia) and this categorization is referred to “identification” in the main manuscript. A country was considered “culturally Christian” when at least half of its population identified as Christian (Catholic, Protestant, Orthodox, or other) (13). A country was considered “culturally Muslim” when at least half of its population identified as Muslim (14). A country was labeled as “Other” when the majority of its population didn’t identify as either Christian or Muslim. In the case of countries that have parts of their territory in both hemispheres, we used the location of the capital as the deciding criteria. Out of the countries identified as Christian, eleven have a majority that follow either the Russian or Serbian Orthodox Churches (namely: Belarus, Bosnia and Herzegovina, Bulgaria, Georgia, Macedonia, Moldova, Montenegro, Serbia, Slovenia, Russia and Ukraine). In ten of these countries (Bulgaria being the exception), Christmas is celebrated in early January (of the Gregorian Calendar) and they could have been labeled as Other for the proposes of this analysis.

### 3. Searches for “sex”

We downloaded the time-series corresponding to searches for ”sex” for each of the available countries in GT as long as they had at least two cities contributing data, and had enough time points to analyze at least four consecutive holiday seasons (Christmas and Ramadan), thus starting at least in the last week of 2009. Supplementary Table S2 shows all countries included in the analysis. Because Google does not provide the absolute number of searches and we do not have access to the normalization algorithm, all the analyzed data is relative to the total search volume and it has been noticed by ourselves and by others that there is some variation the output GT provide, from week to week. To limit this variation all of the analyzed data was downloaded on the same week.

For a subset of 50 countries (on all continents) we downloaded GT data for 2 search queries: (1) for the term “sex” and (2) for its translation in the local language. We compared the volume of searches between the two queries and calculated their correlation over time. Supplementary Table S1 shows the 25 countries and languages that retrieved a sufficiently significant search volume in the local language to support our analysis. We then calculated the “Search Volume Ratio”, as the number of searches for “sex” divided by the number of searches for the corresponding translation. We also calculated the Correlation between the two time series (“sex” and the translated word) as the Pearson´s R.

GT also provides and ranks the top words associated with the search term and these are also shown on Supplementary Table S1.

**4. Centered Calendars**

Data were organized into yearly “calendars” centered around the holidays of interest in order to better compare time series across cultures, and to create better summaries of averaged yearly time-series. Five “yearly calendars”, or sets, were constructed:

1) The first, a “Civil Calendar” starts on the first week that includes January 1st and ends on the following December 31st.

2) The second was centered around the weeks that contain Christmas. In this paper we refer to it as the “Christian Calendar”.

3) The third was centered around the weeks that contain the Eid-al-Fitr celebrations. In this paper we refer to it as the “Muslim Calendar”.

4) The fourth was centered around June 21st and is referred to as the “June Solstice Calendar”;

5) The fifth was centered around December 21st and is referred to as the “December Solstice Calendar”.

Each week of each calendar was given an index ranging from 1 to the maximum number of weeks in that year. The first week GT indexes starts at the Jan 1 2004, so all remaining weeks will start seven days from this first index. In our centered calendars, the week containing Christmas and the solstices becomes week 26 and the week containing Eid-al-Fitr becomes week 25. This is because both the “Civil”, “Solstices” and ”Christmas” calendars follow the Gregorian Calendar with 52.177457 weeks per year, but the ”Muslim Calendar” follows a lunar calendar with 29.53 days per month, leading to 354 or 355 days per year. Since the “Muslim Calendar” is consistently shorter than the solar year, it shifts with respect to the Gregorian calendar, necessitating the removal of these extra weeks as they contained no major event or holiday. Thus, Christmas was specified as week 26 in a 52 week calendar (starting from week 1), and Eid-al-Fitr as week 25 in a 50 week calendar. Occasional exception weeks were dropped from analysis if they did not fit into these calendars, without greatly altering the analysis; see Supplementary Tables S4-6 for the complete list. Supplementary Figure S5 shows the plot of all countries, centered around the weeks that contain Christmas, Eid-al-Fitr or January 1st, averaged according to their cultural identification (see above).

**5. Country Classification from sex-searches**

If sex searches correspond to countries' self-reported religions or locations (as described in the Country Selection and Categorization section), we can use sex searches as a feature to classify countries. Here we describe the process by which sex searches were used to measure a country's response to events: Eid al-Fitr, Christmas, the December Solstice, and the June Solstice. These responses were used to evaluate sex searches as a feature in a classification task. The centered time series described before were calculated for all countries in Supplementary Table S2. For each country we obtained between 4 and 9 yearly time series for all years for which data is available. These yearly time-series were averaged in five different ways per country: one following the civil Gregorian calendar, one centered on Christmas week, one centered on Eid- al-Fitr week, one centered on June 21st, representing the June solstice, and the last centered on December 21st, representing the December solstice. Average yearly time-series were created by first normalizing the data by year, such that the highest valued week each year was given a value of 1, and other weeks were expressed as a proportion of that maximum, in order to correct for bias towards years with more searches. To identify weeks with peak sex-search behavior, z-scores for each of these averaged time series were calculated as

$$z=(x-\mu)/\sigma$$

where $\mu$ is the mean and $\sigma$ is the (population) standard deviation

We then pursued a simple classification of countries according to their behavior on the Christmas and Eid-al-Fitr weeks. When the averaged Christmas-centered (Eid-al-Fitr-centered) time-series for a country yields $z > 1$ on the Christmas (Eid-al-Fitr) week, the country was classified as a Christian Country (Muslim Country). If $z < 1$ for both the Christmas- and Eid-al-Fitr-centered time-series, then such a country is classified as Other. If $z > 1$ for both Christmas- and Eid-al-Fitr-centered time-series, the country was culturally associated with largest z. Results can be seen in Supplementary Table S7. A similar procedure was followed to compare countries according to geographical location. See also Supplementary Methods S1.

**6. Birth Data**

There are biases and problems with birth data. This data is particularly uncommon in Muslim and Southern countries and is further confused in Muslim countries both by the fact that religious events do not follow the solar calendar and that registration dates do not accurately match actual birth dates (see Supplementary Materials Fig. 6). Nevertheless, if online sex-searches correspond to an actual increase in sexual activity, it should be possible to see an increase in births for countries where good records exist. Monthly birth rates were collected from the United Nations Database (15) (except for South Africa, retrieved from http://www.statssa.gov.za/publications/P0305/P03052012.pdf), See Supplementary Table S9 for data.

For each country, each month was divided by the number of days in the month (February months were divided by 28.25), then each year was normalized to its maximum value. This removes any bias towards years with more births.

To compare monthly birth rates with GT results we were restricted by the time range constraints of both data sets. We only have GT results from 2004 onwards and we rarely have birth data beyond 2012. In Supplementary Table S9 shows the availability of birth data for all countries used in this study.

There is also no increase in sex-searches or September births in Northern countries that do not celebrate Christmas on December 25th (Supplementary Figures S7). In addition, there is independent evidence that, even within the same country, religiously distinct populations−such as the Muslim and Jewish populations of Israel−have different conception patterns that correlate with their religious holidays (16).

**7. World Map**

Countries were color coded according to the z-scores presented in Supplementary Table S7. The World Map was built using the online tool: http://paintmaps.com, after normalization.

**8. ANEW**

The sentiment in tweets was quantified according to the Affective Norms for English Words (ANEW) lexicon (17,18). The ANEW assigns a number between 1 and 9 along three dimensions to 1034 words. These dimensions are arousal (a), dominance (d), and valence (v). The scores were determined though a survey as the mean score participants assigned each word. The valence scores correspond to whether (from 1 to 9) the word made participants feel sad to happy, arousal from calm to excited, and dominance from controlled to in-control For example, the word “laughter” has a valence score of 8.5, while “leprosy” has a score of 2.1. A basic translation to Spanish and Portuguese was performed through Google Translate and refined by speakers.

**9. Twitter Data**

The source of the twitter data used comes from IU’s twitter garden hose feed, a 10% sample of all tweets. Geo-location data in combination with shape objects (29) allowed the country from which a tweet came to be determined for many tweets. We focus on tweets collected between September 2010, when the collection stabilized, and February 2014, when the tweet collection dropped, complicating homogeneous analysis of the data. We analyzed seven countries that yielded a sufficiently large number of tweets per week (about ten thousand): Argentina, Australia, Brazil, Chile, Indonesia, Turkey, and the USA. This includes countries in both hemispheres, both culturally Christian and Muslim, and with both English and Other official languages. Individual country’s tweets are only examined after their collection had stabilized, starting in September 2010 for the US, Australia, and Chile; May 2011 for Indonesia, Brazil and; June 2011 for Argentina, and September 2011 for Turkey. Days were defined according to Greenwich Mean Time, and weeks from Sunday to midnight Saturday. The overall number of weekly collected tweets are shown in Supplementary Fig. S8, ranging from nearly a million scored tweets per week from the USA and Brazil, to only about ten thousand scored tweets from Turkey and Australia. The proportion of scored tweets to all collected tweets is usually quite small, usually below 5%.

An individual tweet’s sentiment score was determined by finding all words within the tweet that matched the ANEW lexicon, and taking the average of their scores in each dimension. In the case that multiple languages were matched, the scores from the language with the most matched words were used. In case of a tie, the average scores over the tying languages were calculated. To better find the actual sentiment during the holidays without generic seasonal greetings, we don't score words if they appear in generic holiday greetings, such as "happy holidays", and we remove the ANEW words Christmas and Valentine from the lexicon entirely. The list of holidays whose greetings we removed were collected from http://www.officeholidays.com/. The complete list of phrases we removed from score calculation is included in Supplementary Table S13.

**10. Mean Sentiment Correlations with Sex-Search Volume**

To see if sentiment in tweets correlates with sex search volume we computed the ordinary least squares estimate of a multiple linear regression for each country, using the time series of mean tweet sentiment each week along the three ANEW dimensions as independent variables, with the weekly volume of sex searches as the dependent variable. To compute the weekly mean sentiment time series for ANEW dimension, we first calculated the mean tweet sentiment score for each day and then calculated the mean sentiment of the week such that each day has an equal weight in the weekly average.

**11. Singular Value Decomposition for Eigenmood Analysis**

Aggregating all sentiment in tweets into a mean value discards information in the distribution of sentiment across tweets. Therefore, we use binned distributions of sentiment across tweets in the following analysis. We focus on a 25-binned distribution of tweet sentiment between the lowest and highest possible ANEW score as a moderately-grained distribution, with fine enough resolution to capture some detailed structure while aggregating an adequate number of tweets per bin, 400 on average for a collection of $10^4$ tweets.

We applied a singular value decomposition (SVD) (20) to the binned distribution of ANEW scores over time. Our matrix M has columns representing bins, and rows representing weeks. The left and right singular vectors then have an interpretation as the "eigenbins" and "eigenweeks" respectively. We will also refer to the singular vectors as components. The first component explains the vast majority of the variance, and is similar to the base distribution of the language, as expected from the Brown corpus, shown in (19). The second component explains a trend over time, while further components correspond to other fluctuations, including yearly variations for holidays. For more information see also Supplementary Methods S3.

**12. Data Reconstruction**

To analyze how sentiment varies, rather than its basic distribution in language use, we reconstructed the original data without the first component. After recalculating the relative variances, we can remove noise by also removing the components explaining the least variance. Reconstruction, then includes only those components that explain 95% of the remaining variance after the first component is removed. This leaves cyclic patterns and outlier weeks deviating strongly from the baseline sentiment distribution, which we visualize as a heatmap of the distribution over time in. We average over all full years in the data for multiple countries, centered on the week of a strong cultural holiday, to emphasize the change in these distributions, as shown in Supplementary Fig. 14. For more information see also Supplementary Methods S4.

**13. Eigenmood Selection**

To investigate the distribution of sentiment in a country during a holiday, we selected an *eigenmood* composed of the two components that best characterized the valence distribution on the holiday. Supplementary Figure S15 and Supplementary Methods S5. These two components were selected to describe a country's twitter sentiment on a holiday in the following way. First, the average projection of the holiday was found over all years of the data, as well as the standard deviation. The two eigenweeks with the highest absolute value of the holiday's projection minus its standard deviation were selected. The standard deviation is calculated over very few points, but subtracting it from the mean allows us to know how small the magnitude of the projected vector we may expect. This way, the mood of the holiday of interest can be expected to have a strong correlation with the selected components and cluster closely together.

**14. Eigenmood correlations to Sex-search volume in target Holidays**

As a measure of mood similarity between weeks in a space defined by a selected eigenmood, we use the dot product between their coordinates in this space (20). See Supplementary Methods M6 for more information.

## References

1. W. Macdowall, et al., Summer nights: A review of the evidence of seasonal variations in sexual health indicators among young people. *Health Education*. **108**(1), 40–53 (2007).
2. D.R. Cummings, Human birth seasonality and sunshine. *American Journal of Human Biology*. **22**(3), 316–324 (2010).
3. F. H. Bronson, Seasonal variation in human reproduction: environmental factors. *Quarterly Review of Biology*. 141–164 (1995).
4. T. Roenneberg, J. Aschoff, Annual rhythm of human reproduction: II. Environmental correlations. *Journal of Biological Rhythms*. **5**(3), 217–239 (1990).
5. K. Anand, G. Kumar, S. Kant, S.K. Kapoor, et al., Seasonality of births and possible factors influencing it in a rural area of Haryana, India. *Indian pediatrics*. **37**(3), 306–311 (2000).
6. U.M. Cowgill, Season of birth in man. Contemporary situation with special reference to Europe and the southern hemisphere. *Ecology*. 614–623 (1966).
7. K. Wellings, W. Macdowall, M. Catchpole, J. Goodrich, Seasonal variations in sexual activity and their implications for sexual health promotion. *Journal of the Royal Society of Medicine*. **92**(2), 60–64 (1999).
8. Worldbank, Completeness of birth registration. [Online http://data.worldbank.org/indicator/SP.REG.BRTH.ZS; accessed 31-March-2015] (2015).
9. S. Beck, D. Wojdyla, L. Say, A.P. Betran, et al., The worldwide incidence of preterm birth: a systematic review of maternal mortality and morbidity. *Bulletin of the World Health Organization*. **88**(1), 31–38 (2010).
10. http://www.google.com/trends/ and https://support.google.com/trends#topic=6248052
11. P.M. Markey, C.N. Merkey, Seasonal Variation in Internet Keyword Searches: A Proxy Assessment of Sex Mating Behaviors. *Archives of Sexual Behavior*. **42**(4), 515-52 (2012).
12. M.L. Levin, X. Xu, J.P. Bartkowski, Seasonality of Sexual Debut. *Journal of Marriage and Family*. **64**, 871-884 (2002).
13. Wikipedia, Christianity by country — Wikipedia, The Free Encyclopedia. [Online https://en.wikipedia.org/wiki/Christianity_by_country; accessed 4-November-2014] (2014).
14. Wikipedia, Islam by country — Wikipedia, The Free Encyclopedia[Online https://en.wikipedia.org/wiki/Islam_by_country; accessed 4-November-2014] (2014).
15. United Nations, WorldWide Births. [Online; accessed 4-November- 2014] (2014).
16. M. Friger, I. Shoham-Vardi, K. Abu-Saad, Trends and seasonality in birth frequency: a comparison of Muslim and Jewish populations in southern Israel: daily time series analysis of 200 009 births, 1988–2005. *Human reproduction*. **24**(6), 1492–1500 (2009).
17. T. Lee, generating the Affective Norms for English Words (ANEW) dataset. [Online https://tomlee.wtf/search/ANEW, accessed 18-June 2010] (2010).
18. M.M. Bradley, P.J. Lang, Affective Norms for English Words (ANEW): Affective ratings of words and instruction manual(1999).
19. I.B. Wood, J. Gonçalves-Sá, J. Bollen, L.M. Rocha, Eigenmood Twitter analysis: Measuring collective mood variation (In Preparation).

20. M. Wall, A. Rechsteiner, L.M. Rocha, Singular value decomposition and principal component analysis. *A practical approach to microarray data analysis*, eds D.R. Berrar, W. Dubitzky, M. Granzow (Kluwer Academic Publishers) pp 91-109 (2003).
21. P.S. Dodds, C.M. Danforth, Measuring the happiness of large-scale written expression: Songs, blogs, and presidents. *Journal of Happiness Studies*. **11**(4), 441-456 (2010).
22. P.S. Dodds, K.D. Harris, I.M. Kloumann, C.A. Bliss, C.M. Danforth, Temporal patterns of happiness and information in a global social network: Hedonometrics and twitter. *PLoS ONE*. **6**(12) (2011).
23. S.A. Golder, M.W. Macy, Diurnal and seasonal mood vary with work, sleep, and daylength across diverse cultures. *Science* **333.6051**, 1878-1881 (2011).
24. S. Moschos, J.L. Chan, C.S. Mantzoros, Leptin and reproduction: a review. *Fertility and Sterility* **77**(3), 433–444 (2002).
25. A.A. Ammar, F. Sederholm, T.R. Saito, A.J.W. Scheurink, A.E. Johnson, P. Södersten, NPY-leptin: opposing effects on appetitive and consummatory ingestive behavior and sexual behavior. *Am. J. Physiol. Regulatory Integrative Comp. Physiol*. **278**, 1627–1633 (2000).
26. M.M. ter Kuile, S. Both, J. van Uden, The effects of experimentally-induced sad and happy mood on sexual arousal in sexually healthy women. *J. Sex Med.* **7**, 1177–1184 (2010).
27. G. Bodenmann, T. Ledermann, Depressed Mood and Sexual Functioning. *Int. J. Sex Health* **19**(4), 63-73 (2007).
28. L.A. Zadeh, The concept of a linguistic variable and its application to approximate reasoning-I. *Information Sciences* **8**(3), 199-249 (1975).
29. Made with Natural Earth. Free vector and raster map data @ naturalearthdata.com.
30. G.J. Székely, M.L. Rizzo, Brownian distance covariance. *The annals of applied statistics*, **3**(4), 1236-1265 (2009).
31. World DataBank. [Online http://databank.worldbank.org/data/; accessed 19-May-2016] (2016).

## Figure Legends

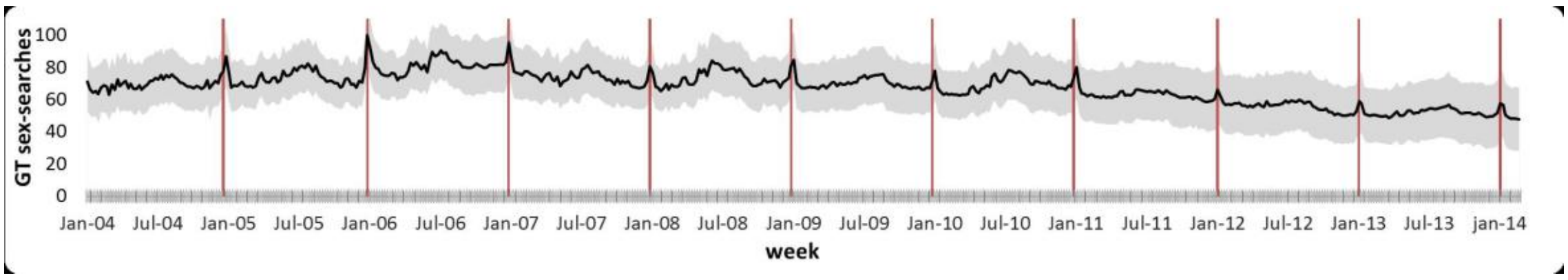

**Fig. 1**. **Weekly queries for the term "sex" for a group of representative western Northern countries.** The black line represents the averaged queries in a 10-year period, obtained from Google Trends, which is normalized by overall search volume. These countries are: Austria, Canada, Denmark, Finland, France, Germany, Italy, Lithuania, Malta, Netherlands, Poland, Portugal, Spain, Sweden and the United States of America. Shaded grey represents the standard deviation. The red vertical line marks Christmas week.

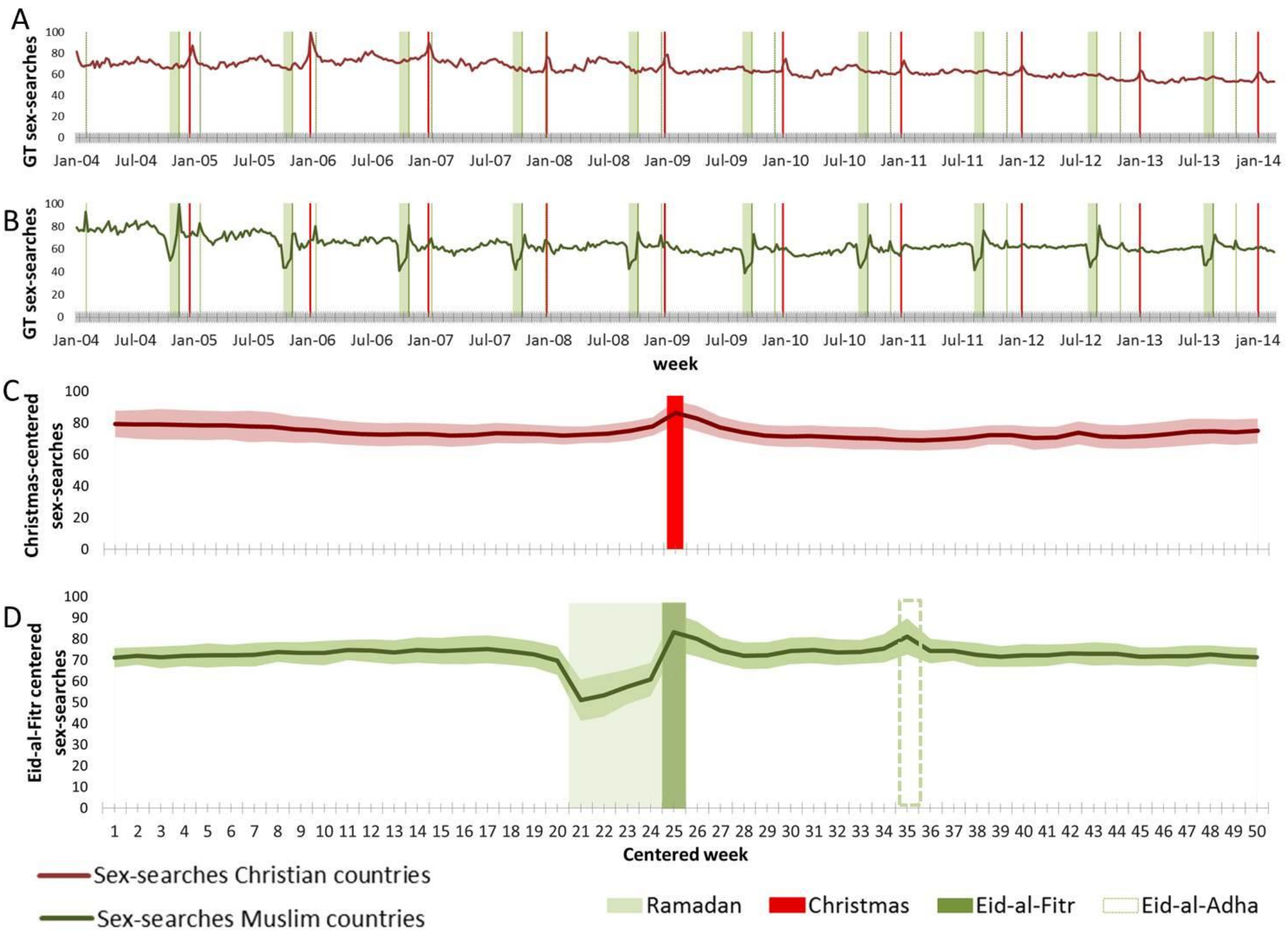

**Fig. 2. Weekly queries for the term "sex" in culturally different countries.** (**A**) Normalized and averaged queries for all available countries identified as Christian (dark red line). (**B**) Normalized and averaged queries for all available countries identified as Muslim (dark green line). (**C**) Searches in all Christian countries centered around Christmas week (26). (**D**) Searches in all Muslim countries centered around Eid-al-Fitr week (25). See Supplementary Table 2 for country identification and availability on GT. The vertical red lines mark Christmas week, the shaded light green area represents Ramadan, with the darker green lines marking Eid-al-Fitr (solid) and Eid-al-Adha (dashed). Shaded areas around the lines in **C** and **D** show the standard deviation.

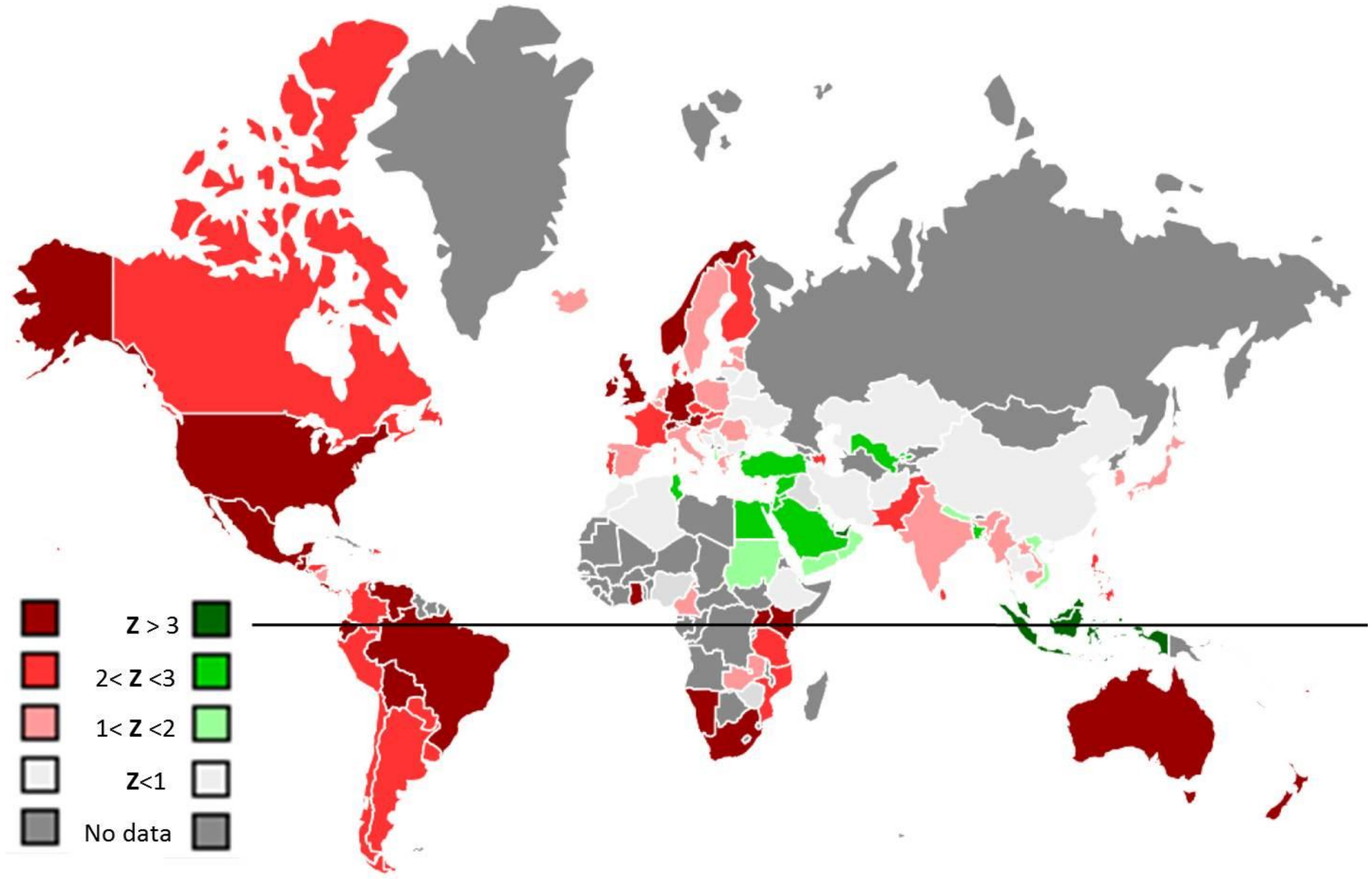

**Fig. 3. World-wide sex-search profiles.** The world map is color-coded according to the z-score of each individual country's sex-search time-series. Shades of red represent a higher z-score (larger increase in searches) during Christmas week (on Christmas-centered data). Shades of green represent a higher z-score (larger increase in searches) during Eid-al-Fitr week (on Eid-al-Fitr ce**ntered** data). White denotes countries with no significant variation above mean in either of these weeks. Dark grey countries are those for which there is no GT data available. Black line represents the equator separating the hemispheres. Built using: http://paintmaps.com.

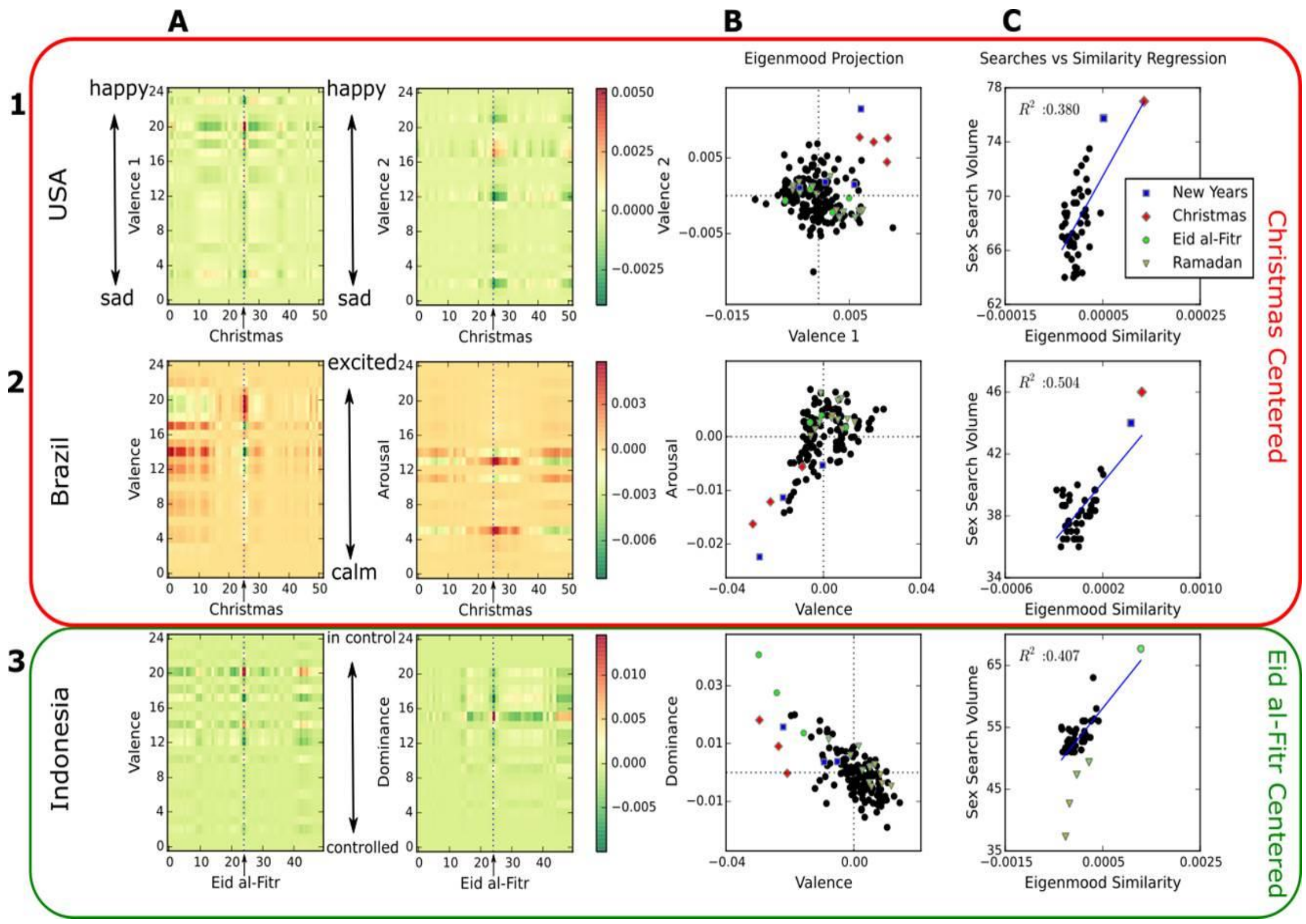

**Fig. 4. Mood distributions and their correlations with sex-searches.** Rows: **1** - USA centered on Christmas, **2** - Brazil centered on Christmas, **3** - Indonesia centered on Eid al-Fitr. Columns: **A** - Heatmaps of sentiment distribution reconstructed from selected eigenmoods. Horizontal axis specifies the week of the centered, averaged year (52 weeks for the Gregorian calendar, 50 for the Muslim Calendar). The dotted line in the center marks the holiday of interest, on week 26 for Christmas, or week 25 for Eid al-Fitr. **B** - Projections of weeks into the space formed by the selected eigenmood components. Each axis specifies the projection of week onto each component that defines the eigenmood. See supplemental materials for more information on component selection. **C** – Linear regressions between sex search volume (vertical-axis) and similarity to holiday center in the eigenmood space depicted in column B (horizontal-axis) for averaged weeks. The weeks of Ramadan are shown with increasing color intensity from more yellow to more green as they approach Eid-al-Fitr. The $R^2$ values for the regressions are 0.380 for Christmas in the USA, 0.504 for Christmas in Brazil, and 0.407 (0.637 without the Ramadan weeks) for Eid-al-Fitr in Indonesia.

**Acknowledgments:** We thank D. Rocha of Proposal Development Services at Indiana University for scientific editing and D. Junk for his work on data processing and collection. R. Correira, A. Gates, A. Kolchinsky, and other members of the CASCI group and CNetS at Indiana University and P. Almeida, M.M. Pita and other members of the S&P group at Instituto Gulbenkian de Ciência for their comments and assistance with this work. This work was supported in part by PTDC IVC ESCT 5337 2012, funded by the Portuguese Fundação para a Ciência e para a Tecnologia (FCT) and by the Welcome DFRH WIIA 60 2011, funded by the FCT and the Marie Curie Actions, both awarded to JGS. Twitter data collection was supported by NSF Award No. IIS-0811994. Google Trends data is publicly available (10). WorldBank data is publicly available (31).

**Data Availability:** Twitter data is subject to contractual limits and the original content of the tweets cannot be shared. However, all quantitative details that support the analysis are included in the Supplementary Materials, as is the dataset created from the Google Trends searches.

**Author Contribution:** J.G.-S. and P.L.R. performed most of the Google Trends data analysis. L.M.R. and J.B. contributed to the design of the Twitter data analysis. I.B.W. performed most of the Twitter data analysis. Methods for Google Trends data analysis was written primarily by P.L.R. and J.G-S, with contributions from the other authors. Methods for the Twitter data analysis was written primarily by I.B.W. with contributions from the other authors. All authors contributed to the writing of the manuscript.

**Author Information:** The authors declare no competing financial interests. Correspondence should be addressed to J.G.-S. (mjsa@igc.gulbenkian.pt) and L.M.R (rocha@indiana.edu)

# Title: Human Sexual Cycles are Driven by Culture and Match Collective Moods

**Authors:** Ian B. Wood[1], Pedro Leal Varela[2], Johan Bollen[1,3], Luis M. Rocha[1,2]*, and Joana Gonçalves-Sá†[2]

**Affiliations:**
[1] School of Informatics & Computing, Indiana University, Bloomington, IN, USA.
[2] Instituto Gulbenkian de Ciência, Oeiras, Portugal.
[3]Wageningen University, The Netherlands.
Correspondence to: † mjsa@igc.gulbenkian.pt and *rocha@indiana.edu

# Supplementary Materials

## Supplementary Methods

## Supplementary Figures

## Supplementary Tables

# Supplementary Methods

## S1. Notes on "misclassifications" for Country Classification from sex-searches

Some of the countries identified as Christian celebrate the nativity according to Julian calendar, with Christmas falling on January 7th or January 14th of the Gregorian calendar. Such is the case of the Christian countries: Belarus, Bosnia and Herzegovina, Georgia, Macedonia, Moldova, Montenegro, Serbia, Slovenia, Russia and Ukraine. Neither of these countries has a national holiday on December 25th nor shows an increase in sex-searches around December 25th. Had these countries been labeled as "Other", the percentage of countries identified as Christian for which we see a significant increase (z-score>1) in sex-searches would have been of 91%. In addition to not celebrating the Christmas on December 25th, some of these countries also have a sizeable percentage of population that self-identifies as Muslim. Such is the case of Montenegro (29%), Macedonia (39%) and Bosnia and Herzegovina (45%).

From the 30 Muslim countries, Pakistan was classified as Christian and 6 other countries didn't make the threshold. Pakistan is highly related to Christmas, probably due to the fact that there is a public holiday on 25th December, which coincidentally celebrates the birthday of Muhammad Ali Jinnah, founder of Pakistan. The other six countries also correspond to the ones for which the quality of the sex-search data was the poorest.

Keeping in mind that we were looking for countries that culturally relate to a Christian or Muslim religious background, all countries that didn't make the threshold to be labelled as either are classified as Other. Unsurprisingly, there are many countries who are originally labelled as Other and end up classified as either Christian or Muslim. European countries, such as the Czech Republic, Estonia and the Netherlands, whose majority does not identify as religious are classified as Christian, most likely due to the fact that these populations celebrate the holiday as well, even if secularly.

## S2. Mean Sentiment Correlations with Sex-Search Volume

As shown in Supplementary Table S9A, there is a highly significant, moderate fit ($R^2 > 0.1$) across all countries, demonstrating a significant correlation between volume of sex-searches and mean sentiment as measured by the three ANEW dimensions. The coefficient of determination is generally stronger for Christian countries than Muslim Countries. Similarly to the GT data, the multiple linear regression models can be improved by averaging sentiment and sex-search volume across years using the 52-week Christmas centered calendar for the USA, Australia, Brazil, Argentina, and Chile, , and the 50-week Eid-al-Fitr centered calendar for Indonesia and Turkey. This smooths out extraordinary events that are picked up by sentiment analysis. The results of this centered-data regression are presented in Supplementary Table S9B. The fit is highly significant for all countries, and improves for all countries, ($R^2 > 0.26$). In every case, valence is yields a positive coefficient, while dominance a negative coefficient; so the happier but less dominant the sentiment expressed by a country, the more sex-searches tend to increase. As far as significance is concerned, t-tests reveal that the valence dimension is most often significant, followed by dominance, with arousal the least likely to be a significant factor.

Interestingly, as shown in Supplementary Table S10, when we computed the ordinary least squares estimate of a standard linear regression on each ANEW dimension independently, we obtained very poor (but significant) goodness of fit, as measured by $R^2$. Therefore, the mean value of each ANEW dimension on its own is a poor predictor of sex-search volume in all countries (with few exceptions such as Arousal in Brazil). We can thus say that mean sentiment correlates with sex-search volume (Supplementary Table S9) but the timeseries of mean weekly values of each ANEW dimension do not yield a nuanced characterization of sentiment correlated with interest in sex.

## S3. Singular Value Decomposition

Singular value decomposition (SVD) is a method by which a matrix can be linearly decomposed into ordered orthonormal components, each explaining as much of the linear variation as possible, after the components that came before it. The SVD of any m × n matrix M of real or complex numbers can represented as follows in Equation 2:

$$M=USV^T$$

Where U is an m× n matrix with orthonormal columns, V is an n× n matrix with orthonormal columns, and S is an n× n diagonal matrix. The columns of U and V are referred to as the left and right singular vectors of M respectively. These singular vectors are eigenvectors of the matrices $MM^T$ and $M^TM$ respectively. The diagonal entries of S, called the singular values of M, are the square roots of the eigenvalues of the matrices $MM^T$ and $M^TM$. By convention, the singular values are ordered from greatest to least. The columns of U form a basis for the column space of M and the columns of V form a basis for the row space of M. The right singular vectors are also known in principal component analysis (PCA) as the loadings of the original variables (bins) onto the new coordinate system. The relative variance explained by each component can then be calculated for each component k as $s_k^2/\sum_i(s_i^2)$ where $s_k$ is the kth diagonal component of S. It is important to note that matrices can be reconstructed with a lower rank by setting elements of S to zero. Typically only the top *l* singular values are kept in order to reduce noise and create the closest rank-*l* approximation of the original matrix[20].

## S4. Data Reconstruction

It can be clearly seen from the data reconstruction averages in Extended Data Fig. 8 and Supplementary Fig. S6 that the distribution of sentiment shifts towards higher bins during holidays, represented by redder high bins and greener low bins on holidays. Christmas stands out in the USA (US), Australia (AU), and Brazil (BR). Eid-al-Fitr stands out in both Turkey (TR) and Indonesia (ID), and in Turkey the beginning of Ramadan is emphasized a few weeks before. The centering performed only looks at weeks within the surrounding cultural year, such that Christmas is week 26 of a 52 week year (starting with a first week 1), while Eid-al-Fitr is week 25 of a 50 week year. Other weeks are averaged in this range according to their displacement from the holiday week (e.g., a week two weeks before the Christmas week in 2012 is averaged with weeks two weeks before Christmas in all other years). This obscures the emphasis on holidays using another calendar, such that Indonesia also has a strong signal on Christmas, but these signals are averaged over multiple weeks when the calendars are misaligned. The heatmaps for all countries centered on all holidays are included in Supplementary Fig. S6.

## S5. Eigenmood Selection and Characterization

The mean value of a holiday's projection on various components for different countries are shown in Supplementary Figures S2 and S3 for Christmas and Eid-al-Fitr respectively, with the two components selected for each country highlighted in red. As described, since the first component corresponds to the basic distribution of sentiment in the language and overwhelms projections because of how much it explains, and the last few components are mostly noise, we only look at the components explaining 95% of the variance after the removal of the first. The second component usually describes a variation over the whole time series of out data, thus it tends to have a large standard deviation.

To better understand how the selected components describe the mood, we define an interpretable linguistic variable[28]. The linguistic variable can take five fuzzy values, "low", "medium-low", "medium", "medium-high", and "high" with membership functions defined over the 25 bins of the original twitter sentiment distribution. These membership functions are shown in Supplementary Fig. S4 and were chosen such that each original bin's membership in all values sums to one, and the area under each membership function is the same.

The response of the linguistic variable to the holiday in each selected eigenmood is shown in Supplementary Figure S5 for the selected relevant holiday for each country. These responses were calculated by reconstructing the distribution bins with only the eigenmood selected for the country and holiday, multiplying the reconstructed bin value by its memberships, and summing over all bins for each linguistic value. These responses can be interpreted as the change from the language's base sentiment distribution on the holiday contributed by the selected eigenmood. The response characterized by the Christmas eigenmood in the USA is an increase in medium-high happiness, with decreases in other levels of happiness, low and medium happiness in particular. How mood changes on a major holiday varies between countries but generally we see that the selected eigenmood describes increases medium-high or high valence on the holidays, with decreases in low, medium-low, and medium valence, as well as lower or more moderate dominance and arousal. The behavior of the dominance mood dimension in the week of Eid-al-Fitr in Indonesia highlights the importance of the more nuanced mood measurement that eigenmoods afford. While the ANEW mean value measurement above suggested a dominance decrease towards a less "in-control" mood, what we have at Eid-al-Fitr is a shift away from the extremes to a collective mood state that is neither very "in-control" nor very "controlled" – coherent with a happier and calmer mood scenario typically found in these holidays for all countries. In other words, during most weeks of the year, there is increased bimodal dominance activity in higher and lower bins (simultaneously high "in-control" and "controlled", respectively), but in the week of Eid-al-Fitr, the dominance mood converges to a mid-level dominance (Figure 4 column A, row 3, dominance panel).

## S6. Eigenmood correlations to Sex-search volume in target Holidays

As a measure of mood similarity between weeks in a space defined by a selected eigenmood, we use the dot product between their coordinates in this space[20]. This measure increases between weeks with similar (positive or negative) projections onto the eigenweeks forming the space, becomes negative with opposite projections, and decreases in magnitude with weeks that are not correlated with the eigenweeks and are thus projected near the origin. Due to the properties, it is important to select an eigenmood that strongly corresponds to a week or weeks of interest, by containing high-magnitude values in the corresponding eigenbins. The similarity can then be expressed as $w \cdot c$ where w and c are weeks projected into the eigenmood, which is equivalently the vector of corresponding weighted eigenbin values. In comparison between weeks and a holiday averaged over years, these vectors are the element-wise averages of the week's projection coordinates over the years. We report results with these averages, but these results are robust to yearly, non-averaged data, as well as different selection criteria for the eigenmoods (for example, allowing a greater number of components). The projection spaces for each eigenmood are shown in Supplementary Fig. S7.

In general, weeks close in proximity in time will be more similar in eigenmood, but certain weeks, often other holidays, more distant in time can have a high similarity in eigenmood to the selected holiday. In the USA, for example, the weeks closest in eigenmood to Christmas are, in order, the week of New Year's Day, the other weeks of December, and the weeks following July 4th, Father's Day, and Memorial Day. National Day in Chile is similar in eigenmood and sex searches to Chile's Christmas. New Year's Day and Christmas in Indonesia are similar to Eid-al-Fitr's eigenmood and high sex searches. In Turkey, weeks in late June, early July, and the week following Eid-al-Fitr are the most similar in terms of eigenmood and sex search volume to Eid-al-Fitr.

To investigate the relationship between a week's similarity in eigenmood to a holiday and the number of sex searches, we perform an ordinary least squares regression between sex searches as the dependent variable, and similarity as the independent variable. Displayed in Figure 4 and reported in Extended Data Table 2 are the results of this regression as well as Brownian distance correlation statistics, a nonlinear measure of correlation[30]. The plots of all linear regressions are included in Supplementary Fig. S7.

There is a fairly strong correspondence ($R^2 \geq .380$) between similarity in eigenmood to Christmas and sex searches in the C countries: the US, Brazil, Australia, Argentina, and Chile. The southern hemisphere Christian countries Brazil, Argentina, and Chile also have a noticeable correlation with Eid-al-Fitr, however, the slope of the regression is negative, implying that the less like the mood during the winter week of Eid-al-Fitr, the more sex searches are conducted.

In Muslim countries Turkey and Indonesia, we were limited by having less Twitter data and fewer tweets that match. However, there are significant correlations between similarity to Eid-al-Fitr and increased sex searches. The linear correlation is reduced compared to Christmas in Christian countries, since over time the weeks of Ramadan become more similar in eigenmood to Eid-al- Fitr, the festival at Ramadan's conclusion, while the cultural pressure is one of abstinence, such that these weeks have unusually low sex searches. In the case of Turkey in particular, the holiday of Eid-al-Adha, or the Sacrifice Feast, also has high sex searches, but is different in eigenmood from Eid-al-Fitr. The positive correlation between sex searches and Christmas eigenmood in Indonesia is likely caused by the sizable Christian population living there and effects due to summer.

Turkey is an interesting case, since it has a very strong negative correlation between sex searches and similarity to Christmas although the response to Eid-al-Fitr is smaller. In part, this may be due to limitations in our data gathering and method application, since our ANEW is only available in English, Spanish, and Portugese. However, we still have a good number of tweets from Turkey, so we look more closely at its eigenmood. The projection of all weeks into its eigenmoods for Christmas and Eid-al-Fitr is shown in Supplementary Fig. S7, which happen to be same in this case. The regressions between sex searches and the similarity of averaged weeks to Christmas and Eid-al-Fitr are shown in Supplementary Fig. S7. The mood associated with Eid is also associated with Ramadan, which emphasizes abstinence. During the weeks of Ramadan, there are much fewer sex searches than usual, although the weeks are not too far different in mood. In addition, there is a separate holiday, Eid-al-Adha, that is associated with a second peak in sex searches, but with a different mood. Perhaps due to Turkey's small Christian population and winter timing, Christmas and weeks like it in eigenmood have low sex searches and averaging over years decreases the effects of holiday traditions (like Eid-al-Fitr) due to misaligned calendars.

## Supplementary Figures

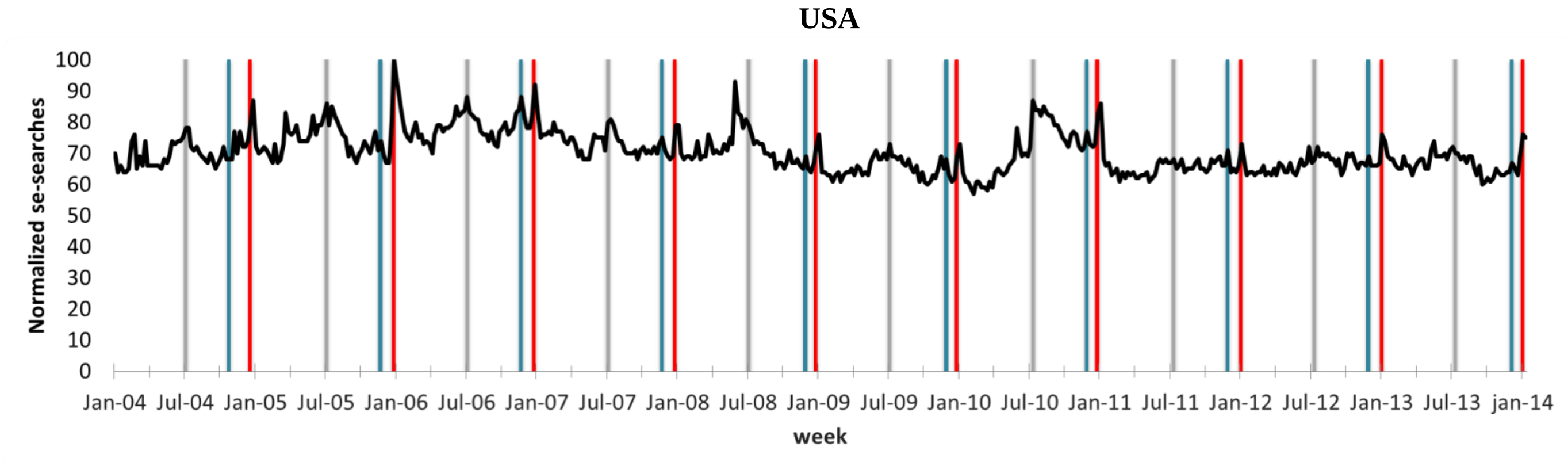

Fig. S1A. **GT query [sex] results for the USA. The weeks containing Thanksgiving day, Christmas and the 4th of July are highlighted in blue, red and grey, respectively.**

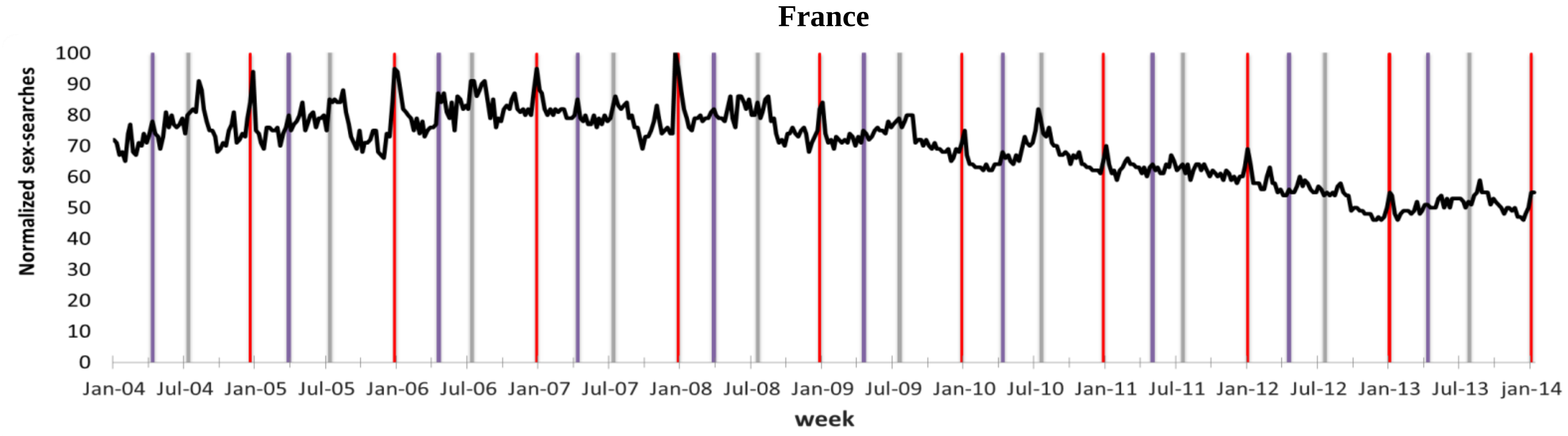

Fig. S1B. **GT query [sex] results for France. The weeks containing Easter Sunday, July 14th and Christmas are highlighted in purple, grey and red, respectively.**

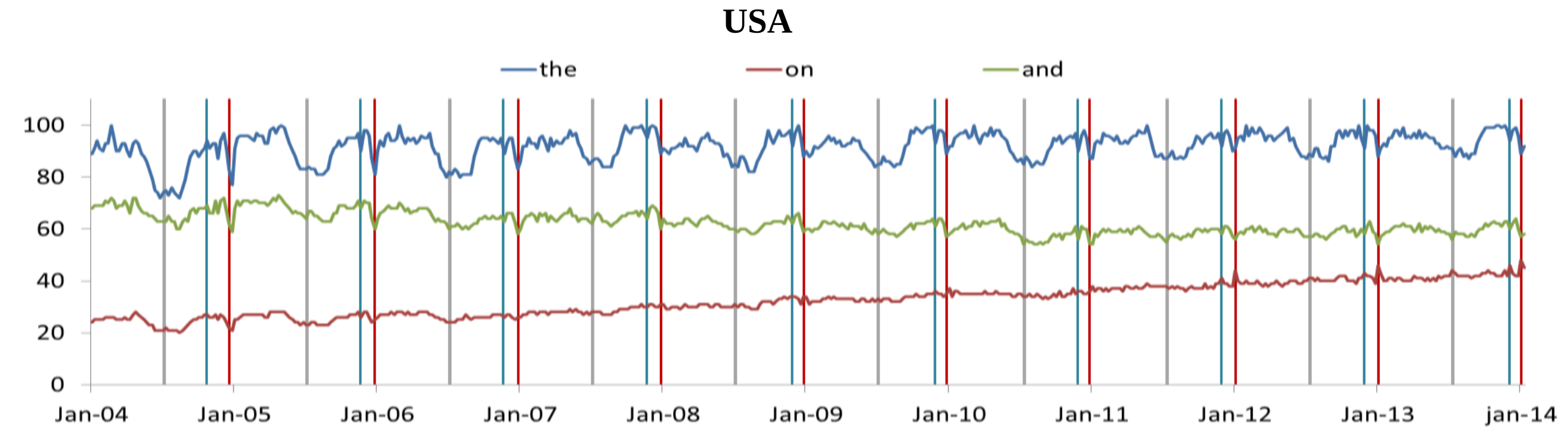

Fig. S2A. **GT queries for "the", "on" and "and", in the USA. The weeks containing Thanksgiving day, Christmas and the 4th of July are highlighted in blue, red and grey, respectively.**

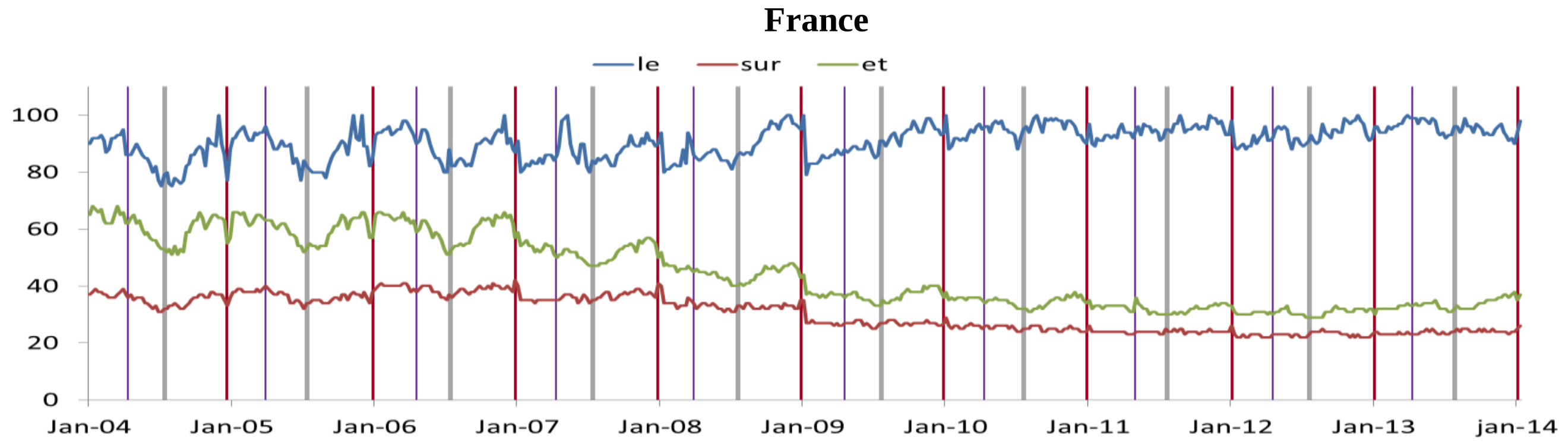

Fig. S2B. **GT queries for "le", "sur" and "et", in France. . The weeks containing Easter Sunday, July 14th and Christmas are highlighted in purple, grey and red, respectively.**

A

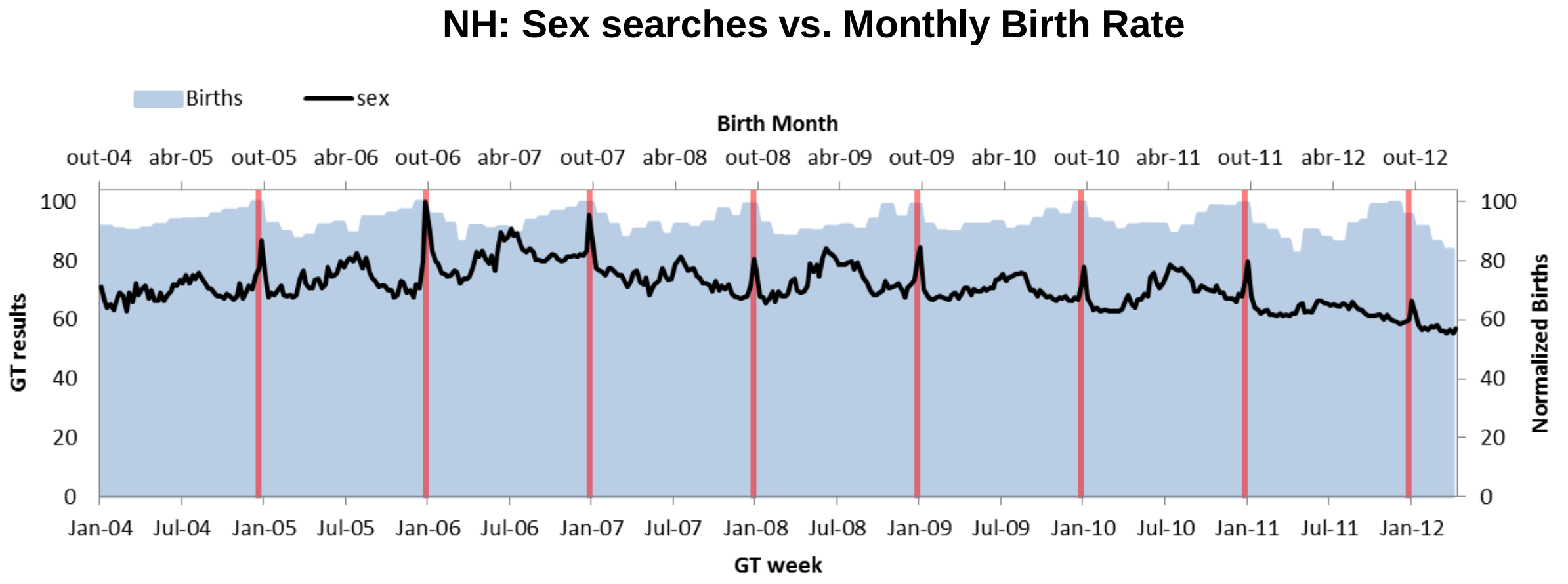

B

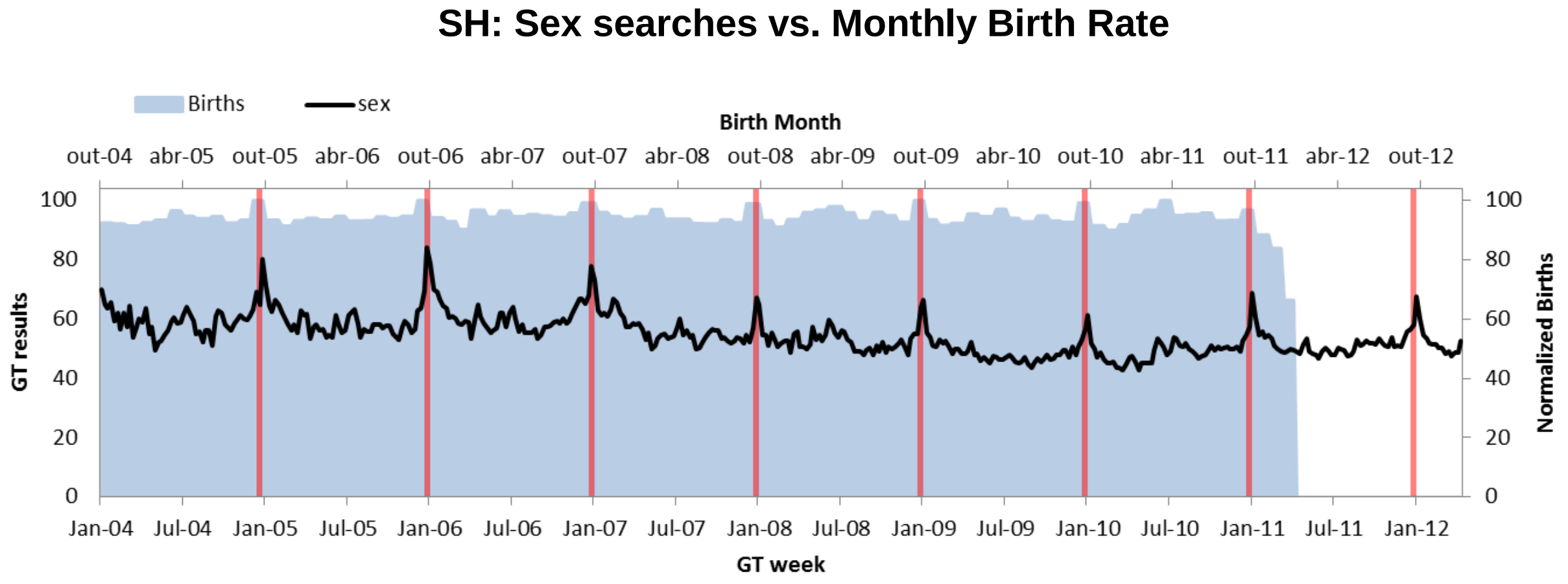

**Fig. S3. Monthly birth data shifted by nine months** (blue shaded area, top and right axis) and weekly averaged Google Trends results for "sex-searches" (black line, bottom and left axis) plotted for:

A) All Western Northern countries for which both birth and GT data exist (Austria, Canada, Denmark, Finland, France, Germany, Italy, Lithuania, Malta, Netherlands, Poland, Portugal, Spain, Sweden and United States of America), also represented in Fig. 1 in the main paper. Births in September are higher than the yearly average in all countries but Lithuania and Sweden, with an average variation of 6%).

B) All Southern countries for which both birth and GT data exist (Australia, New Zealand, Chile and South Africa). Births in September are higher than yearly average in all countries (average variation 5.5%, with the difference being as high at 10% in South Africa and New Zealand.)

Births were shifted nine months to match probable conception month. The red line marks Christmas week.

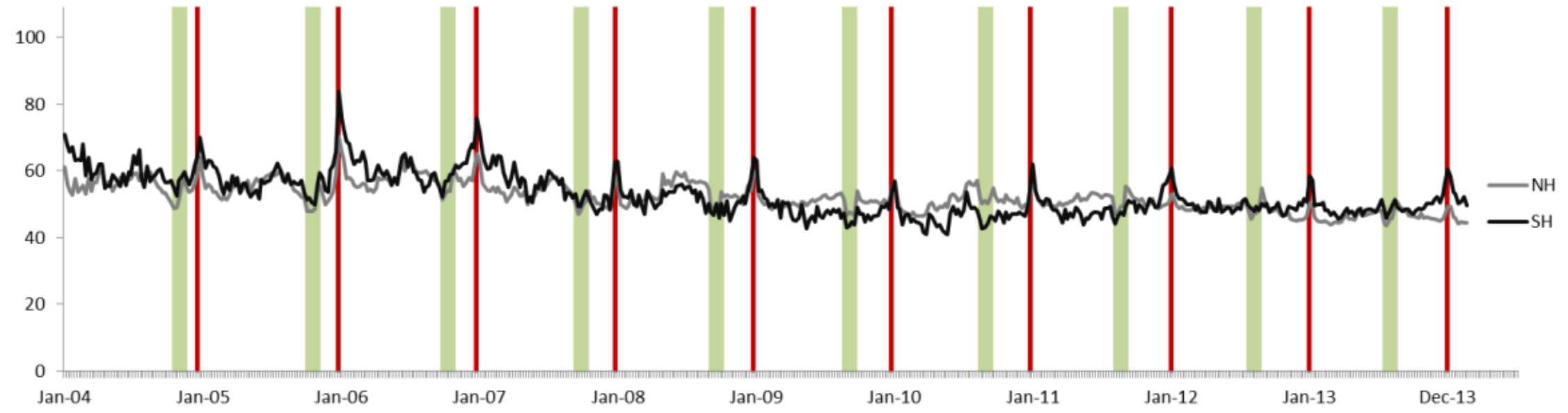

**Fig. S4A. Averaged sex-searches for Northern and Southern countries**. $R^2$ is 0.54 with a p-value of 2E-41. The weeks containing Ramadan and Christmas Day are highlighted in green and red, respectively.

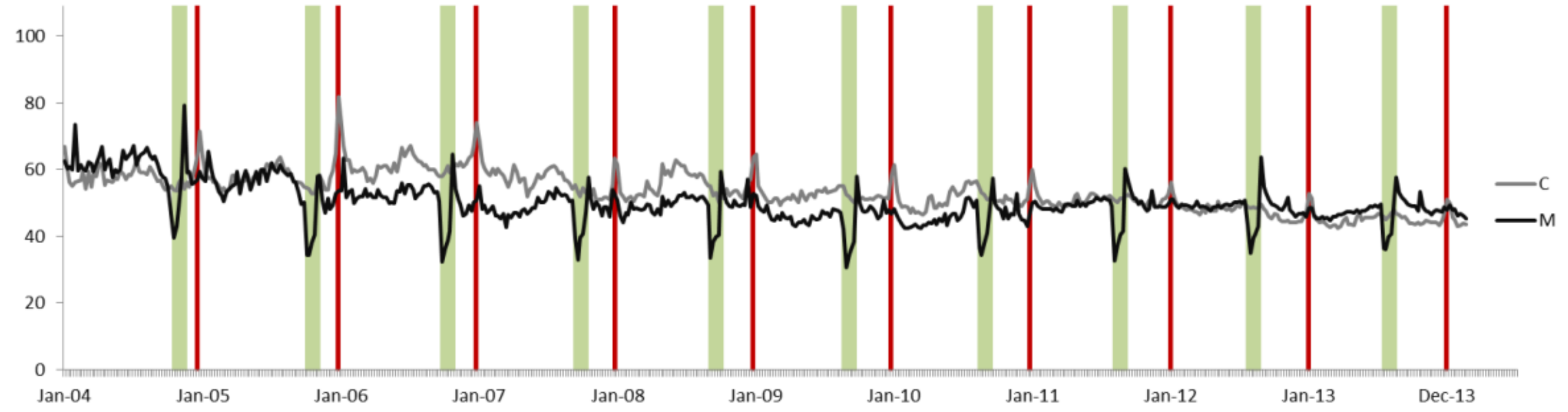

**Fig. S4B. Averaged sex-searches for all Christian and Muslim countries**. $R^2$ is 0.19 with a p-value of 3E-26. The weeks containing Ramadan and Christmas Day are highlighted in green and red, respectively.

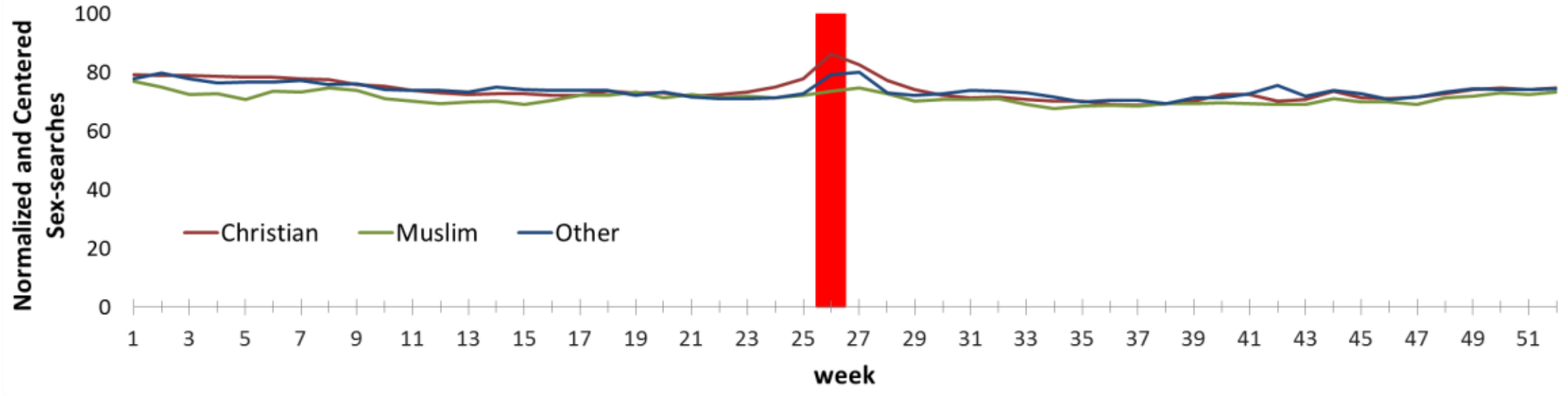

**Fig. S5A. Averaged Christmas-centered results** for the Christian (red), Muslim (green) and Other (dark blue) country sets. The red vertical bar represents the Christmas week, centered on week 26.

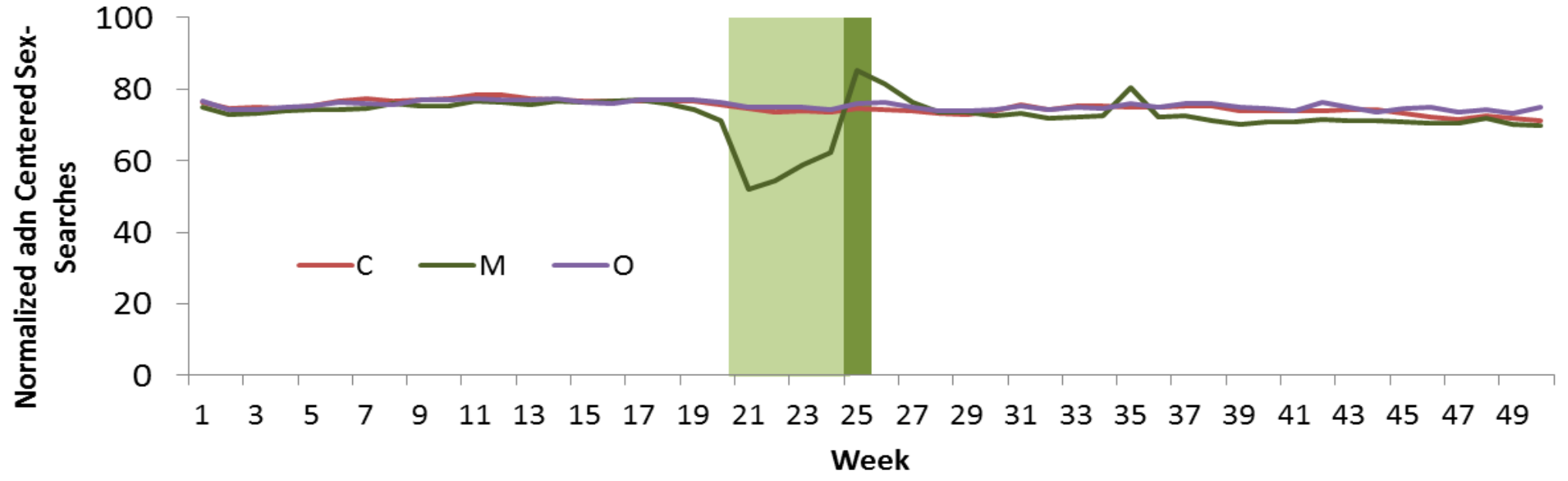

**Fig. S5B. Averaged Eid-al-Fitr-centered results** for the Christian (red), Muslim (green) and Other (dark blue) country sets. The darker green vertical bar represents the Eid-al-Fitr week, centered on week 25. The light green area represents the remaining Ramadan weeks.

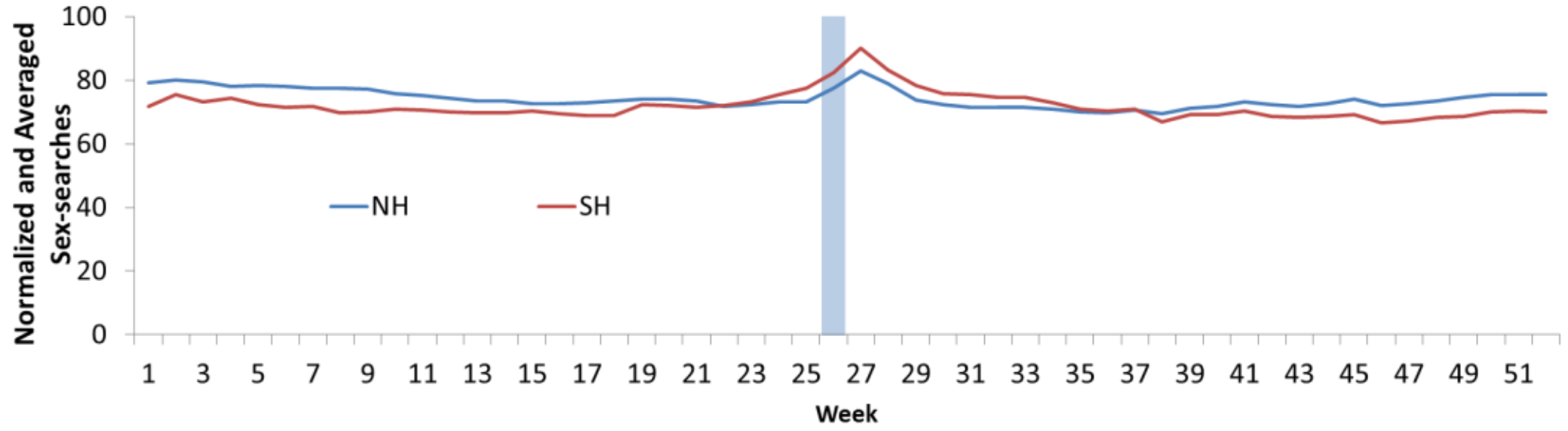

**Fig. 5C. Averaged December Solstice-centered results** for the Northern Hemisphere (blue) and Southern Hemisphere (red) country sets. Light blue vertical bar represents the week of the December-Solstice.

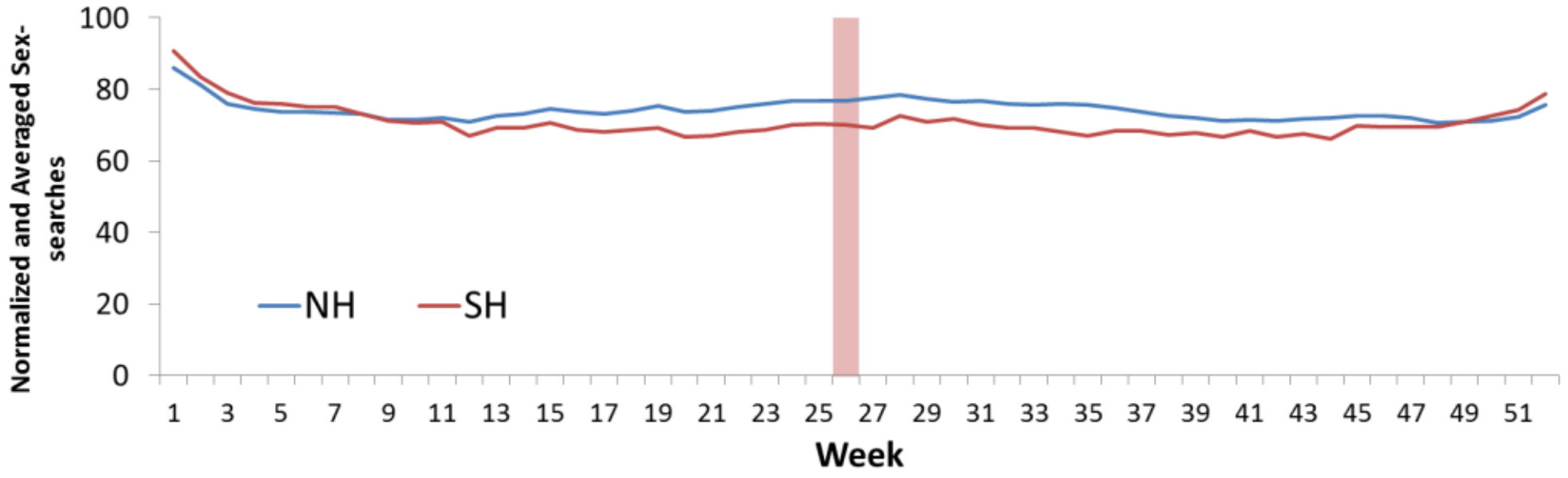

**Fig. S5D. Averaged June Solstice-centered results** for the Northern Hemisphere (blue) and Southern Hemisphere (red) country sets. Light pink vertical bar represents the week of the June-Solstice.

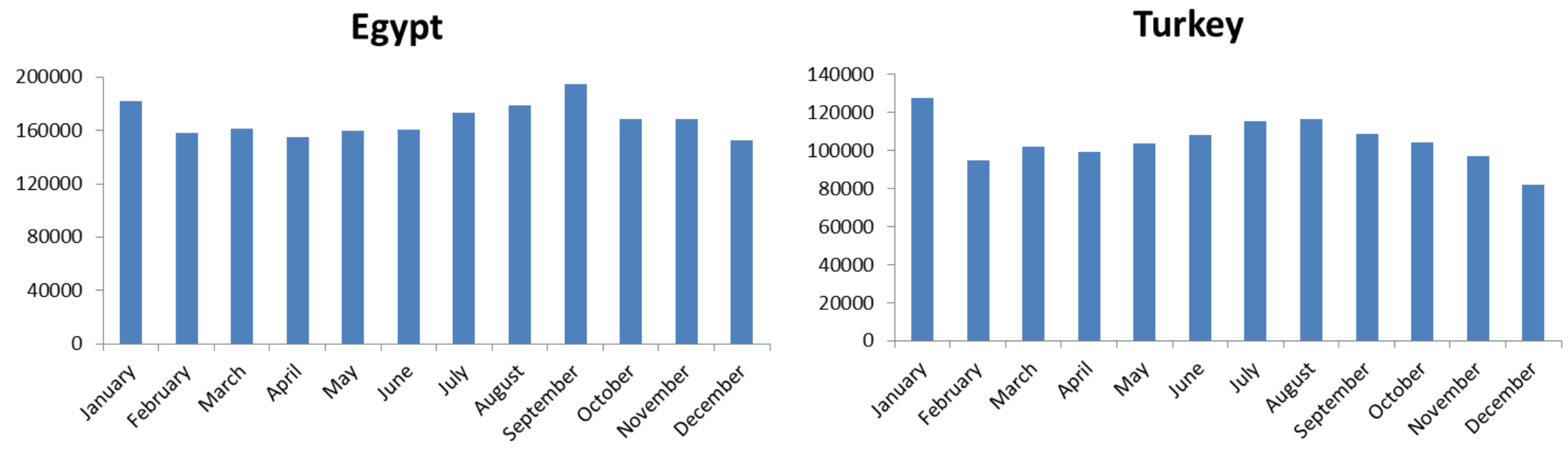

**Fig. S6A. Averaged monthly births (for all available years) for Turkey and Egypt**. In some Muslim countries, as in these examples, birth records are artificially at their lowest in December (in the case of Turkey, 22% below average) and peak in January (in the case of Turkey, 202% above average) , as parents prefer to have their children registered in the New Year.

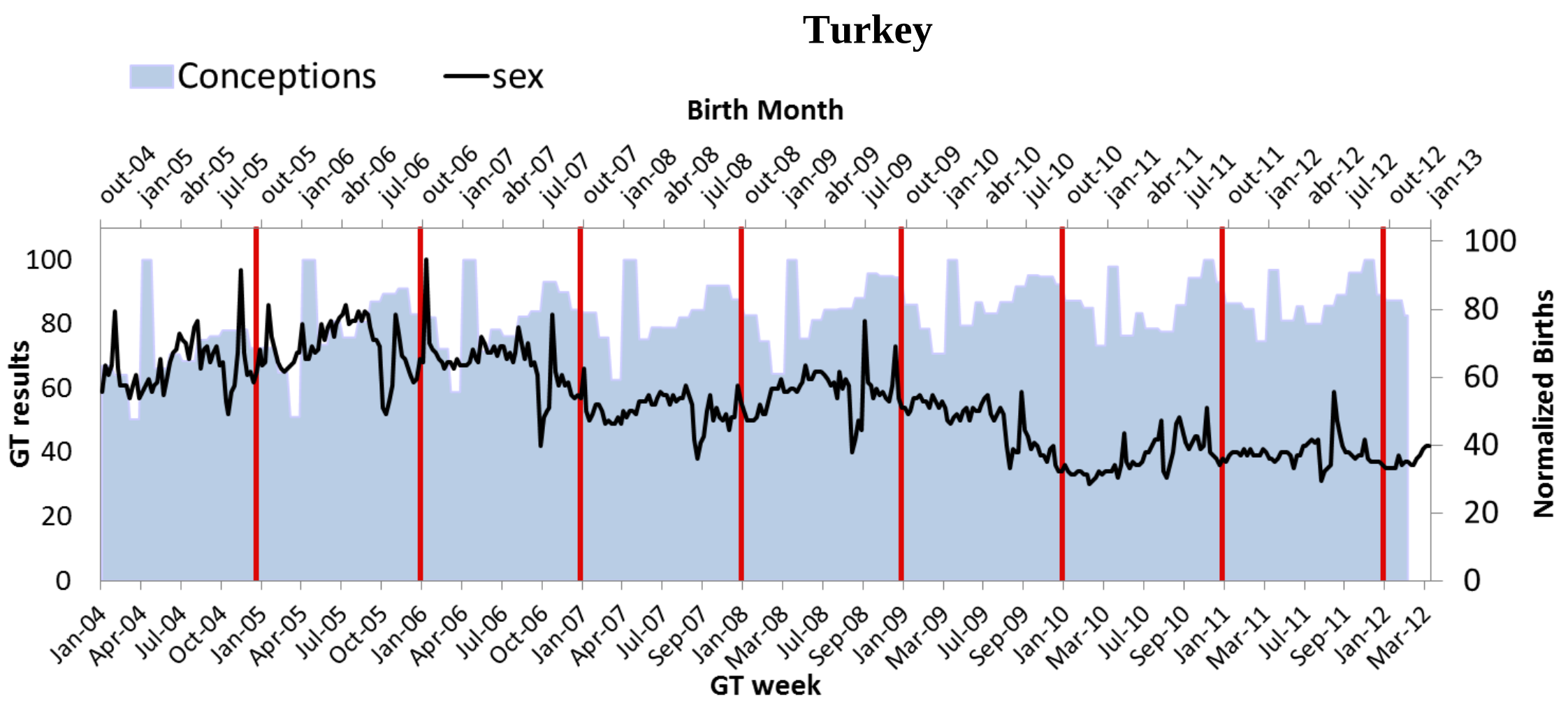

**Fig. S6B** Normalized monthly birth data (shaded blue, top and right axis) and Google Trends results of “sex”-searches (black line, left and bottom axis) for Turkey. Births were normalized so that each year’s maximum becomes 100 and shifted nine months to match with probable conception month. The red line represents Christmas week, which was very close to Eid-al-Ada in 2005, 2006 and 2007. (It is obvious that the major registration peak happens in January of each year and it’s not matched by an increase in sex-searches).

**A) Russian and Serbian Orthodox Countries**

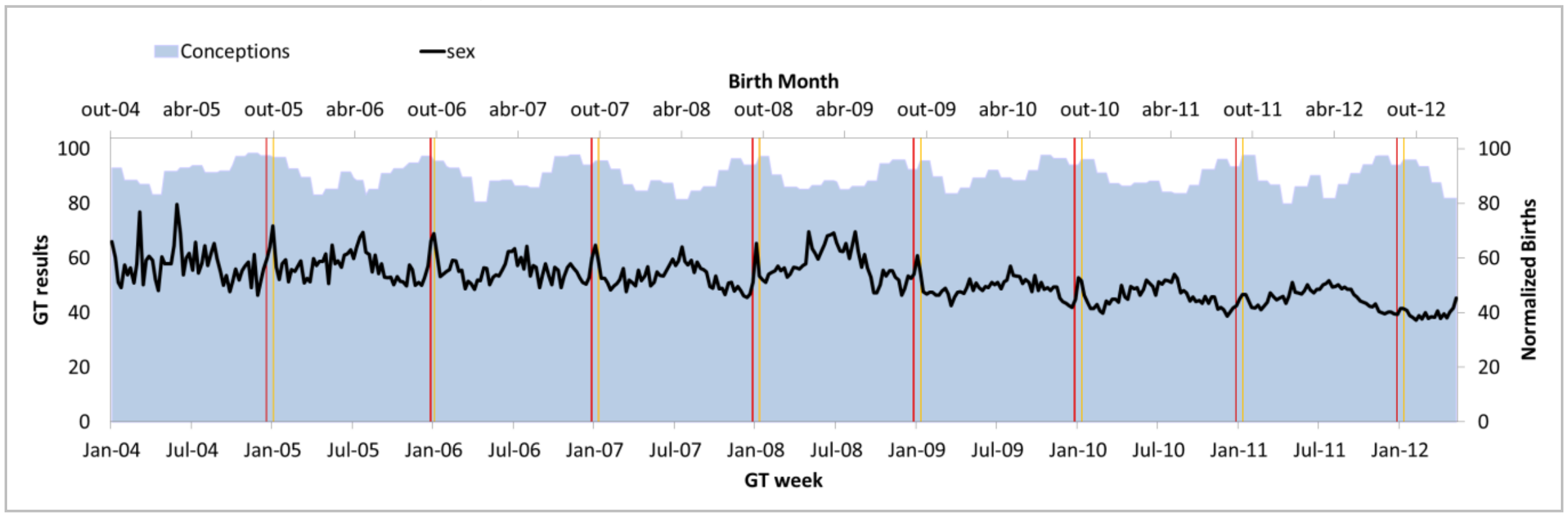

**B) South Korea**

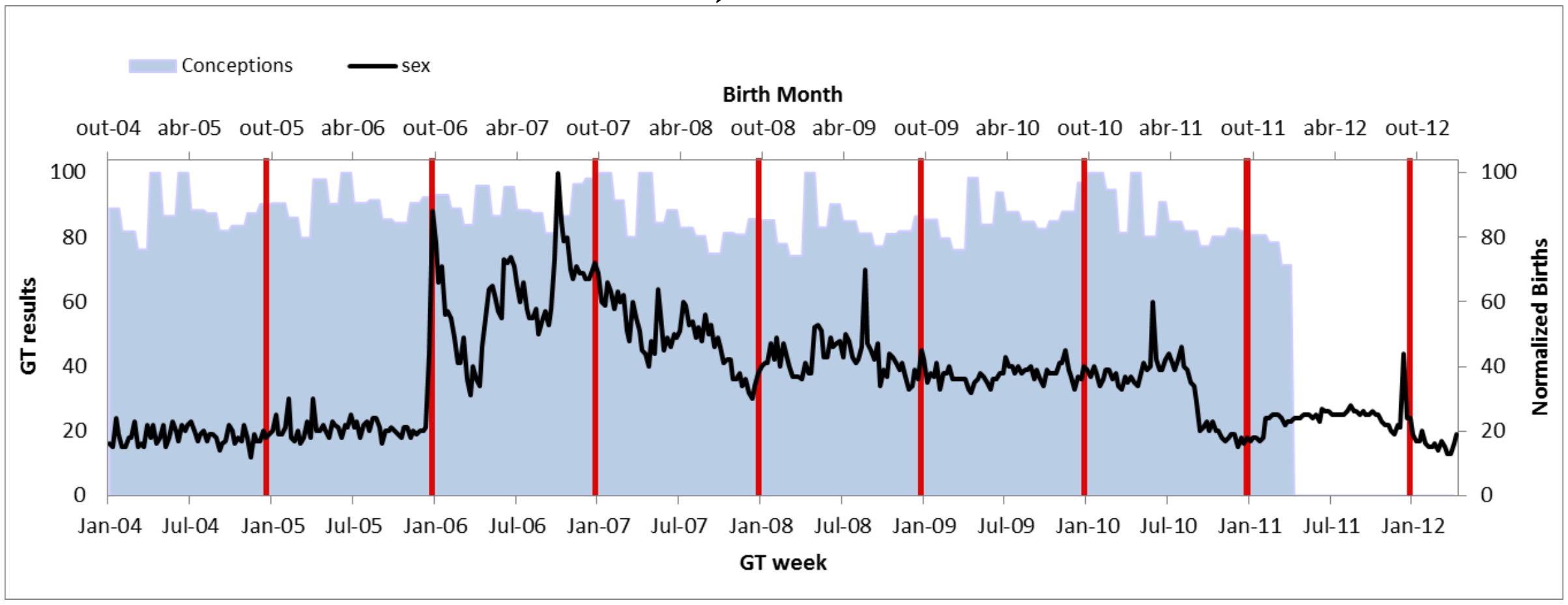

**Fig. S7.** Normalized monthly birth data (shaded blue, top and right axis) and Google Trends results of ”sex”-searches (black line, right and bottom axis) for
A) All Northern and Christian countries for which both birth and GT data exist that Celebrate Christmas on January 6th (Belarus, Bosnia and Herzegovina, Georgia, Macedonia, Moldova, Montenegro, Serbia, Slovenia, Russia and Ukraine). Births in September are higher than yearly average in all countries and the difference is as high as10% in South Africa and New Zealand
B) South Korea, as an example of an Northern Other country, for which both birth and GT data exists.

Births were shifted nine months to match with probable conception month. Vertical lines represent Christmas week with red marking the week of December 25$^{th}$ and orange marking the week of January 6$^{th}$.

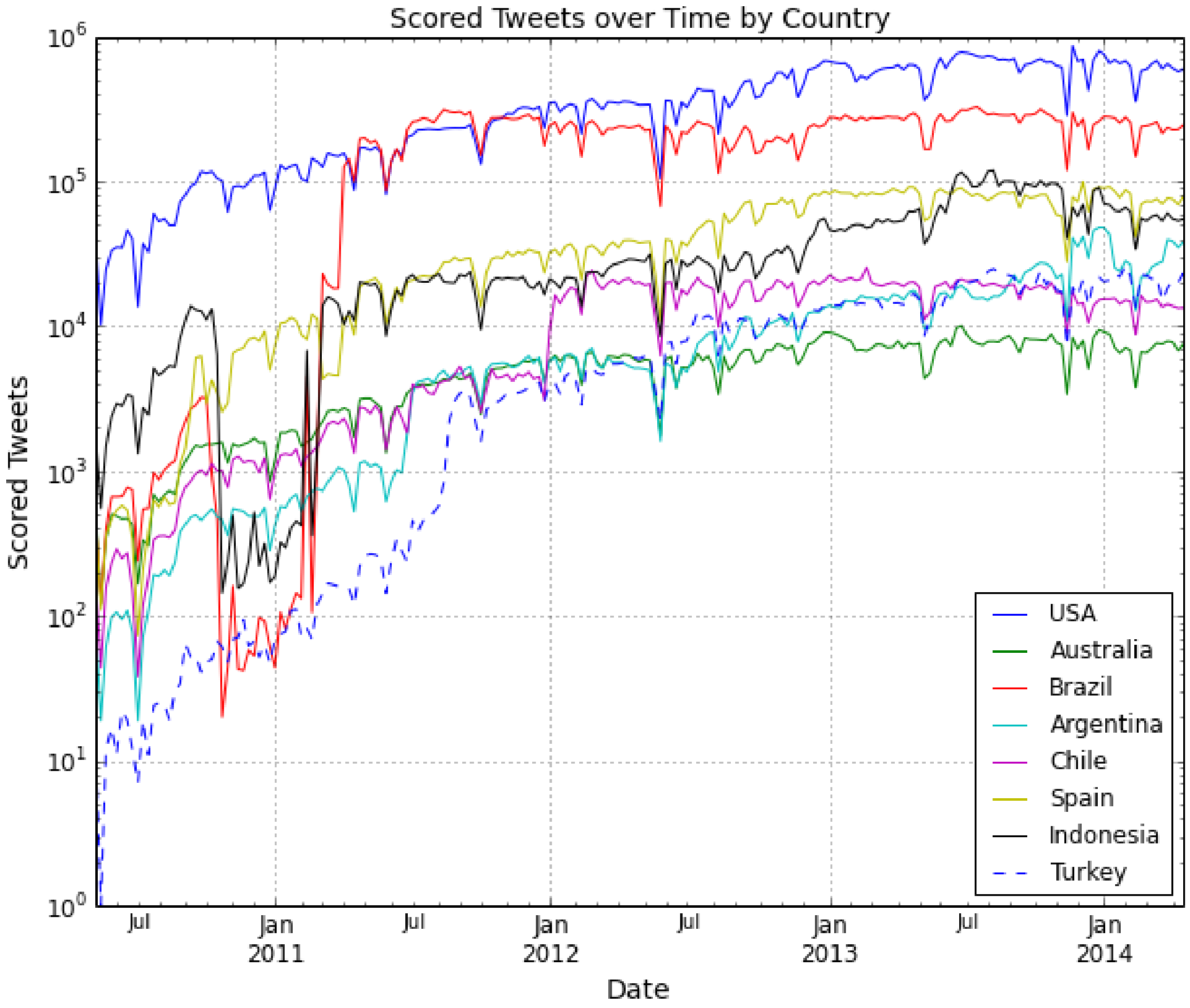

**Fig. S8.** Total number of weekly geolocated tweets matching ANEW for countries selected for *Eigenmood* analysis.

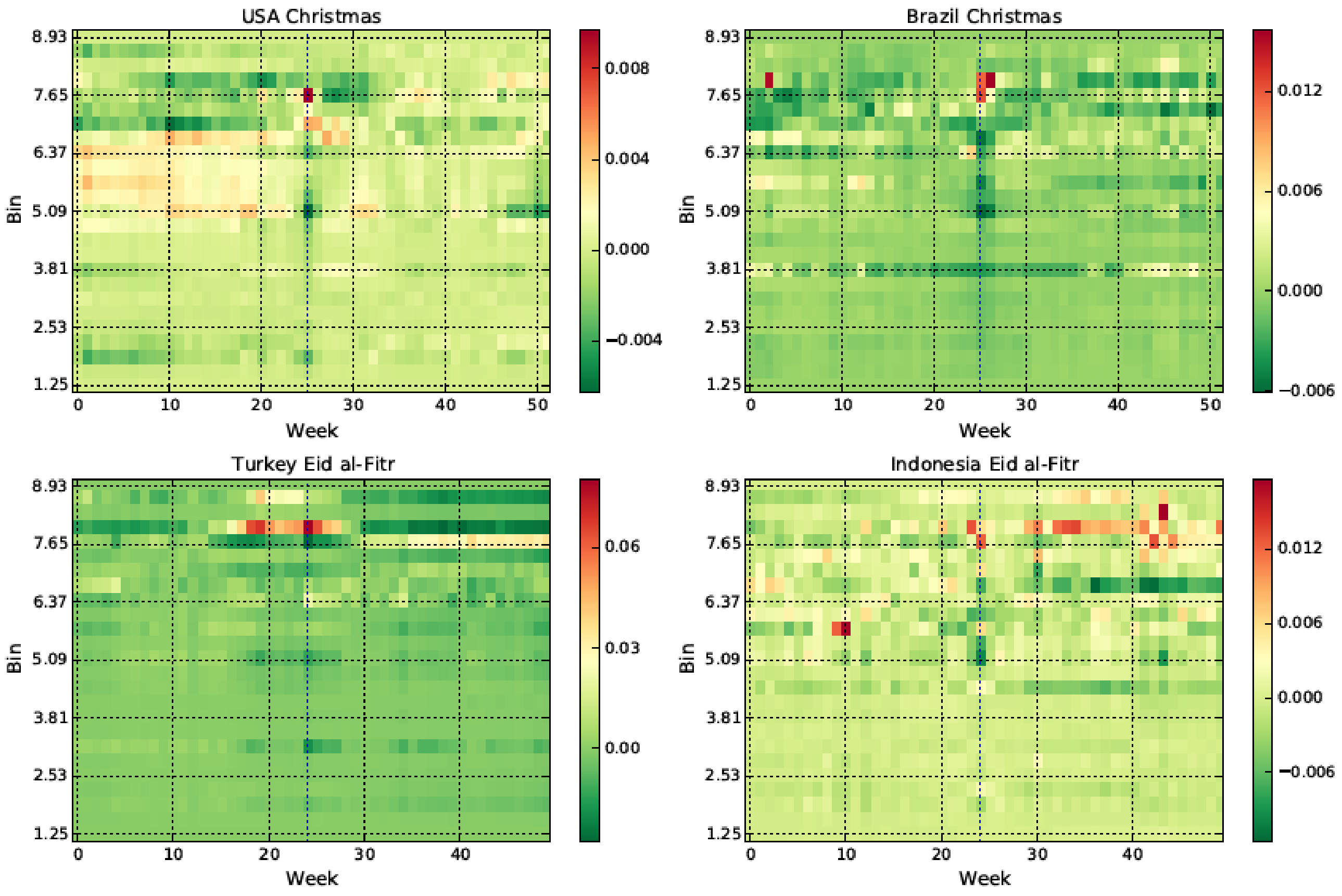

**Fig. S9.** Reconstructed valence heatmaps for multiple countries, centered on cultural holidays. Probability distributions of tweet valence were arranged in 25 bins (y-axis) each week (x-axis) for each country. Years were centered on a chosen holiday, marked by a central, vertical line. These data were averaged over all years, so each cell contains the average probability of a tweet's valence falling into a bin during a week. The data were reconstructed by removing the first component and components explaining less than 95% of the remaining variance.

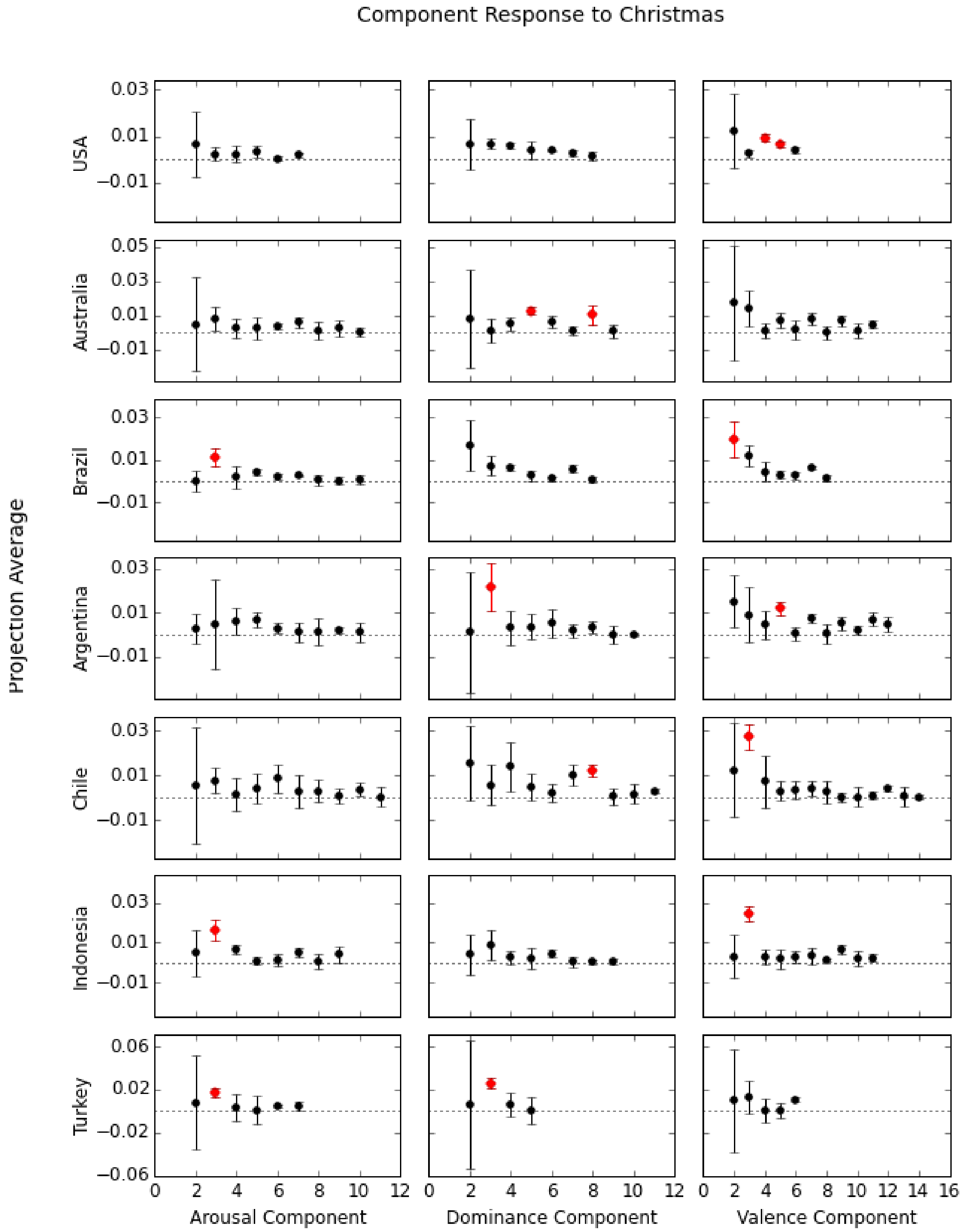

**Fig. S10.** ANEW component response to Christmas by country. Selected components highlighted in red.

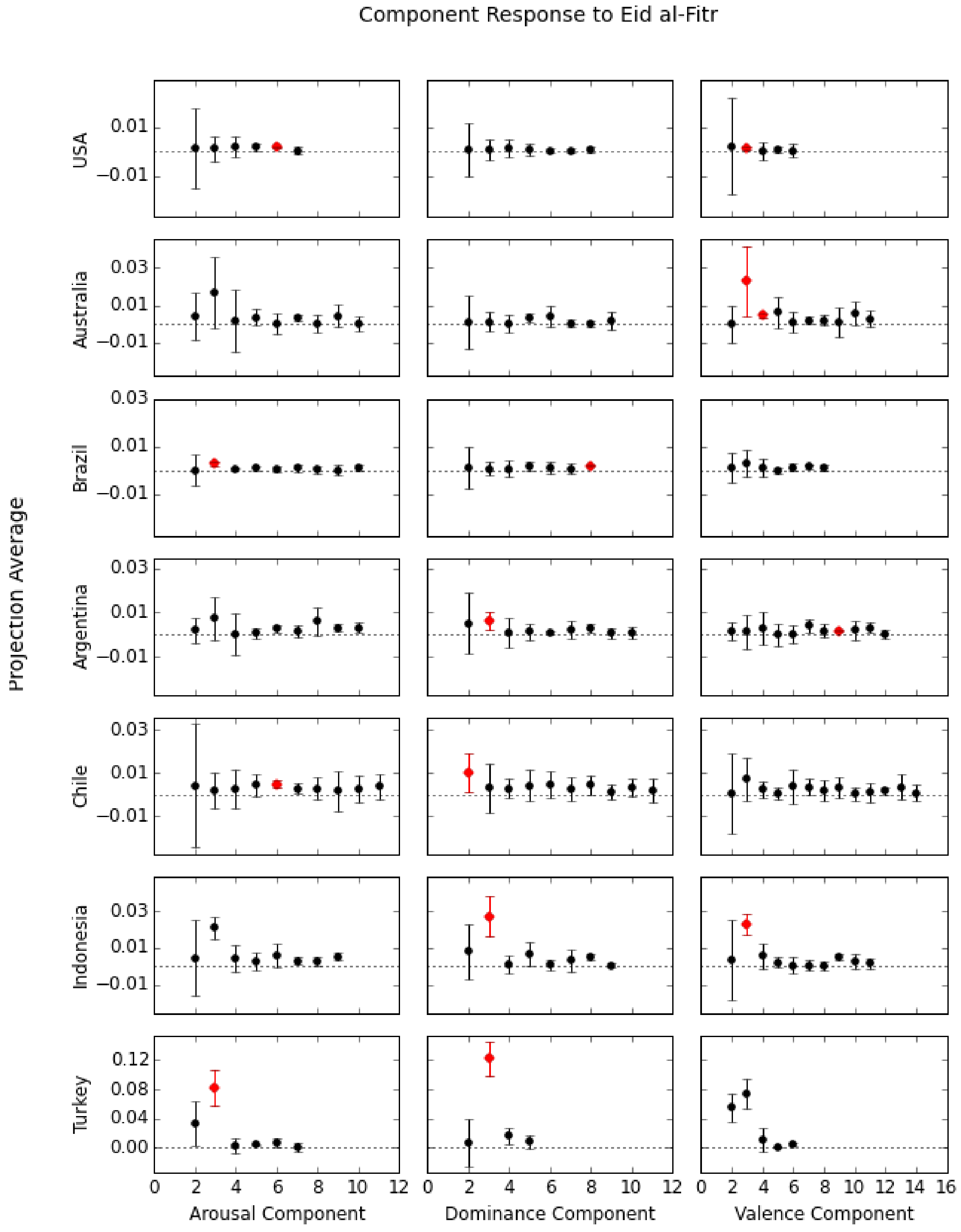

**Fig. S11.** ANEW component response to Eid-al-Fitr by country. Selected components highlighted in red.

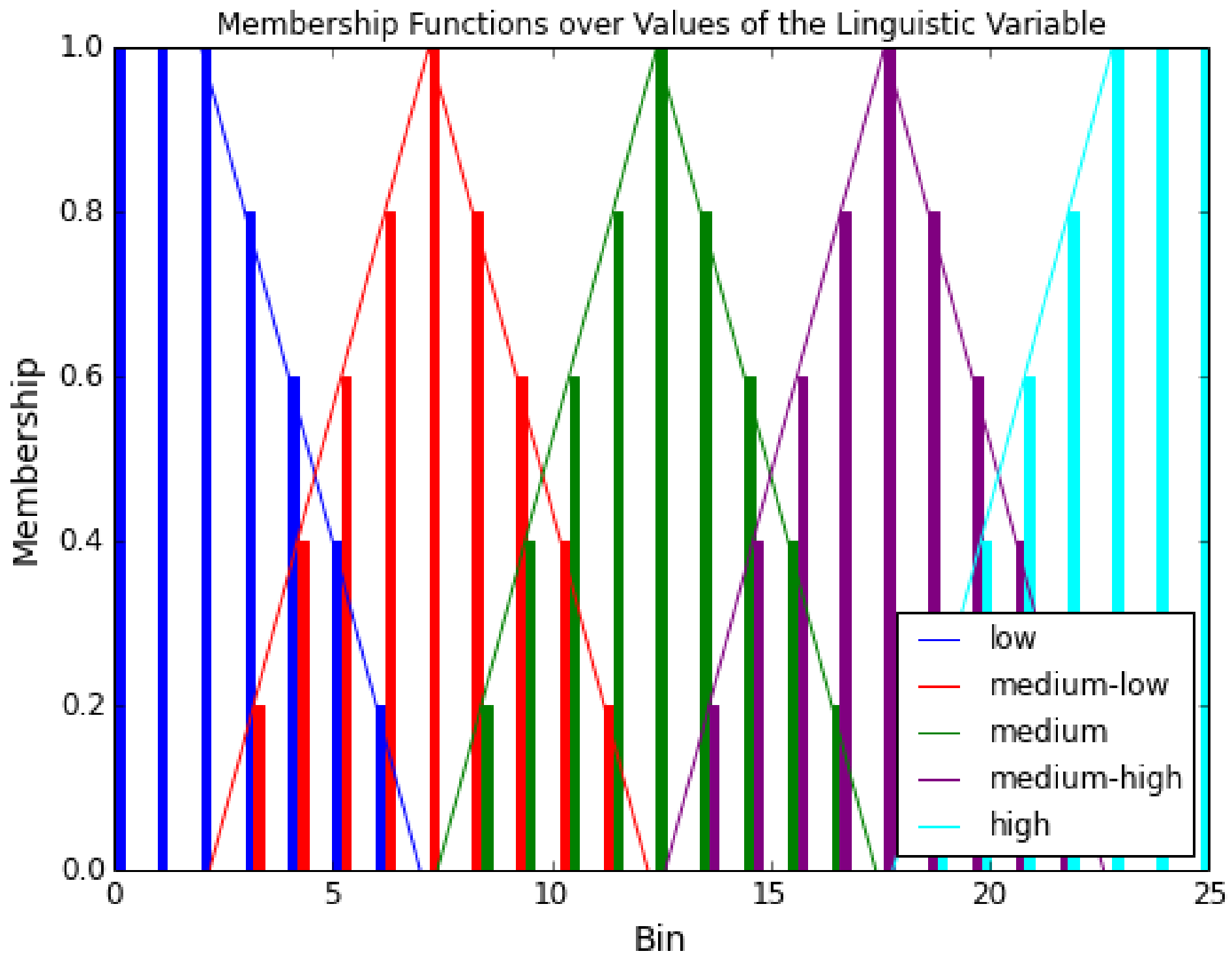

**Fig. S12.** Linguistic Variable value membership functions over 25 bins. The original bins belong to the values of the linguistic variable ("low", "medium-low", "medium", "medium-high", "high") to different extents. The membership functions are mappings from the original bins to a value between 0 and 1, representing membership fuzzy value of that the linguistic variable can take. The membership functions were chosen such that the sum of a bin's membership across all functions is 1, and the area under each membership function's curve is equal.

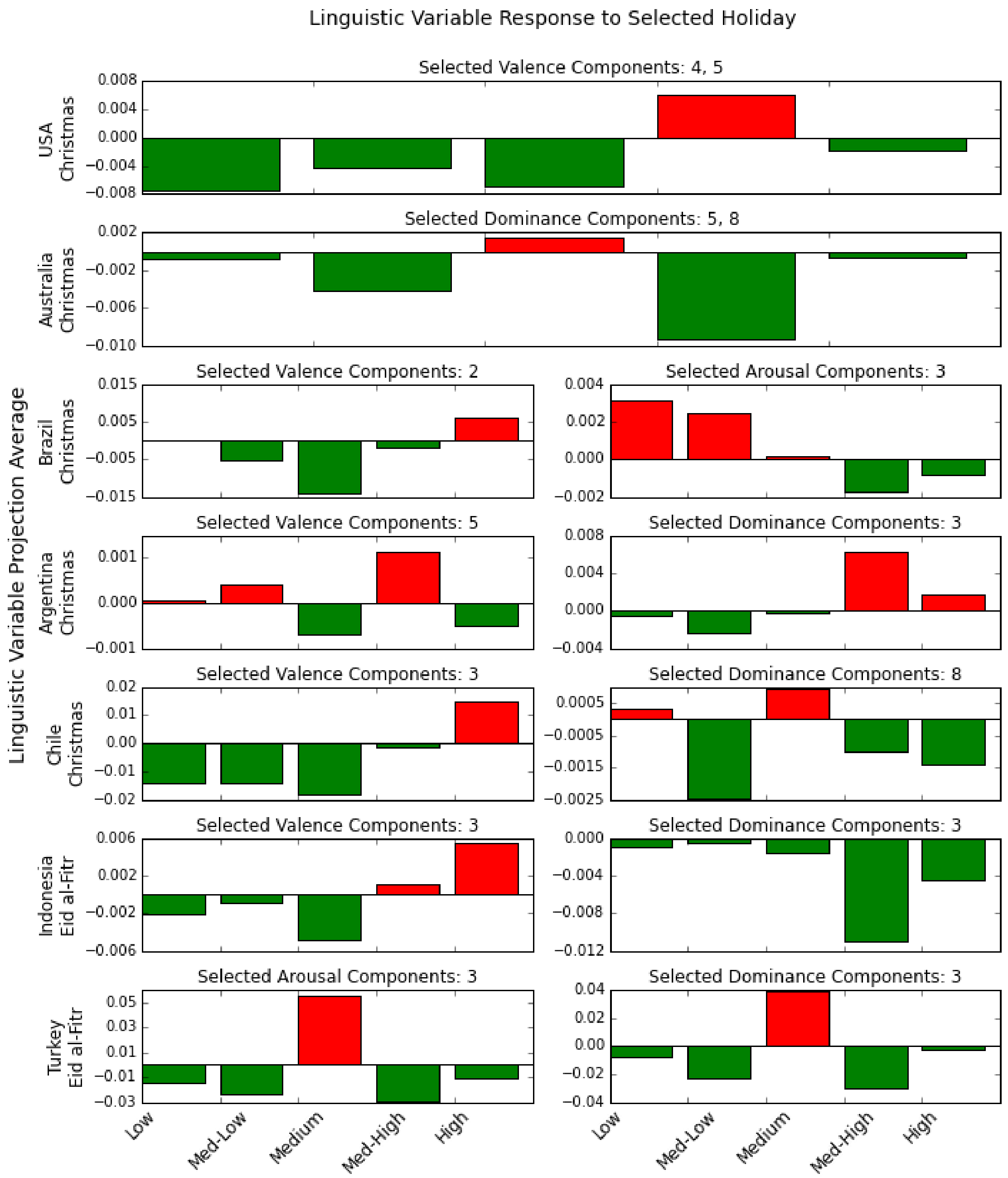

**Fig. S13.** Linguistic Variable Response to relevant holidays selected for each country, as an aid to interpret the effect of chosen *eigendays* during the holidays. A positive value (in red) means that the members of that value of the linguistic variable had increased weight on the holiday, while negative (in green) means they had decreased weight on the holiday.

**Fig. S14.** Average year reconstructed heatmaps. Reconstructed valence heatmaps for each country's average year centered on different holidays. Distributions over time are reconstructed from the components that explain 95% of the variance in the data after the first component is removed. Green represents a decrease in the bin compared to the full distribution, red represents an increase, and yellow represents no change. Center dotted line is the holiday of interest. Left: Christmas, Right: Eid-al-Fitr. Countries top to bottom: USA, Australia, Brazil, Argentina, Chile, Indonesia, Turkey

### *Arousal*

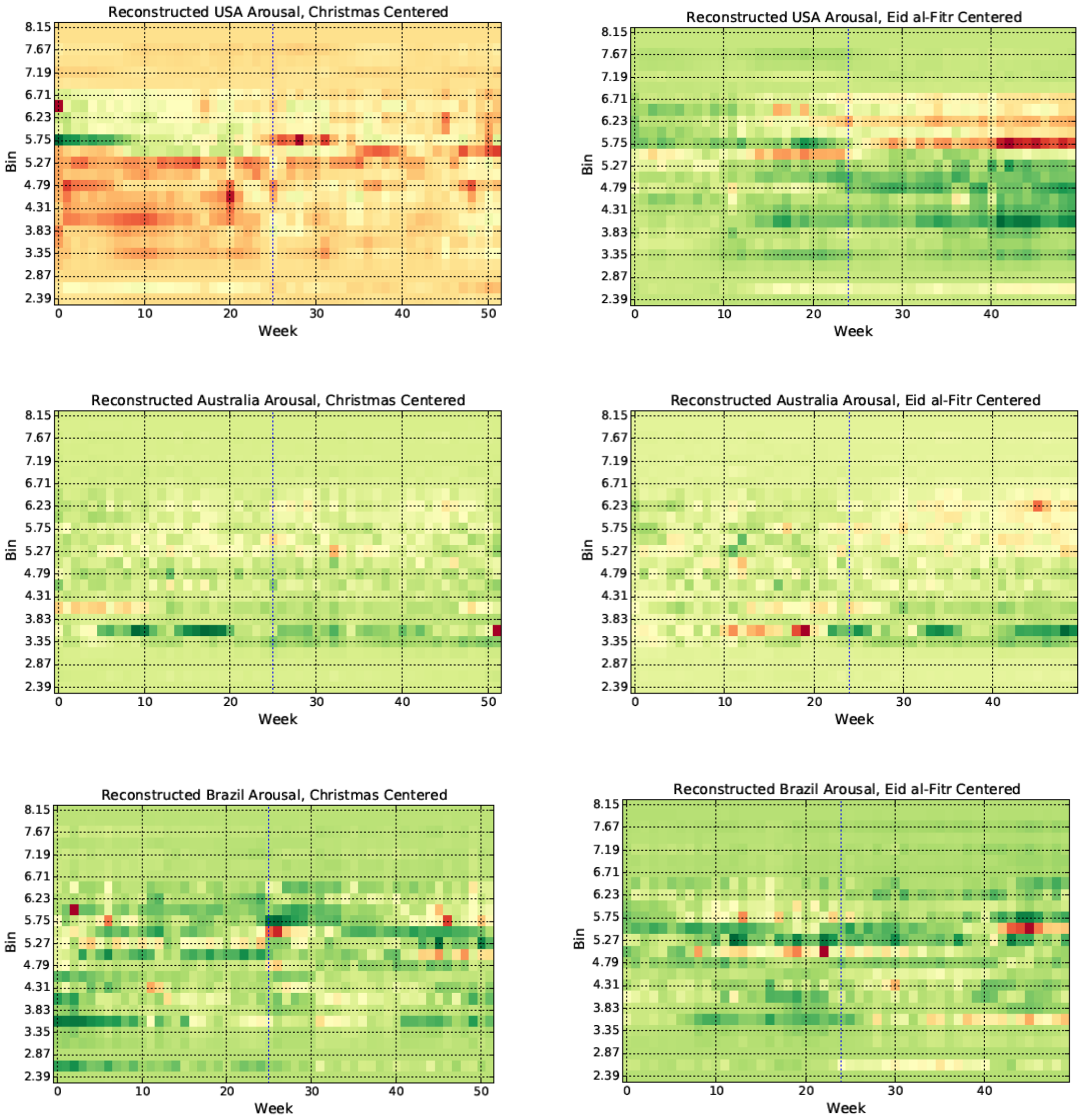

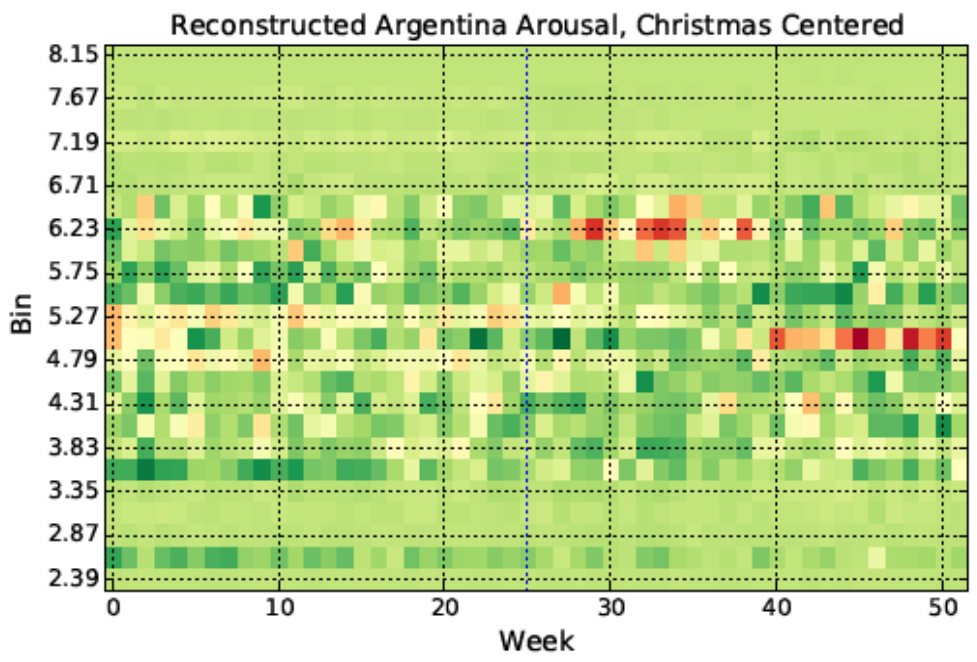
Reconstructed Argentina Arousal, Christmas Centered
Bin
Week

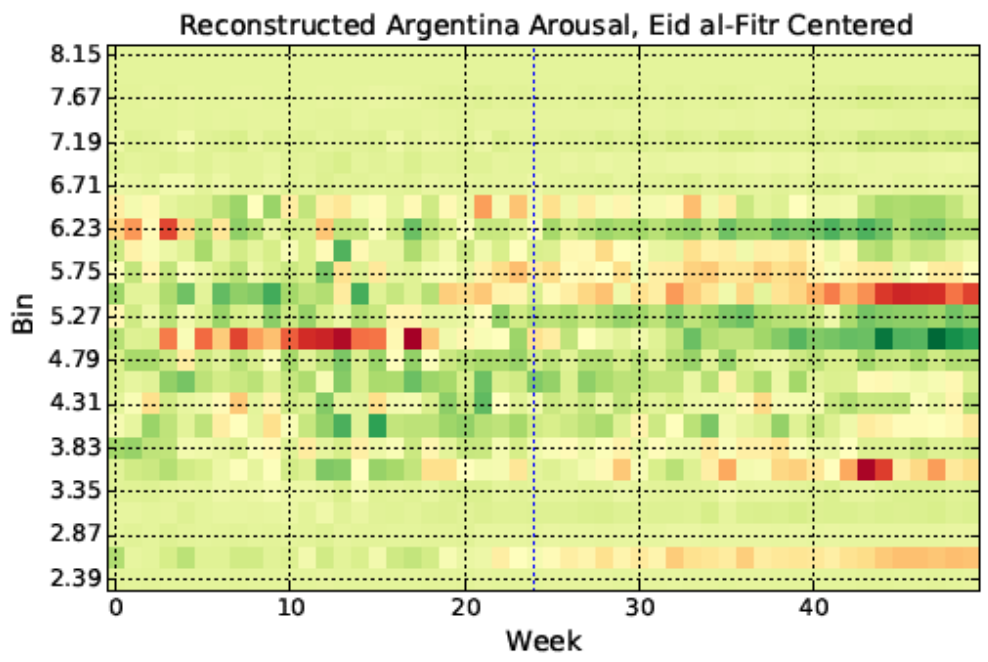
Reconstructed Argentina Arousal, Eid al-Fitr Centered
Bin
Week

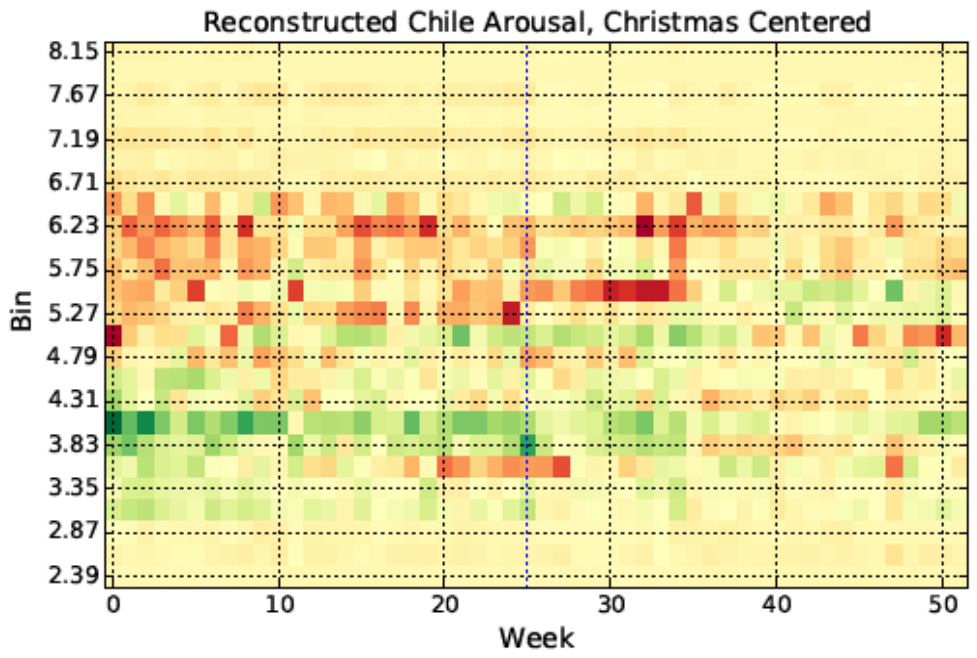
Reconstructed Chile Arousal, Christmas Centered
Bin
Week

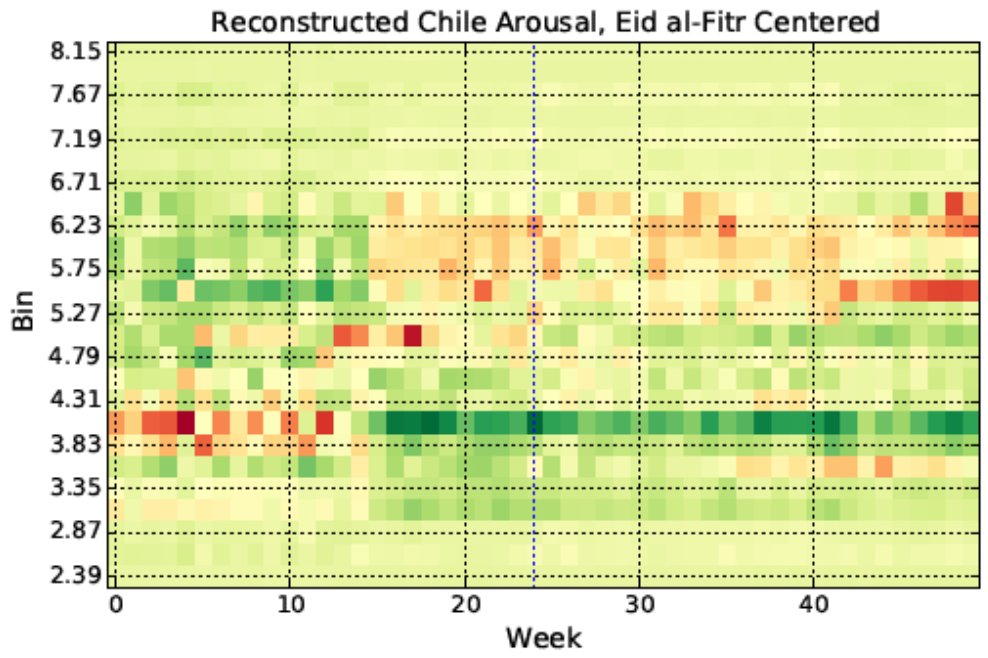
Reconstructed Chile Arousal, Eid al-Fitr Centered
Bin
Week

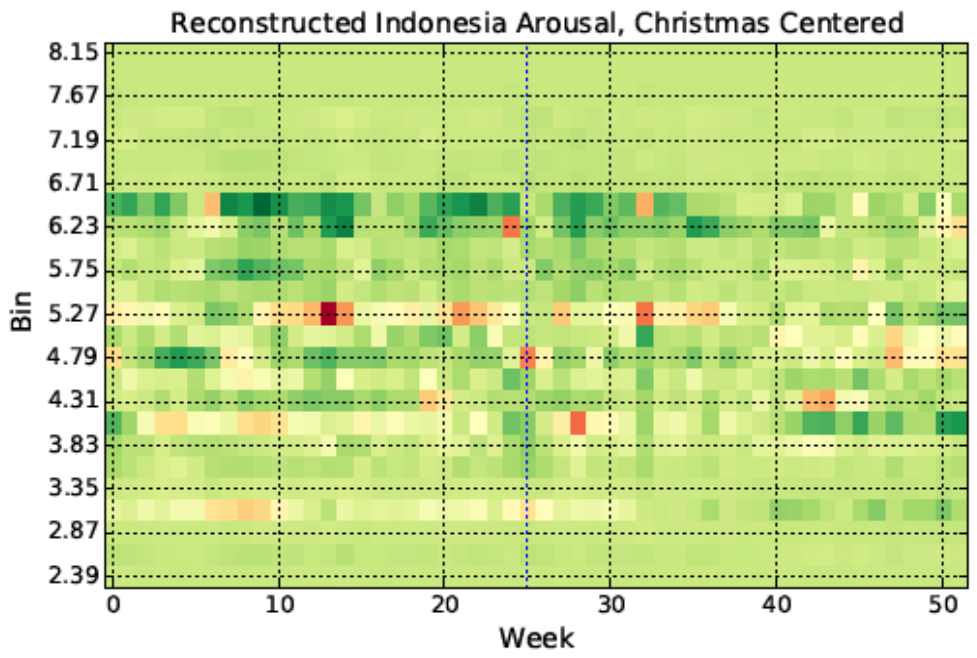
Reconstructed Indonesia Arousal, Christmas Centered
Bin
Week

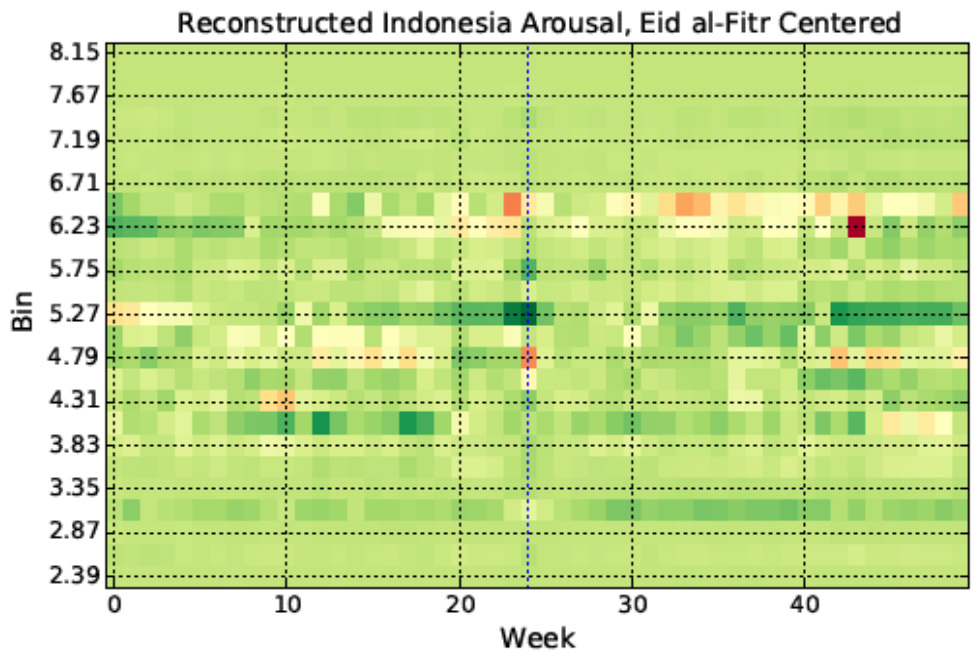
Reconstructed Indonesia Arousal, Eid al-Fitr Centered
Bin
Week

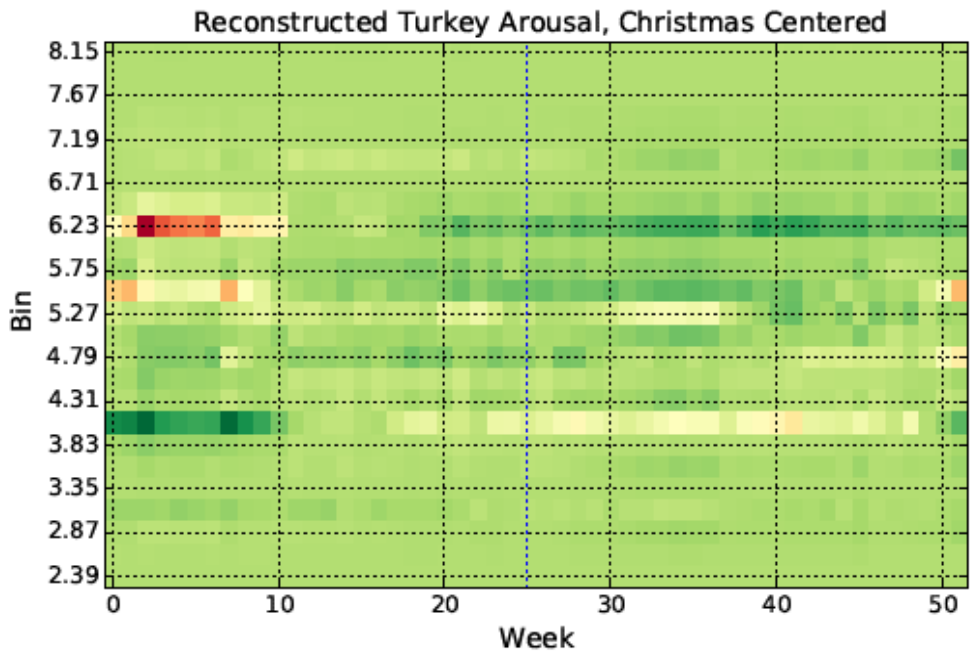
Reconstructed Turkey Arousal, Christmas Centered
Bin
Week

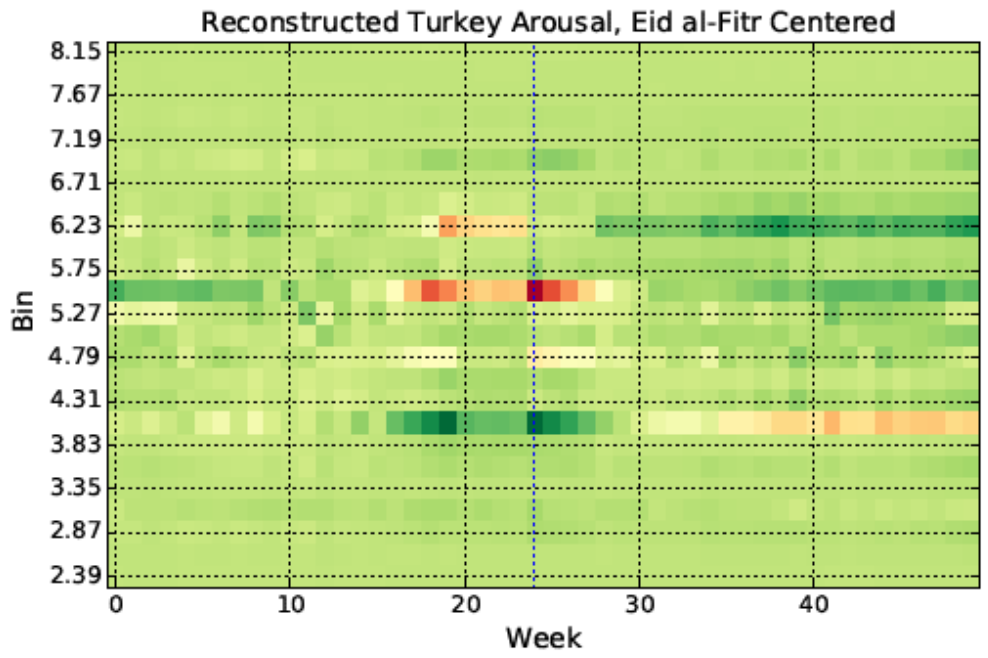
Reconstructed Turkey Arousal, Eid al-Fitr Centered
Bin
Week

### *Dominance*

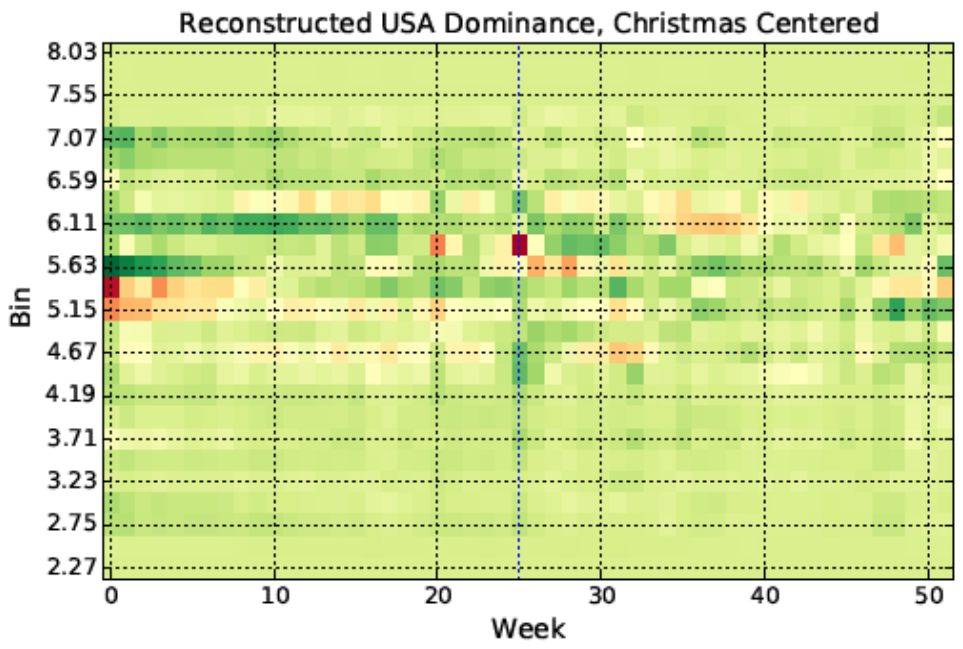

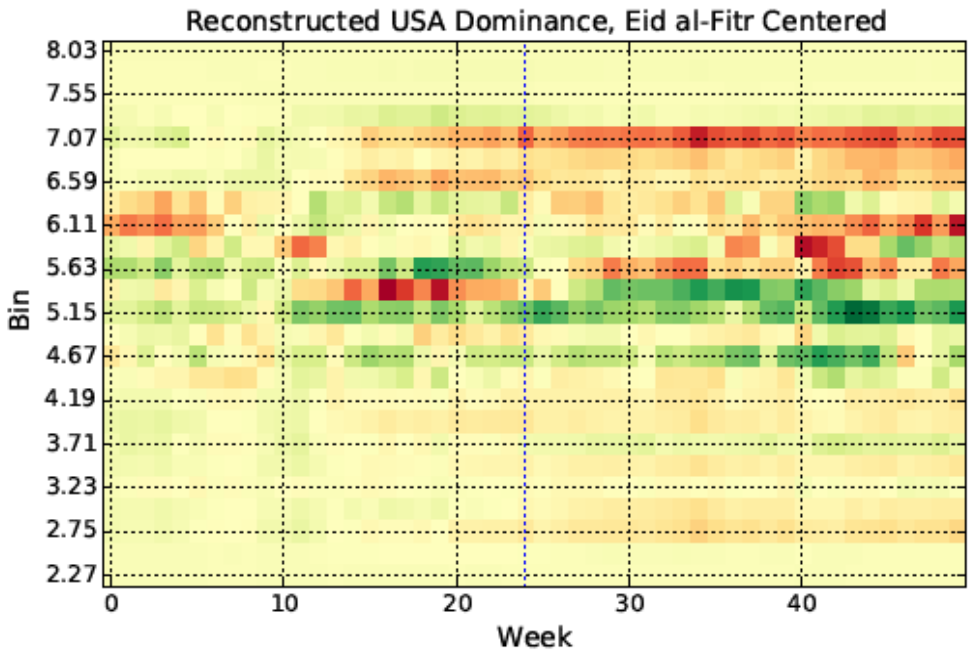

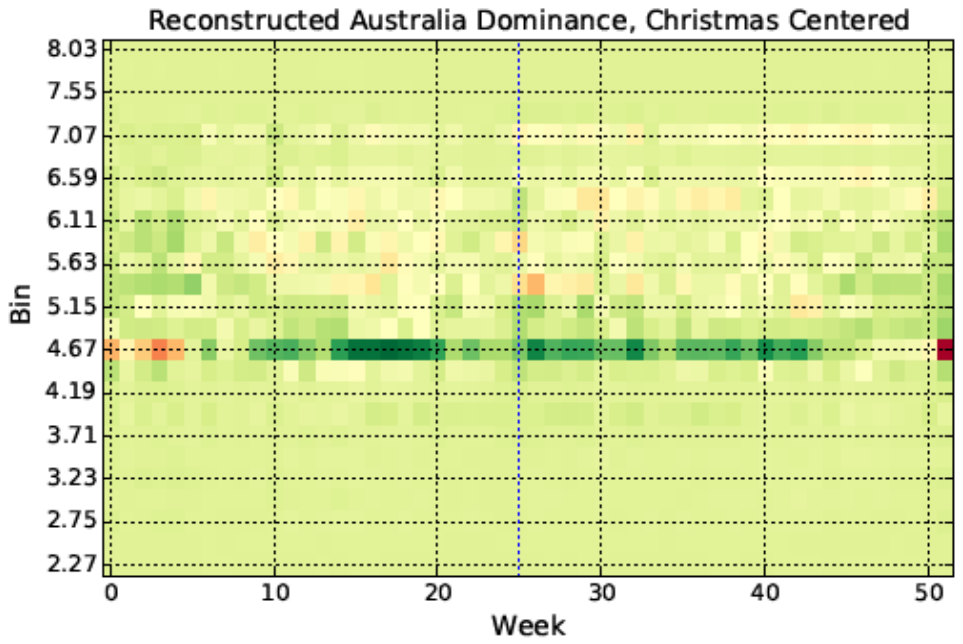

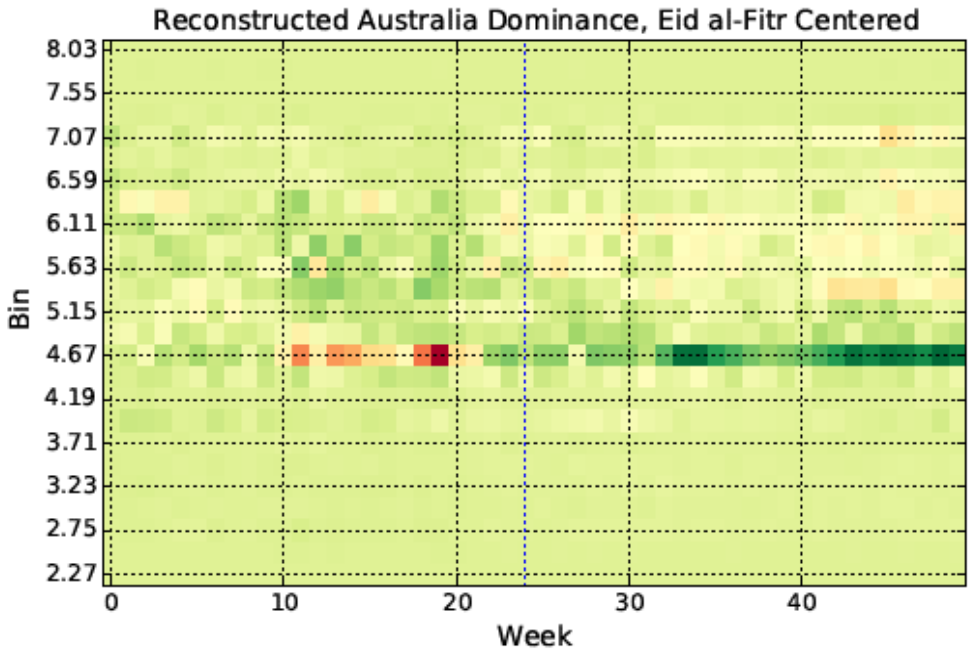

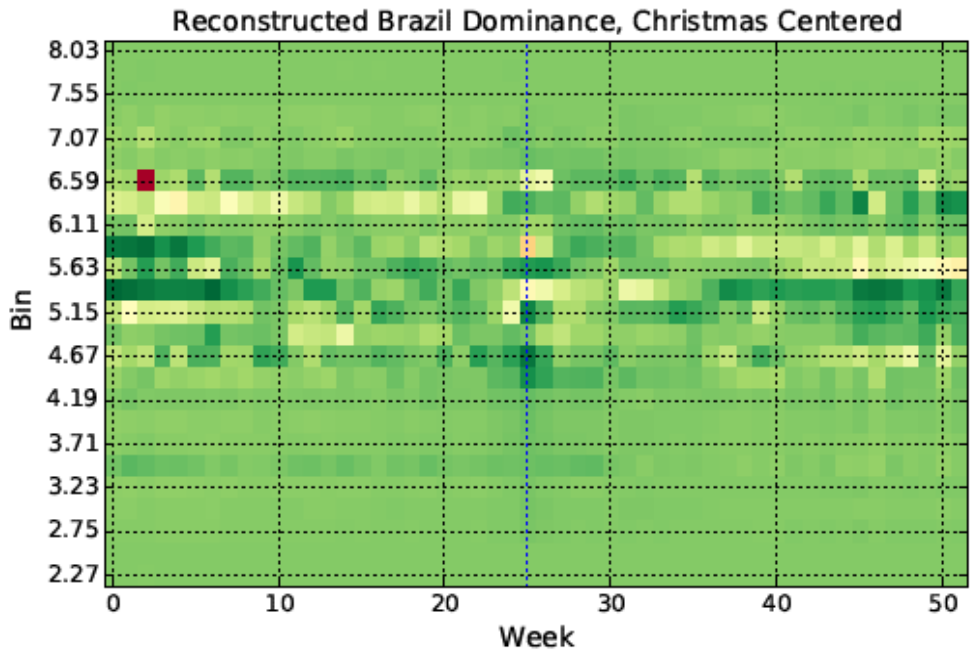

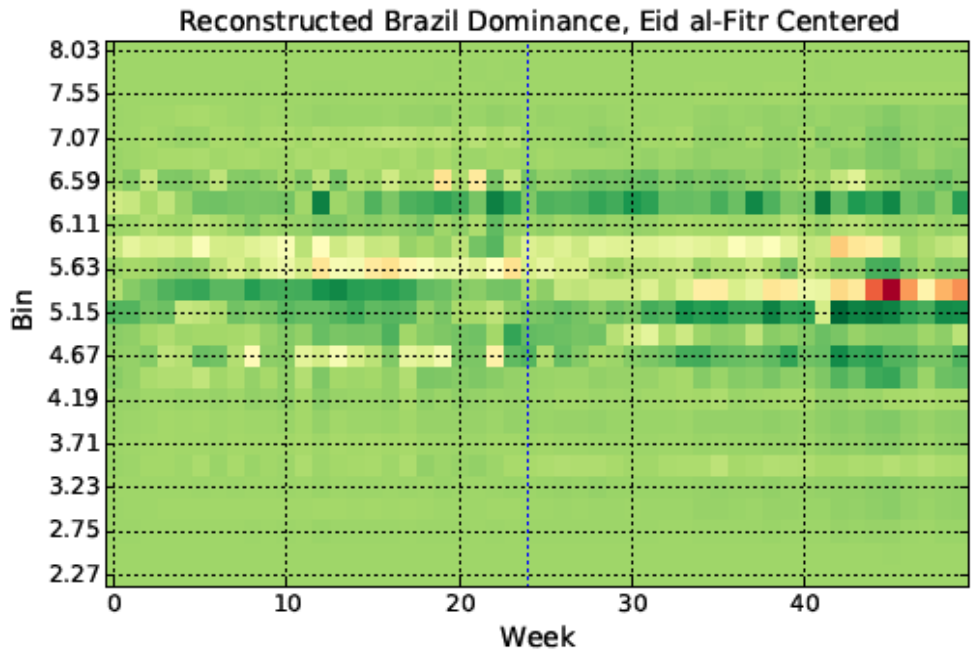

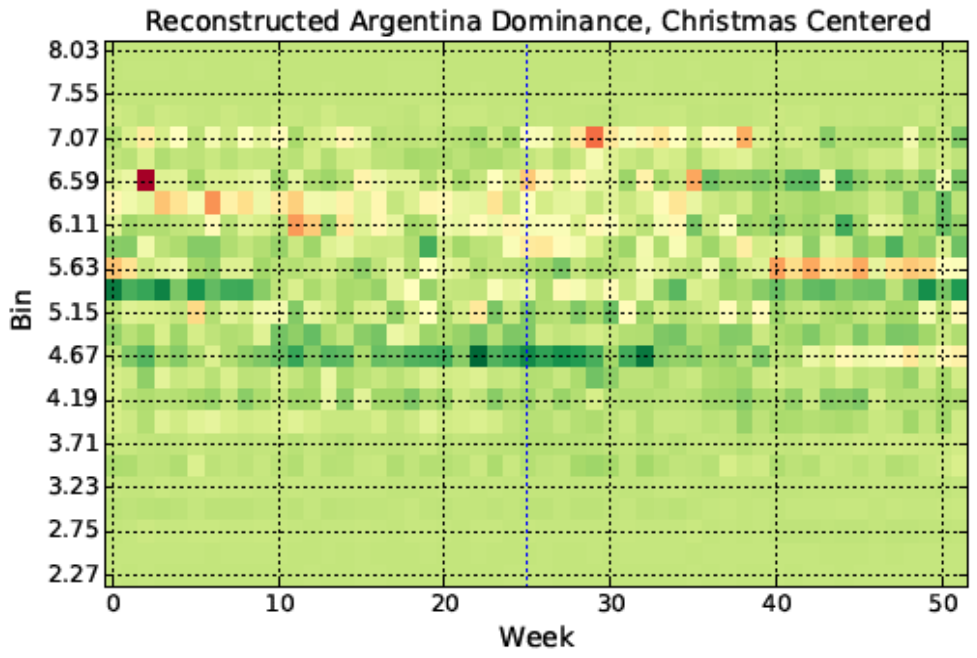

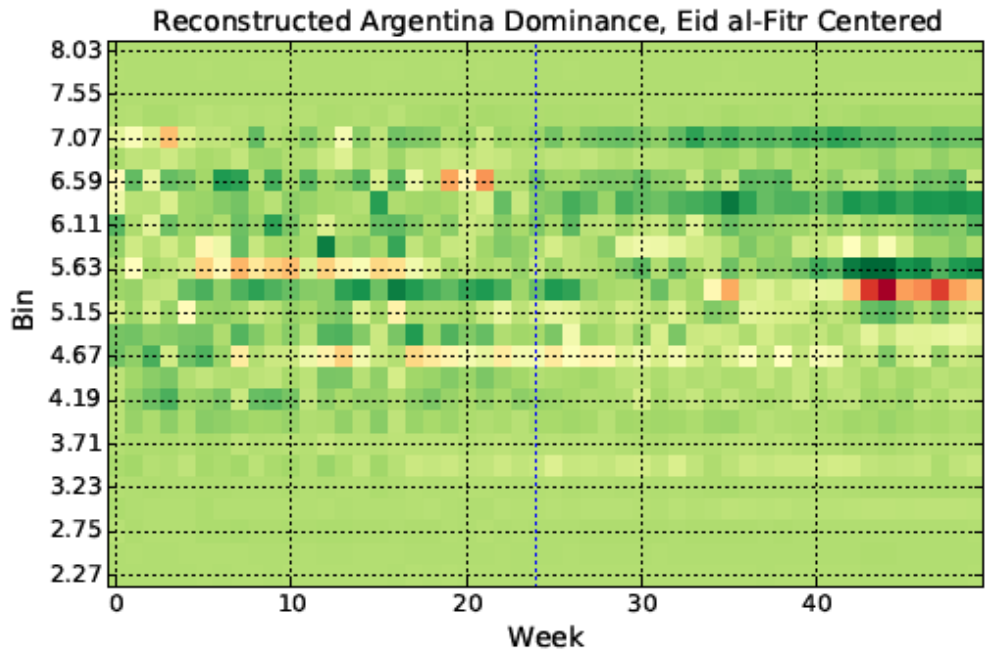

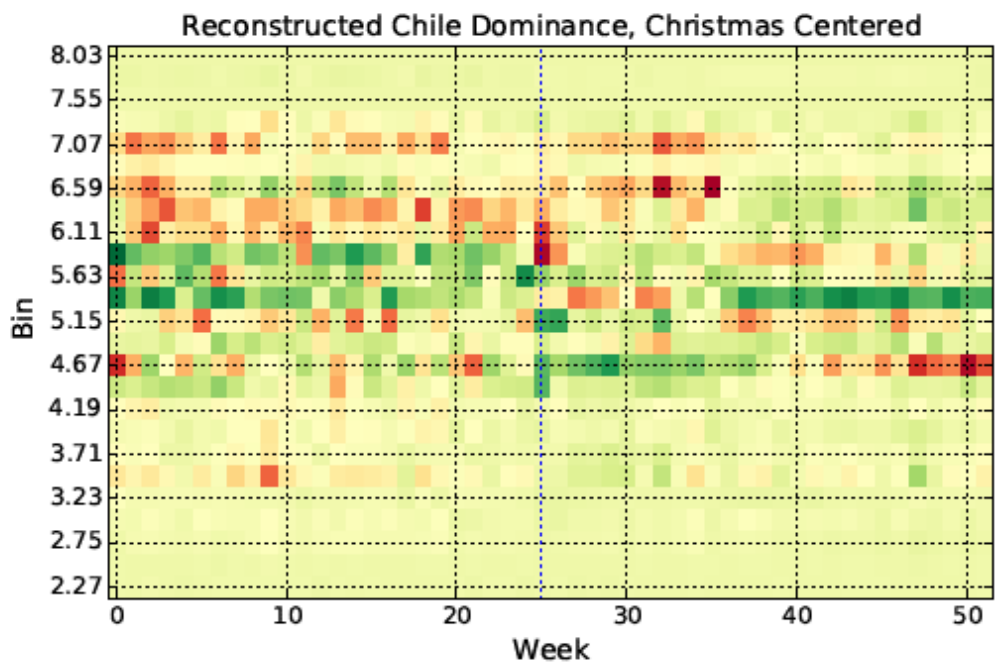
Reconstructed Chile Dominance, Christmas Centered
Bin
Week

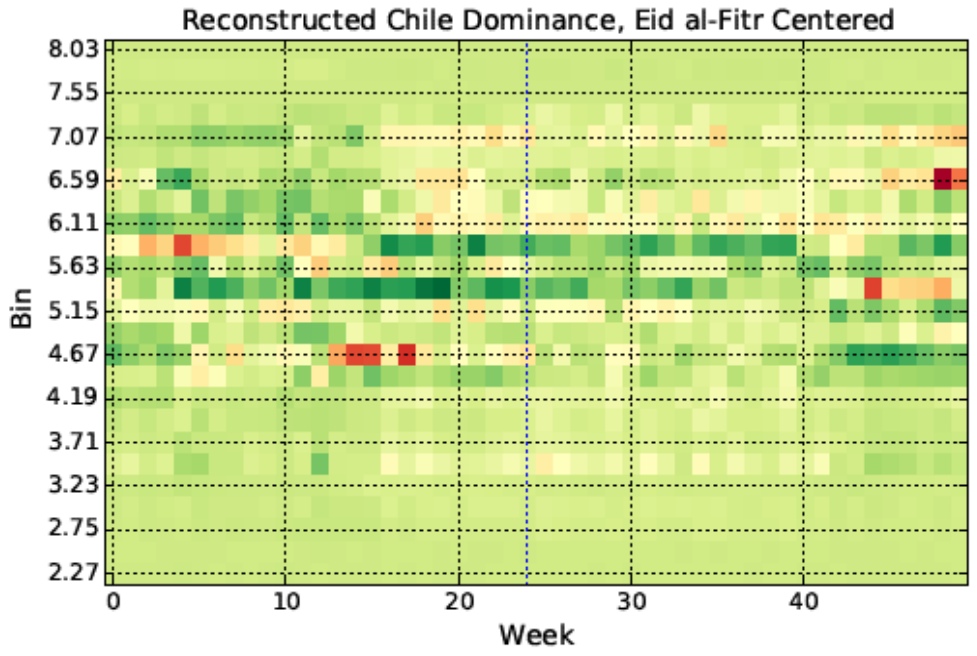
Reconstructed Chile Dominance, Eid al-Fitr Centered
Bin
Week

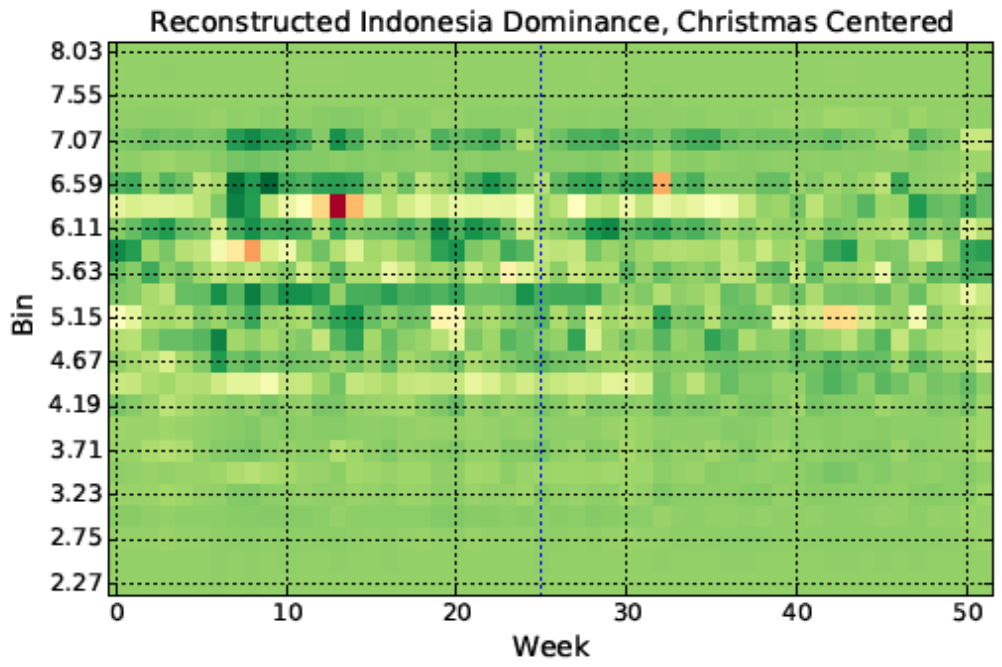
Reconstructed Indonesia Dominance, Christmas Centered
Bin
Week

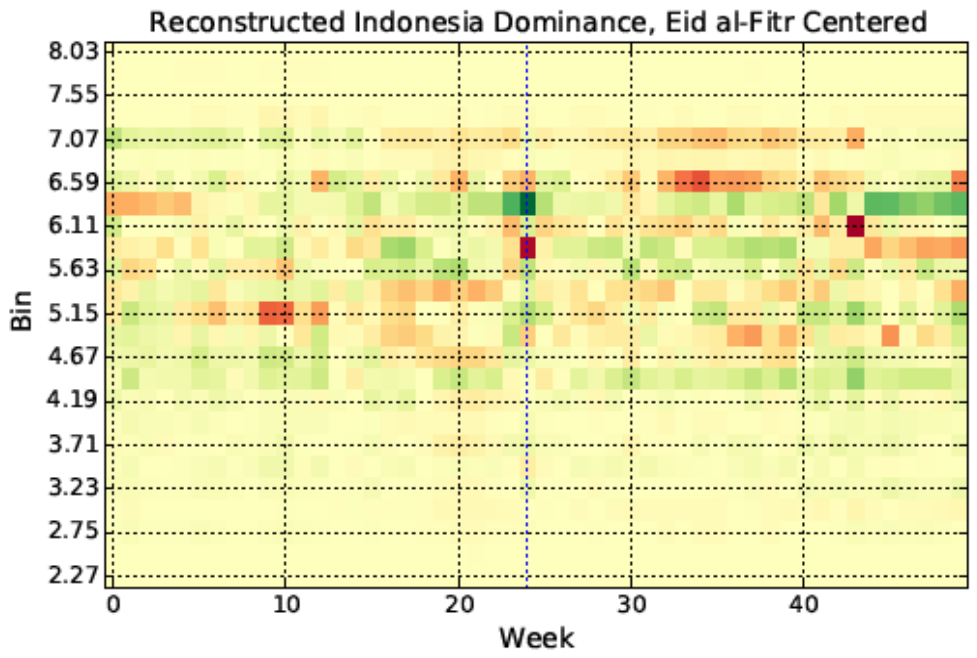
Reconstructed Indonesia Dominance, Eid al-Fitr Centered
Bin
Week

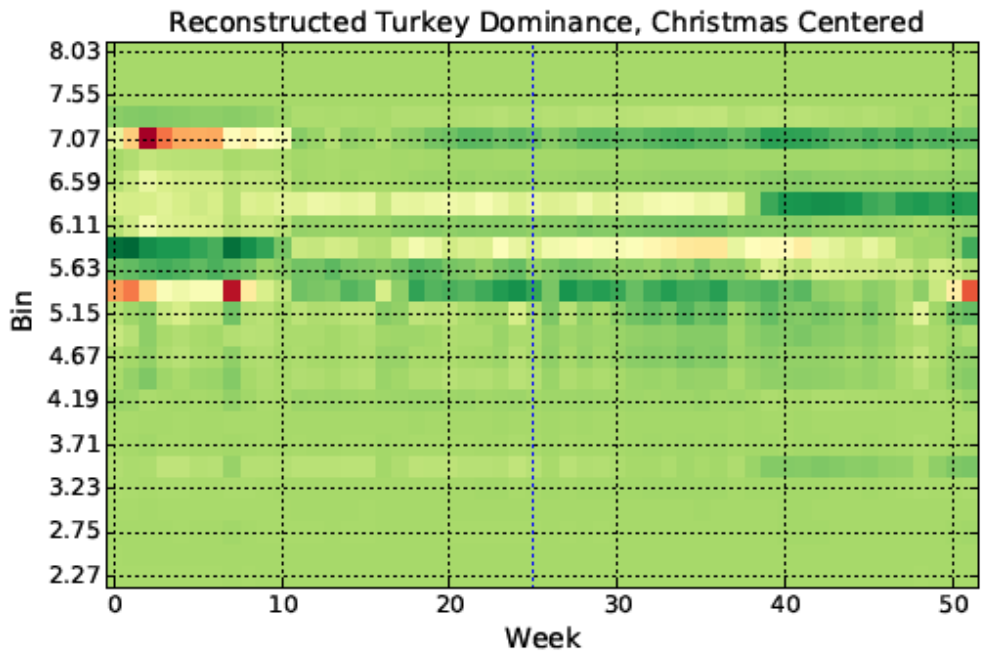
Reconstructed Turkey Dominance, Christmas Centered
Bin
Week

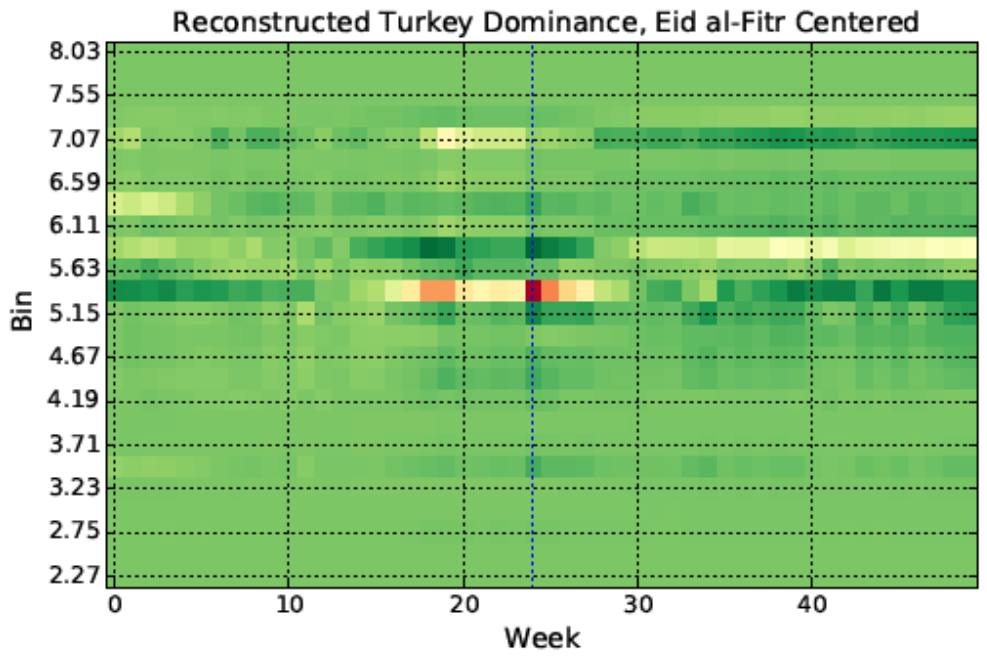
Reconstructed Turkey Dominance, Eid al-Fitr Centered
Bin
Week

## *Valence*

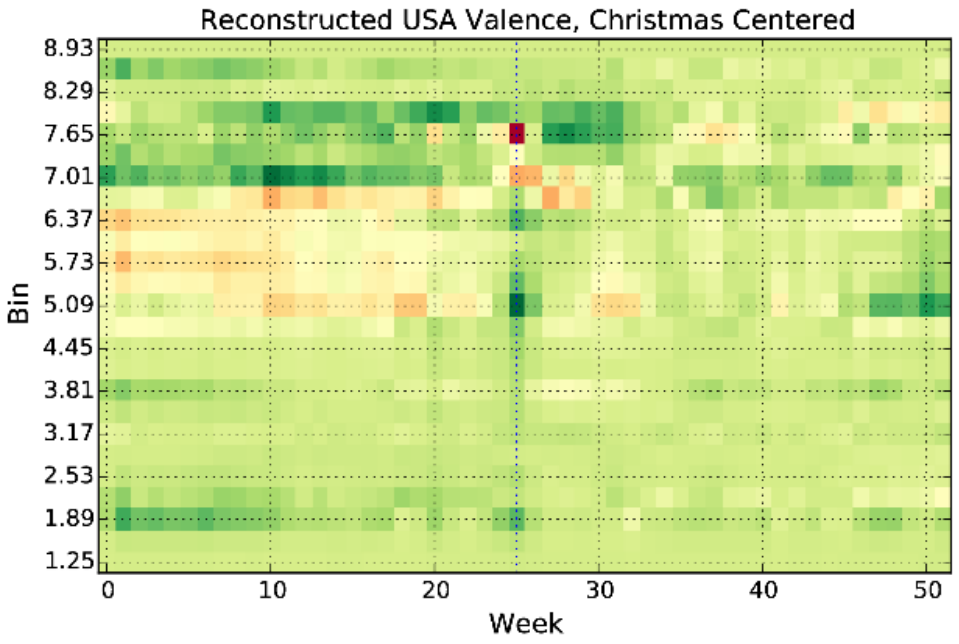
Reconstructed USA Valence, Christmas Centered
Bin
Week

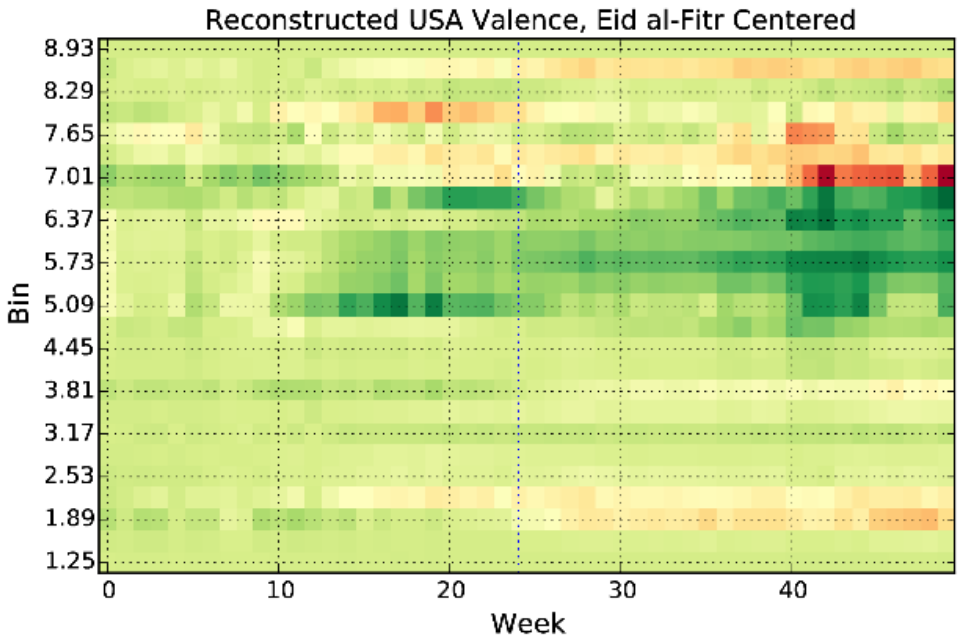
Reconstructed USA Valence, Eid al-Fitr Centered
Bin
Week

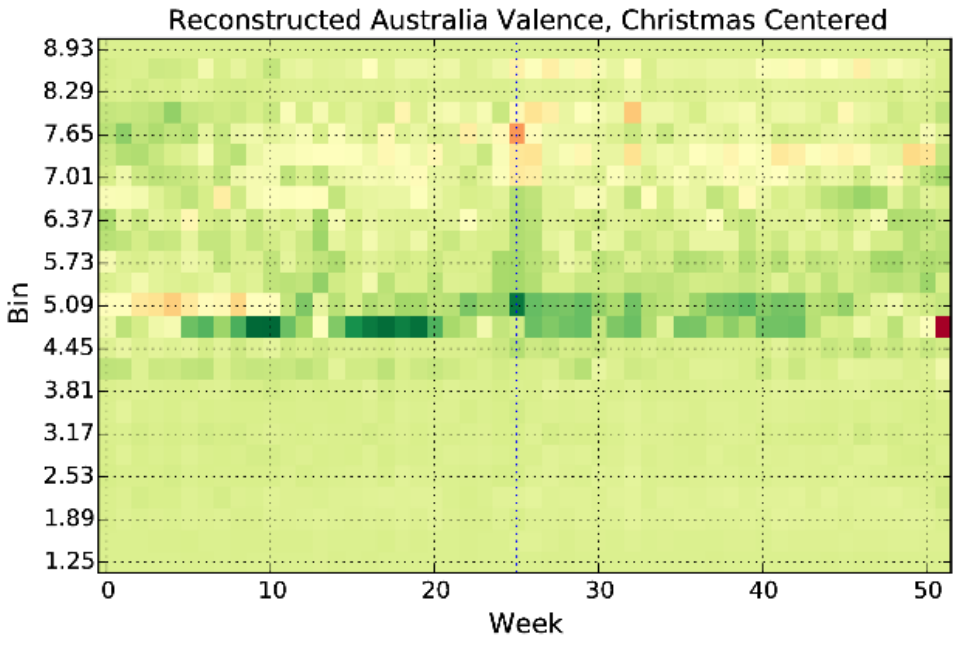
Reconstructed Australia Valence, Christmas Centered
Bin
Week

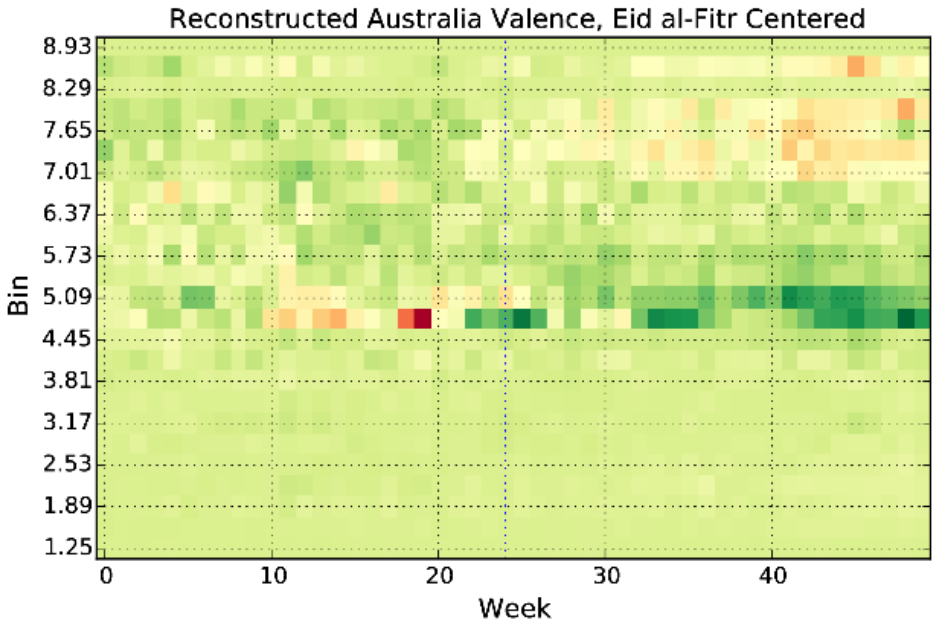
Reconstructed Australia Valence, Eid al-Fitr Centered
Bin
Week

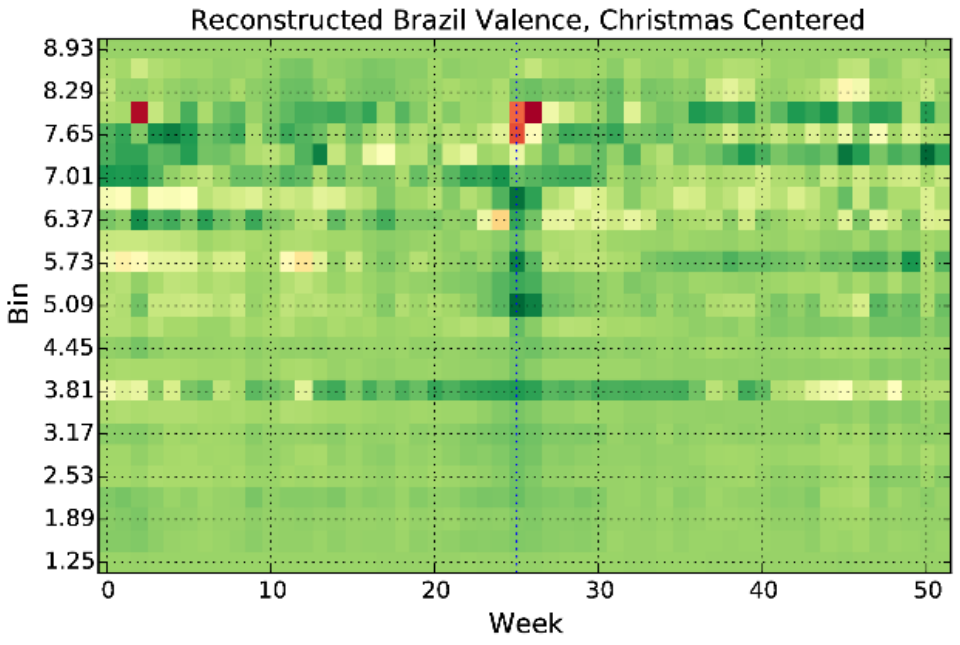
Reconstructed Brazil Valence, Christmas Centered
Bin
Week

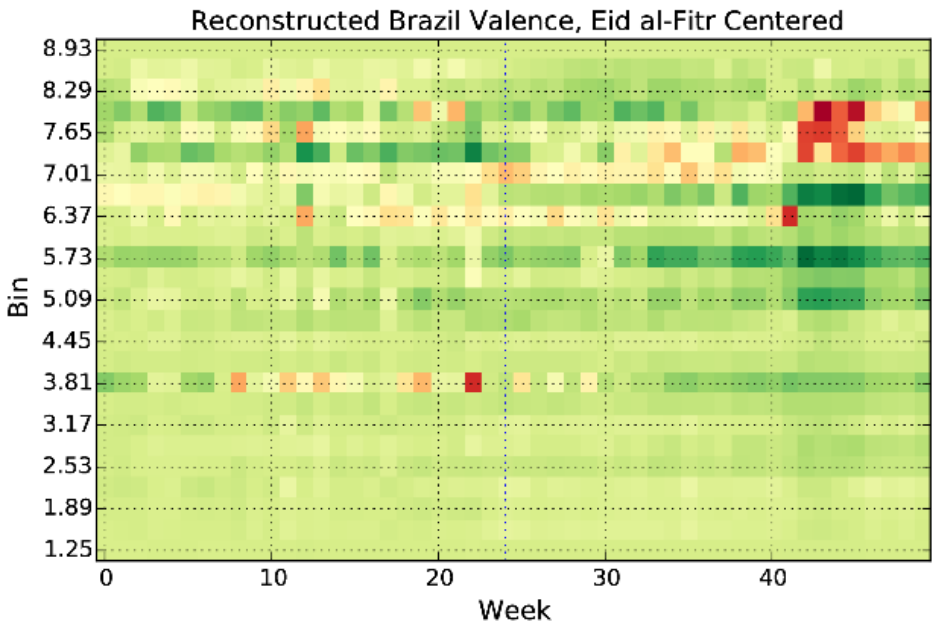
Reconstructed Brazil Valence, Eid al-Fitr Centered
Bin
Week

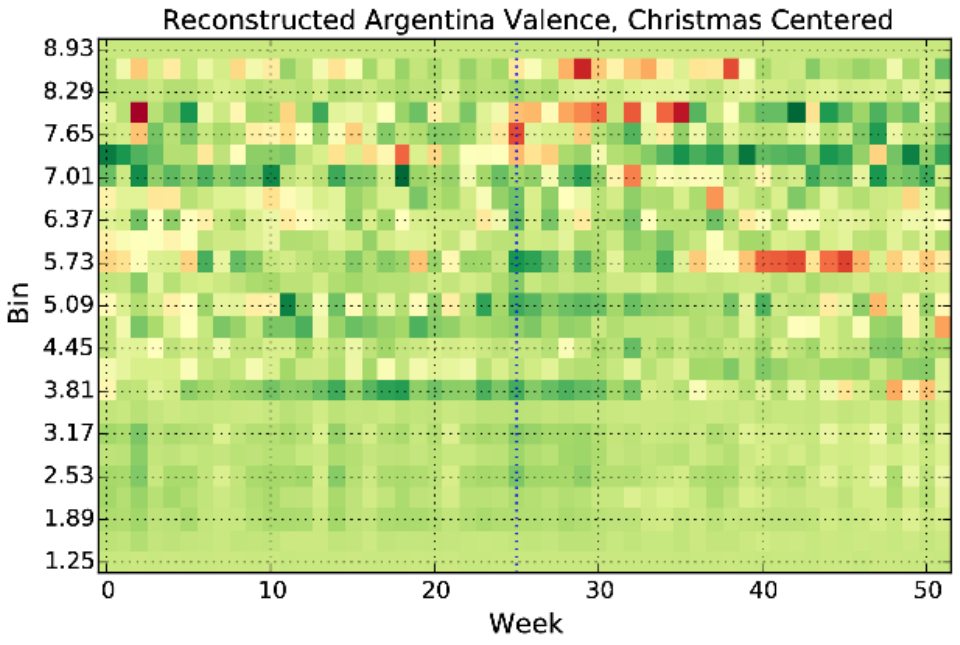
Reconstructed Argentina Valence, Christmas Centered
Bin
Week

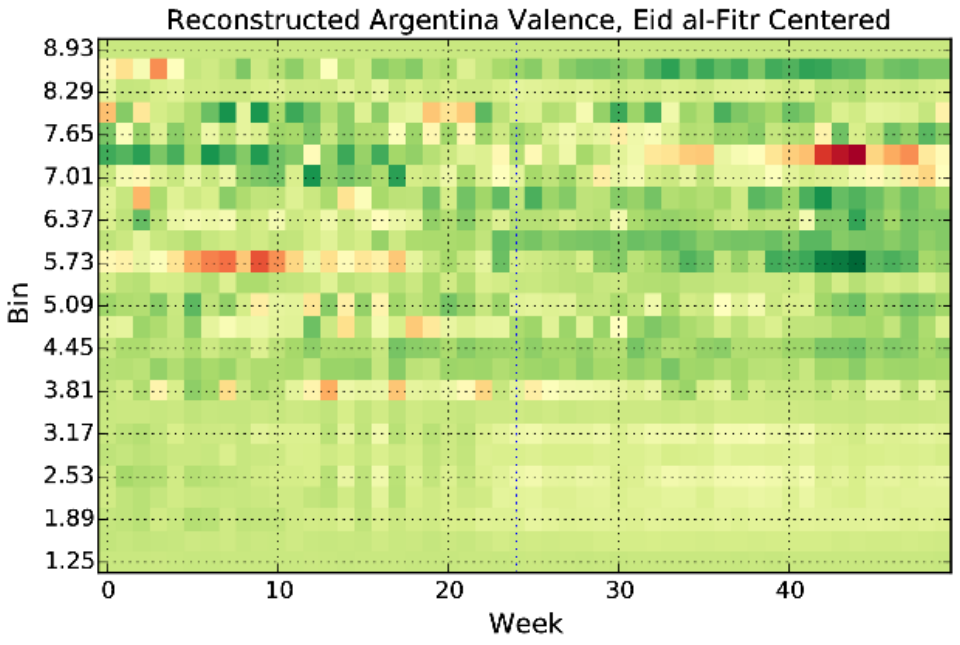
Reconstructed Argentina Valence, Eid al-Fitr Centered
Bin
Week

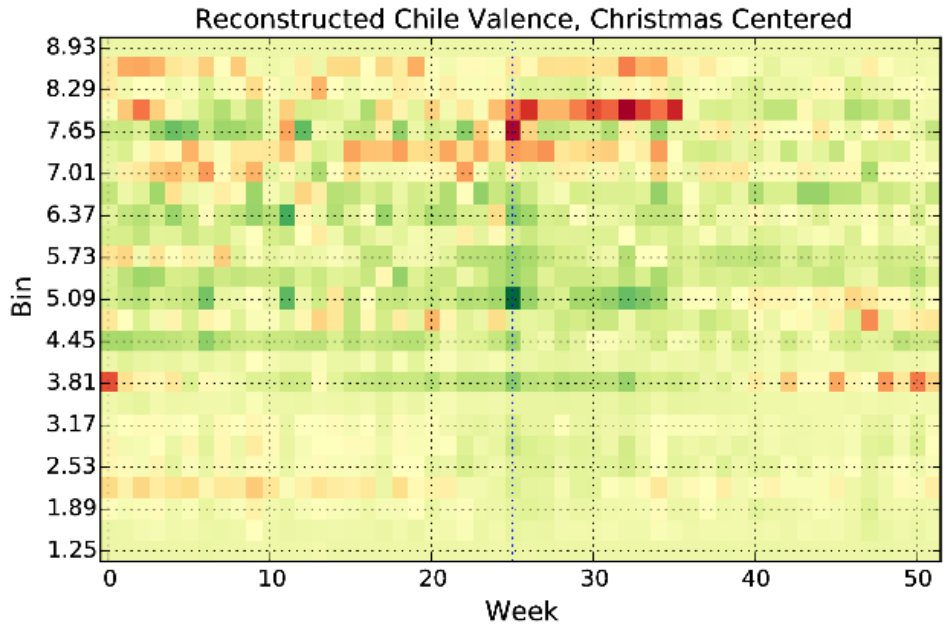
Reconstructed Chile Valence, Christmas Centered
8.93 8.29 7.65 7.01 6.37 5.73 5.09 4.45 3.81 3.17 2.53 1.89 1.25
Bin
0 10 20 30 40 50
Week

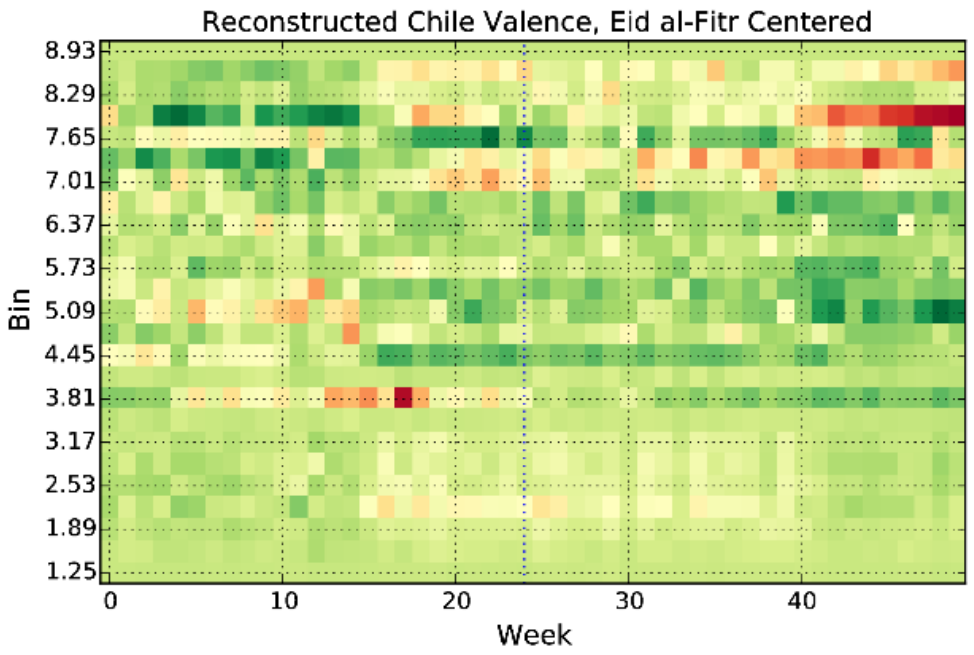
Reconstructed Chile Valence, Eid al-Fitr Centered
8.93 8.29 7.65 7.01 6.37 5.73 5.09 4.45 3.81 3.17 2.53 1.89 1.25
Bin
0 10 20 30 40
Week

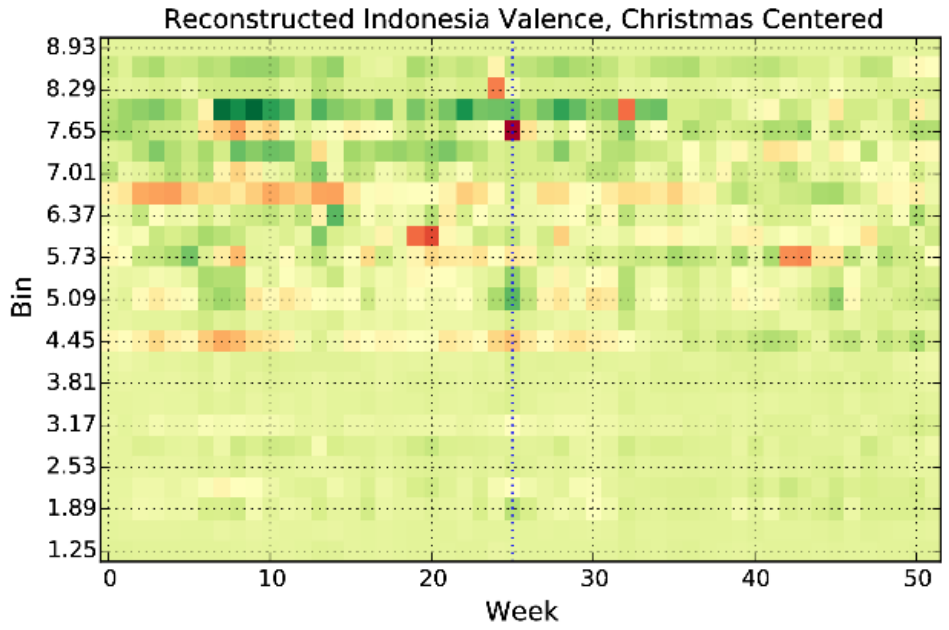
Reconstructed Indonesia Valence, Christmas Centered
8.93 8.29 7.65 7.01 6.37 5.73 5.09 4.45 3.81 3.17 2.53 1.89 1.25
Bin
0 10 20 30 40 50
Week

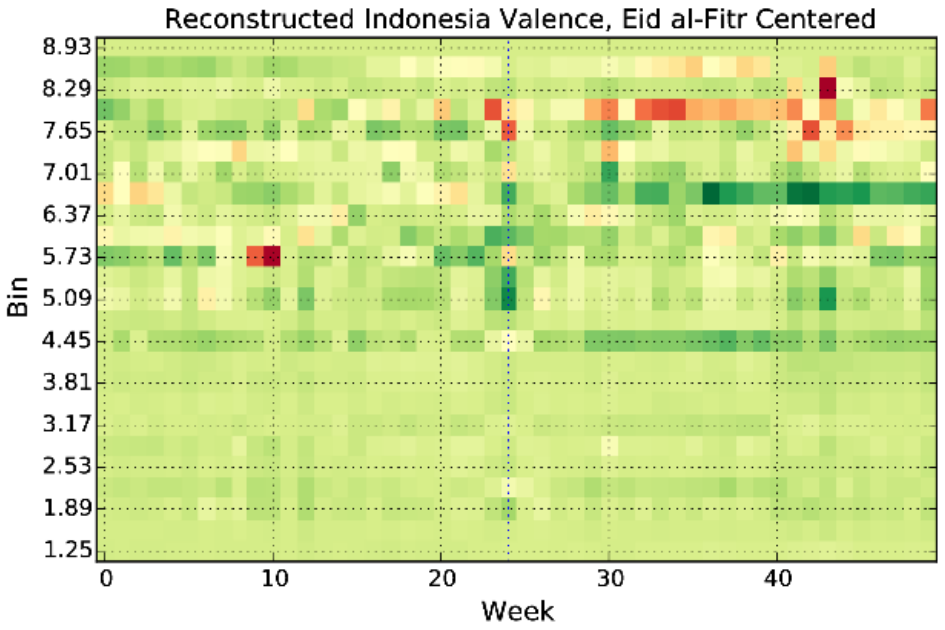
Reconstructed Indonesia Valence, Eid al-Fitr Centered
8.93 8.29 7.65 7.01 6.37 5.73 5.09 4.45 3.81 3.17 2.53 1.89 1.25
Bin
0 10 20 30 40
Week

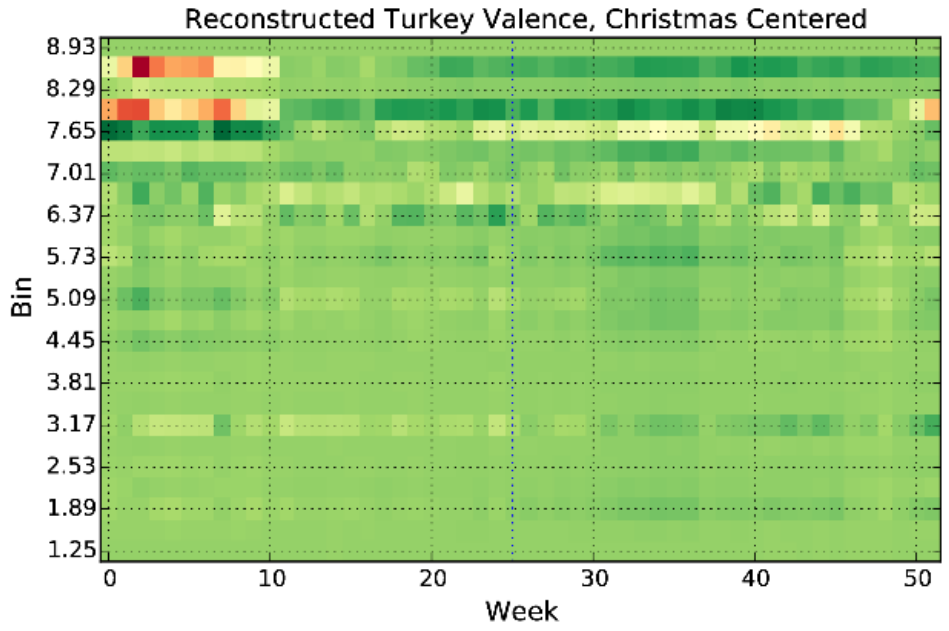
Reconstructed Turkey Valence, Christmas Centered
8.93 8.29 7.65 7.01 6.37 5.73 5.09 4.45 3.81 3.17 2.53 1.89 1.25
Bin
0 10 20 30 40 50
Week

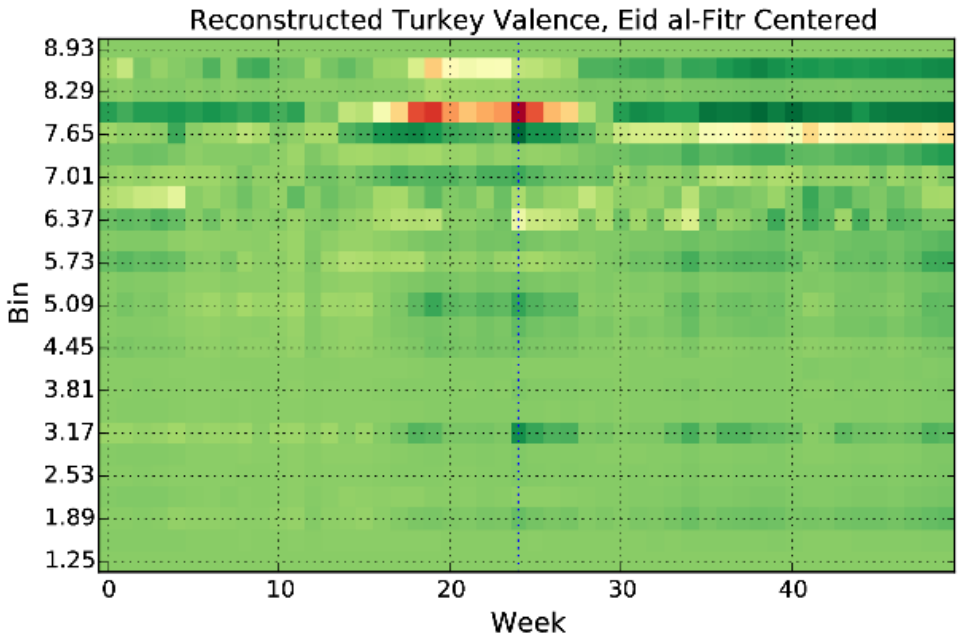
Reconstructed Turkey Valence, Eid al-Fitr Centered
8.93 8.29 7.65 7.01 6.37 5.73 5.09 4.45 3.81 3.17 2.53 1.89 1.25
Bin
0 10 20 30 40
Week

**Fig. S15.** Eigenmood projections and regressions**.** Projections show all yearly data points, projected into the space formed by the selected eigenweeks; regressions show the average year's sex searches and similarity to the holiday center.

## *USA Christmas*

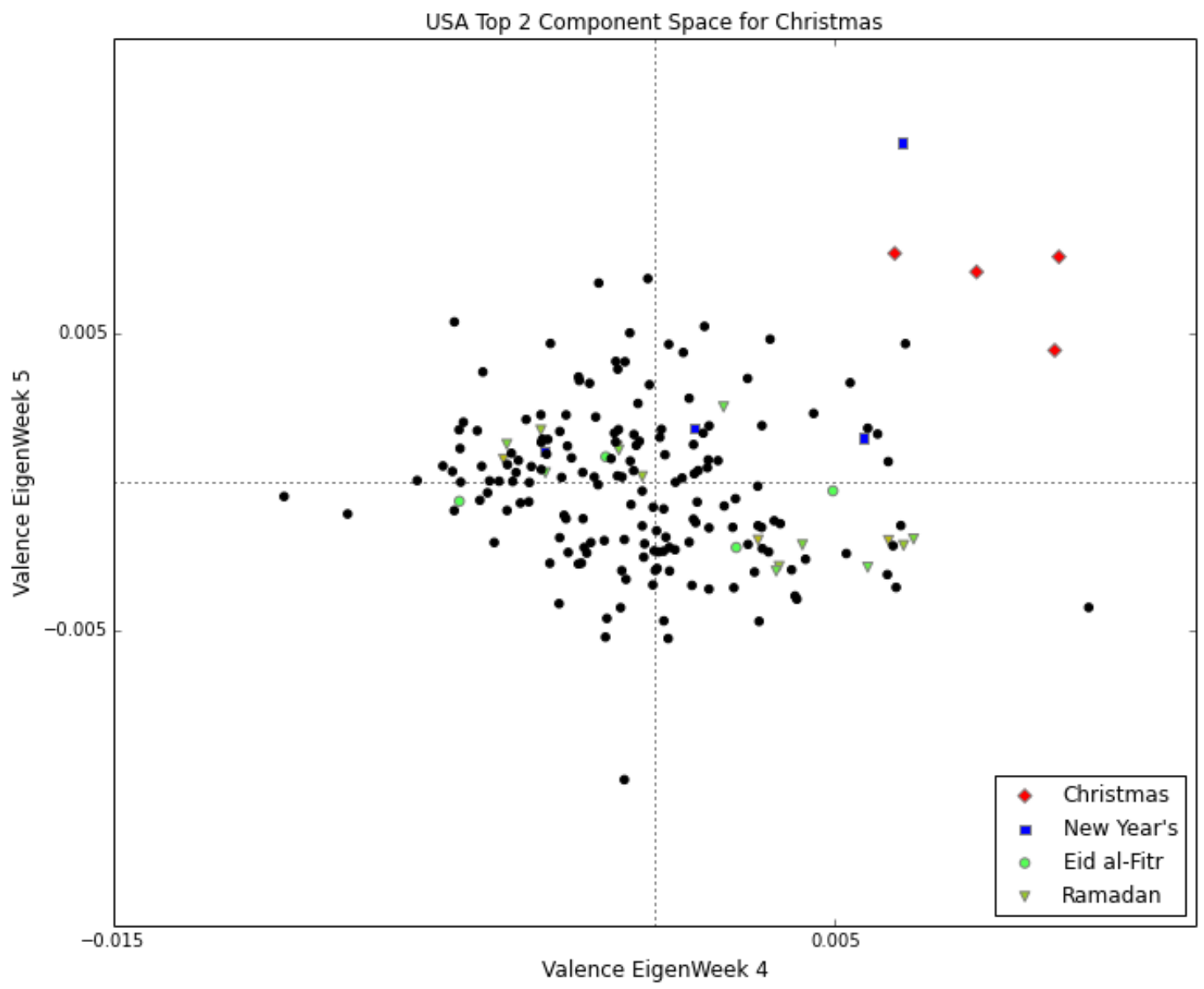

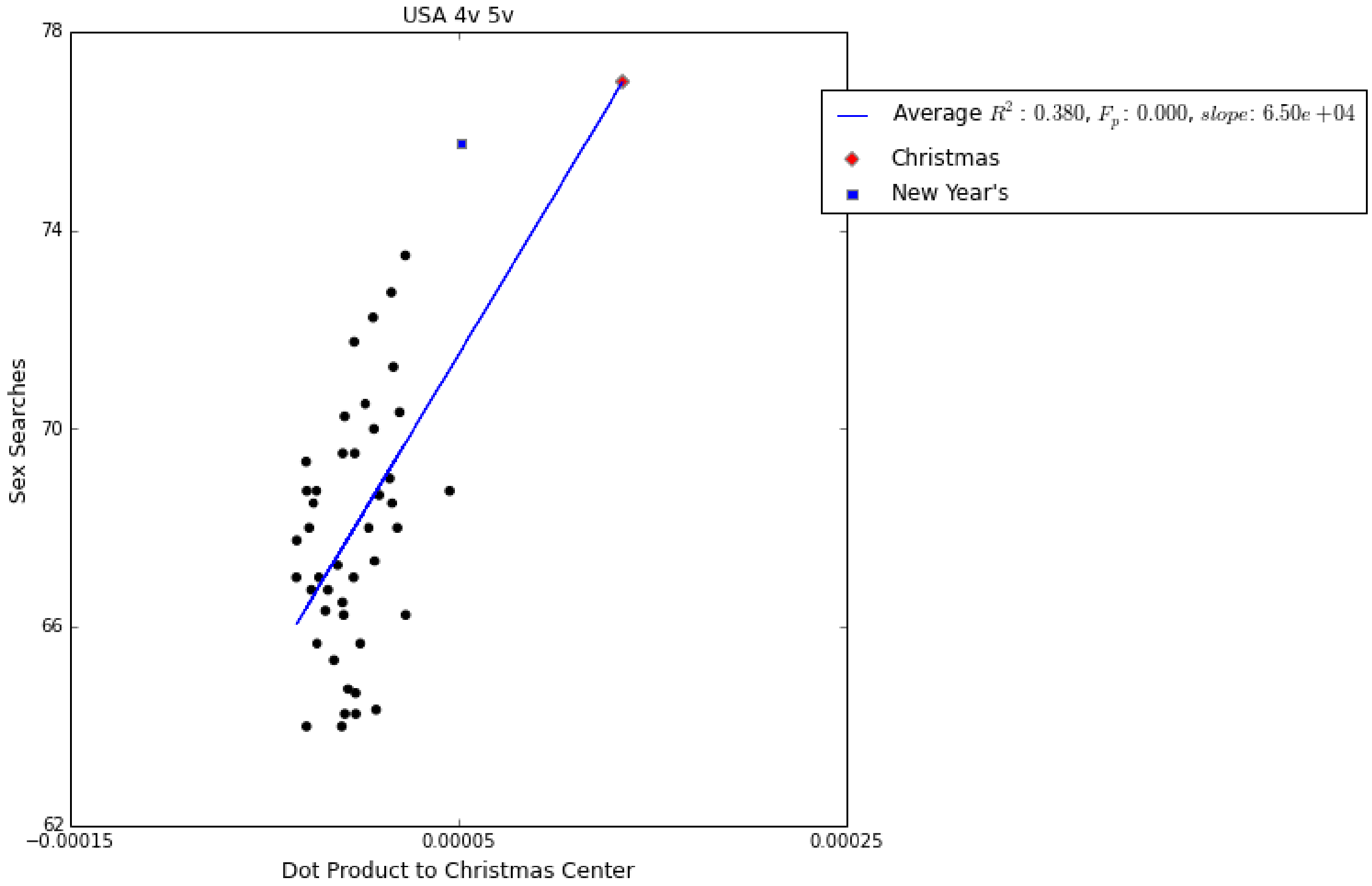

## *USA Eid-al-Fitr*

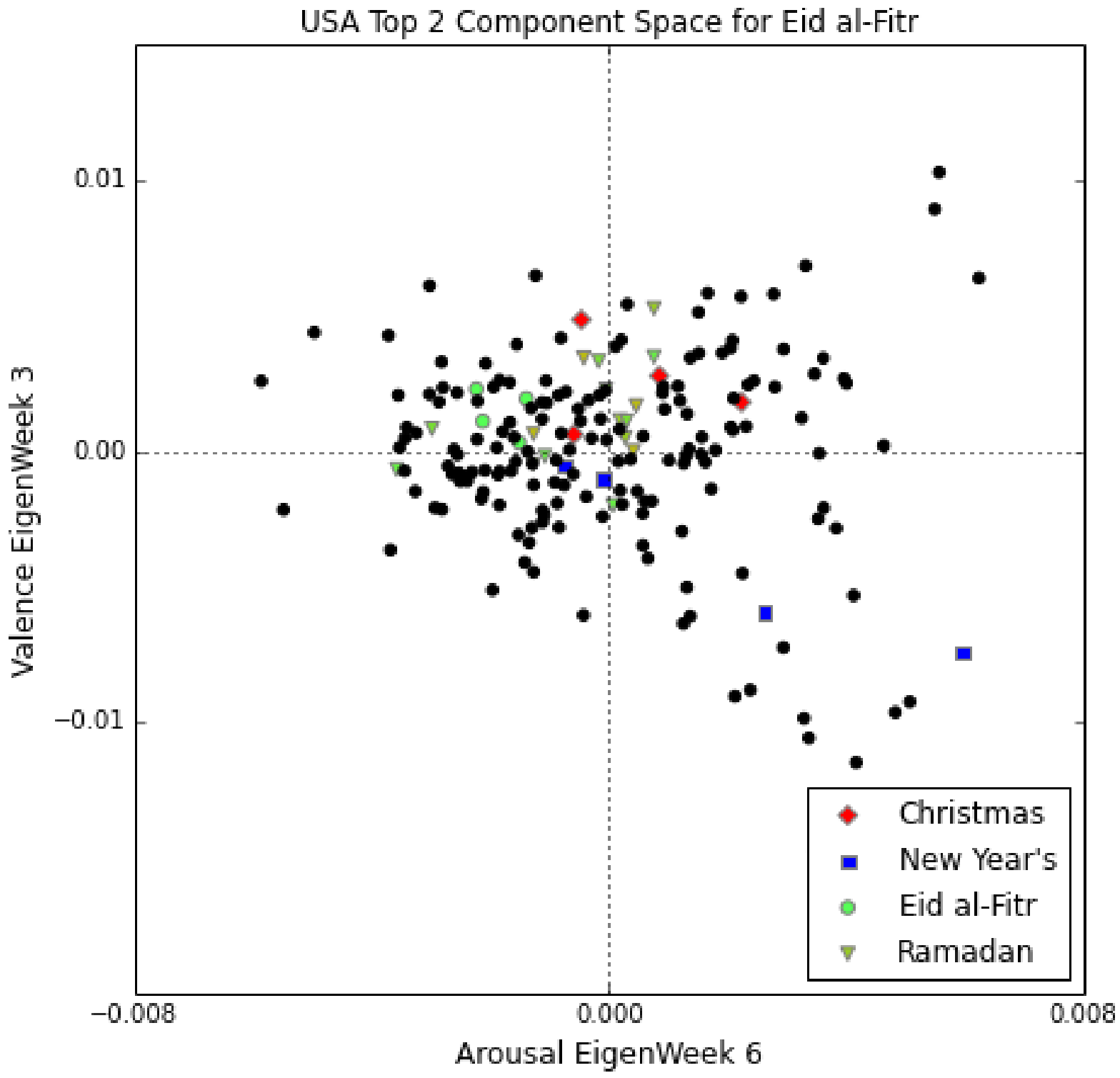

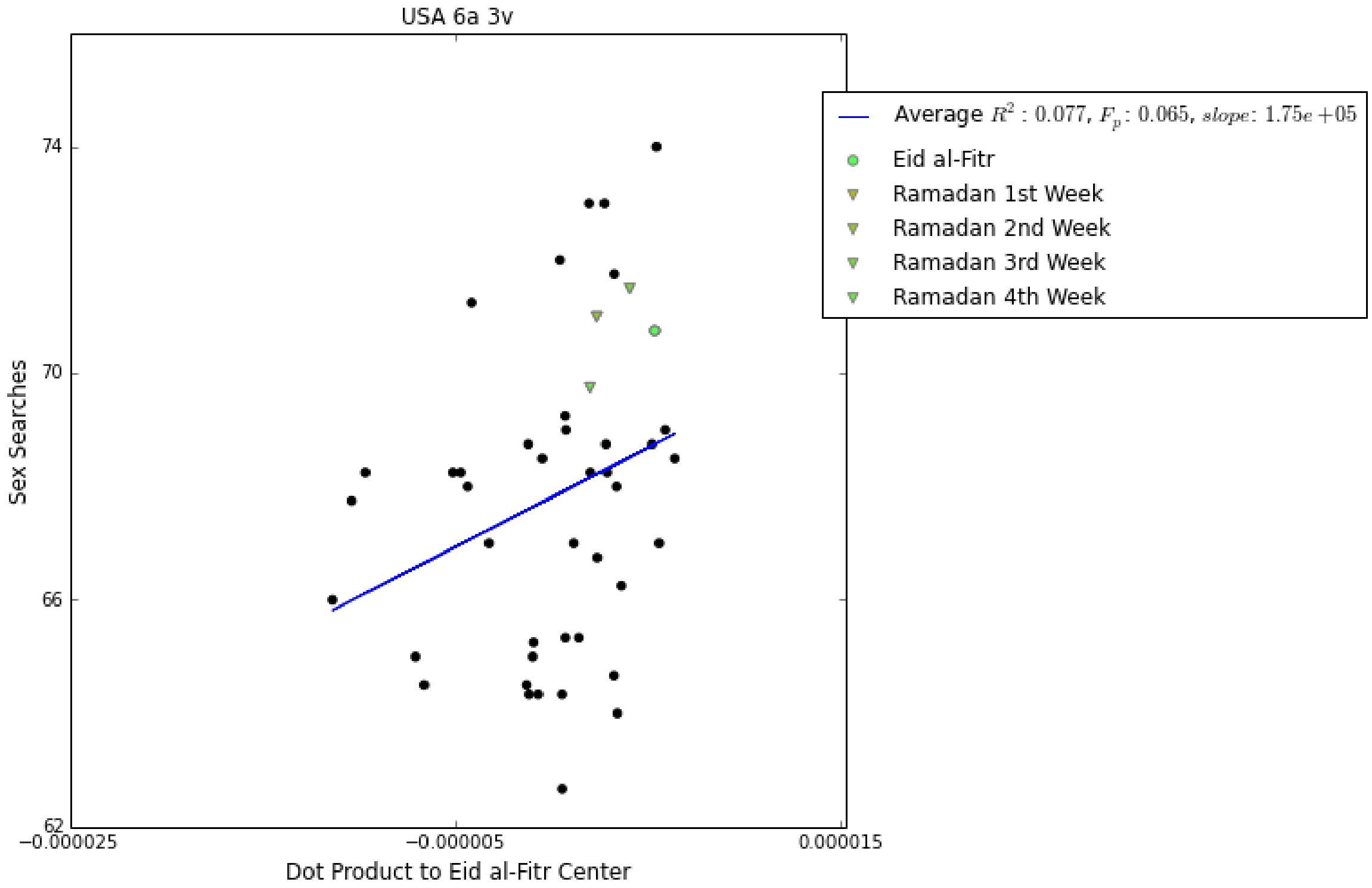

## *Australia Christmas*

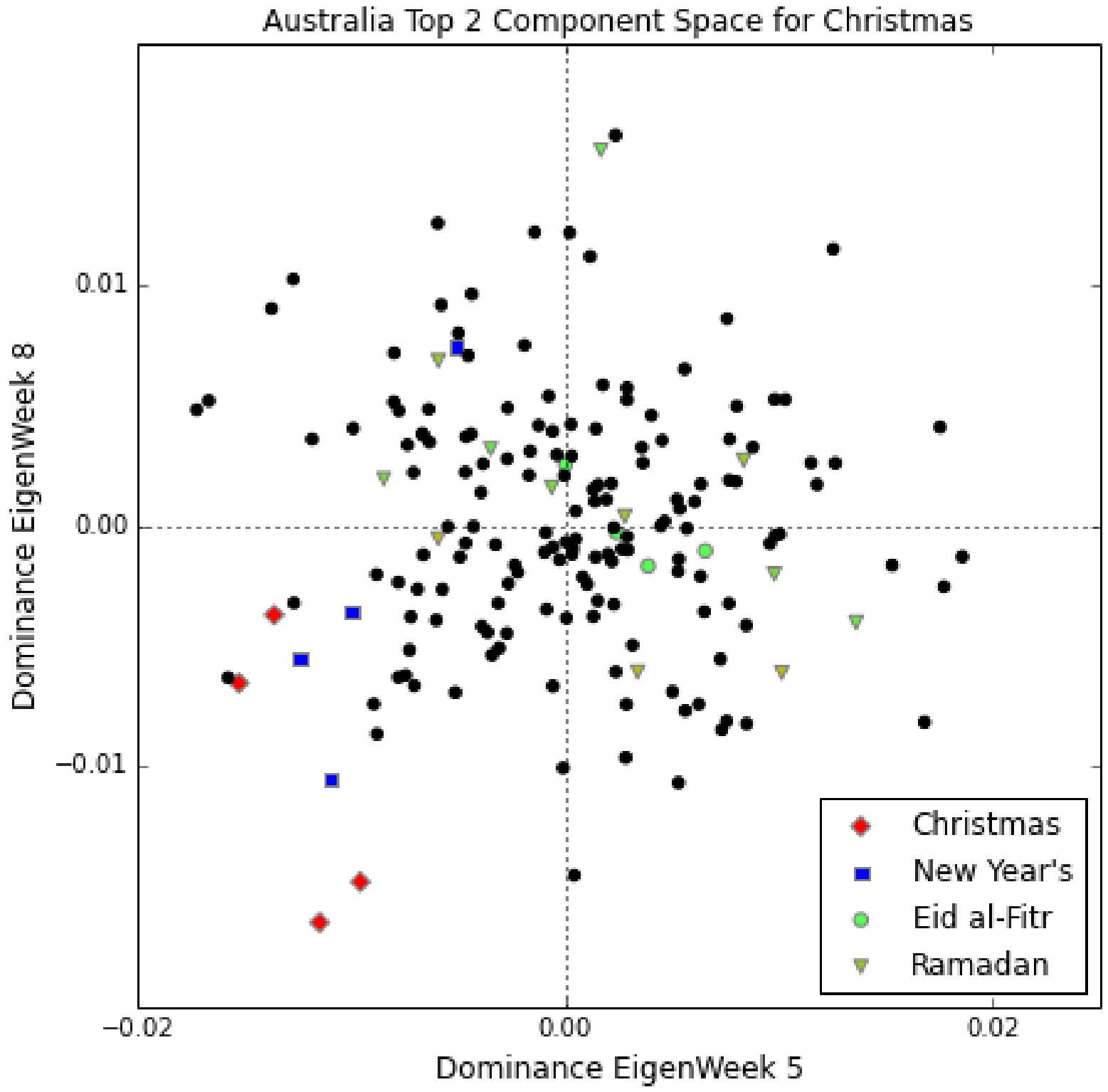

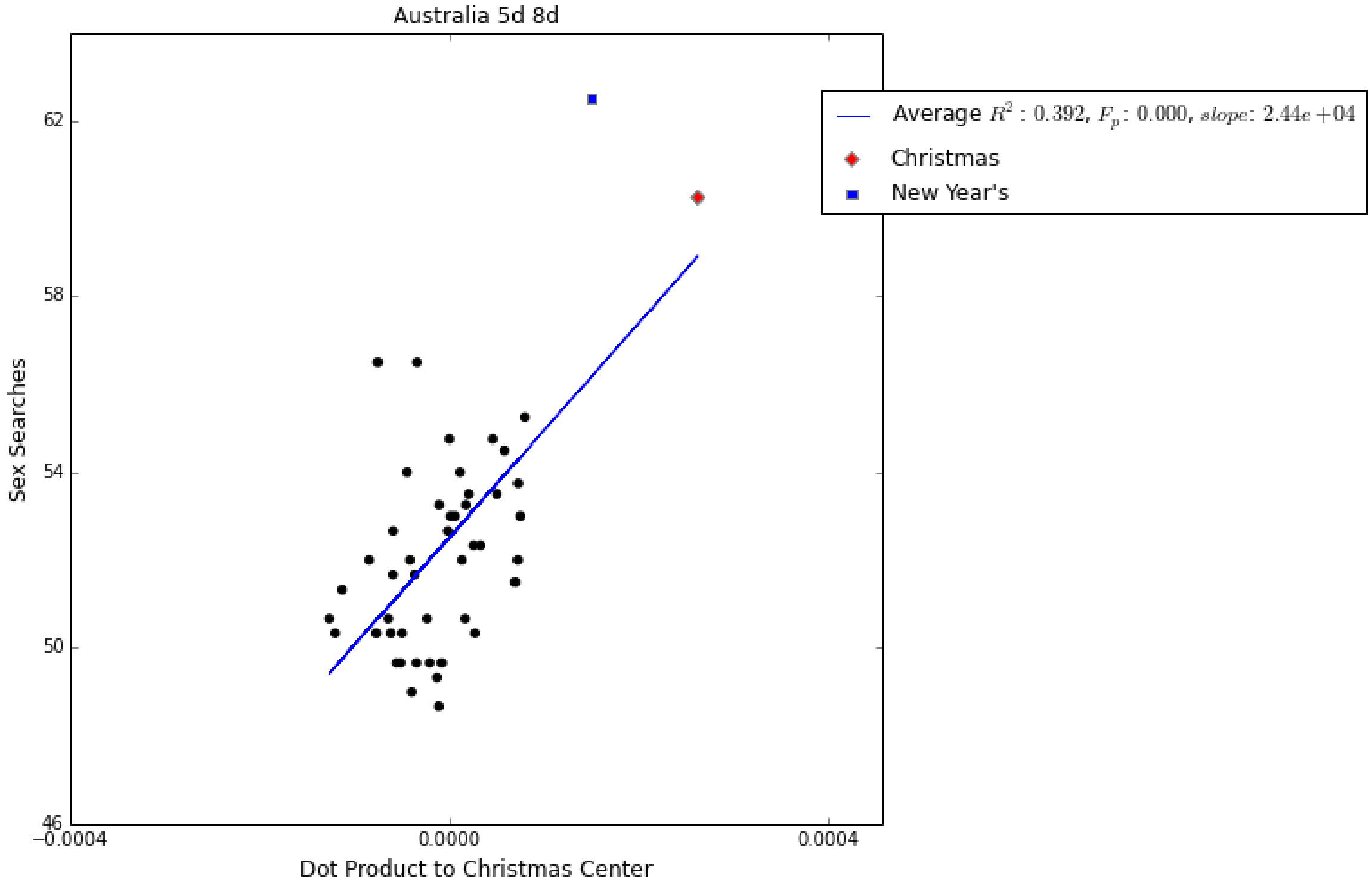

## *Australia Eid-al-Fitr*

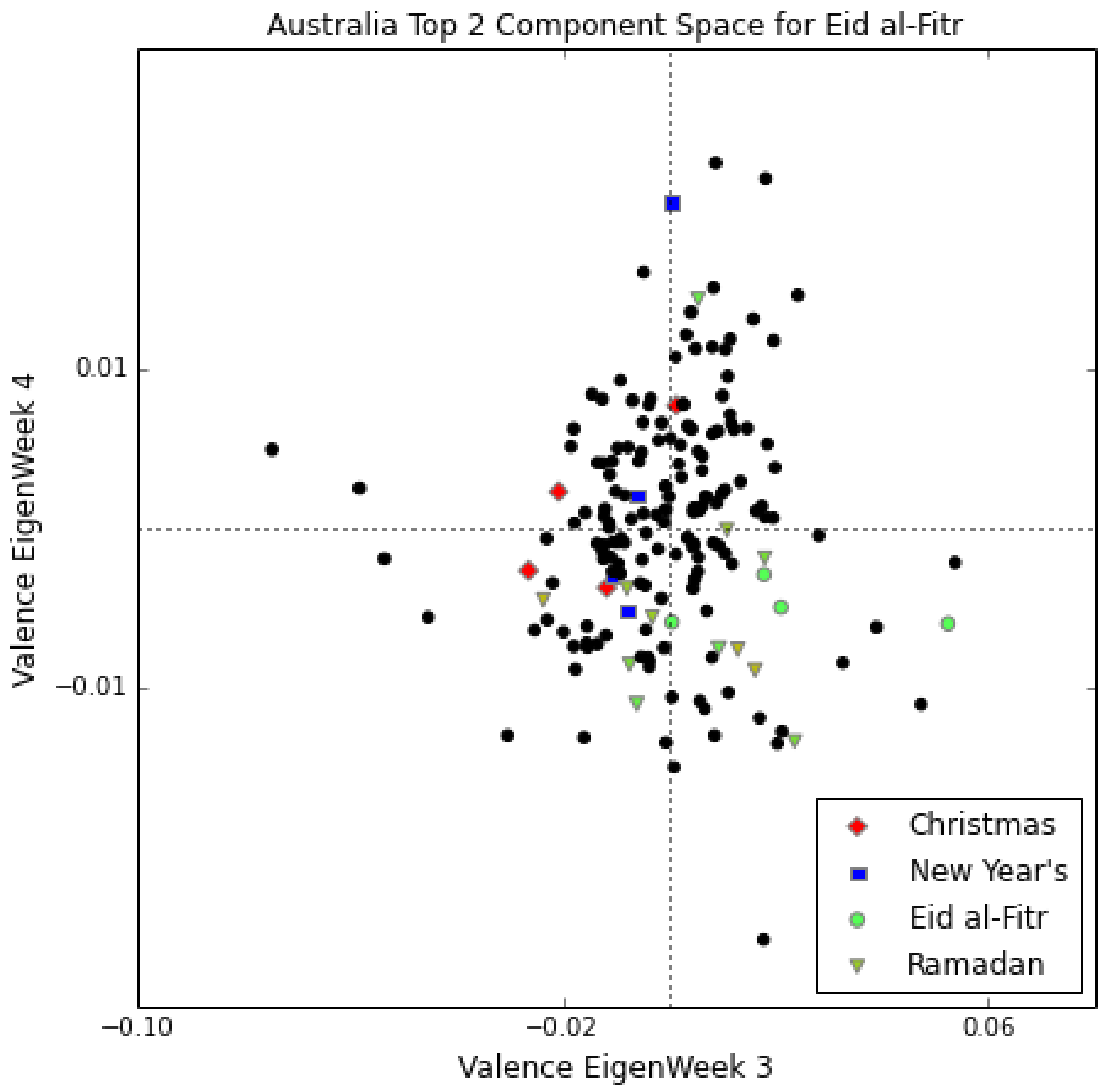

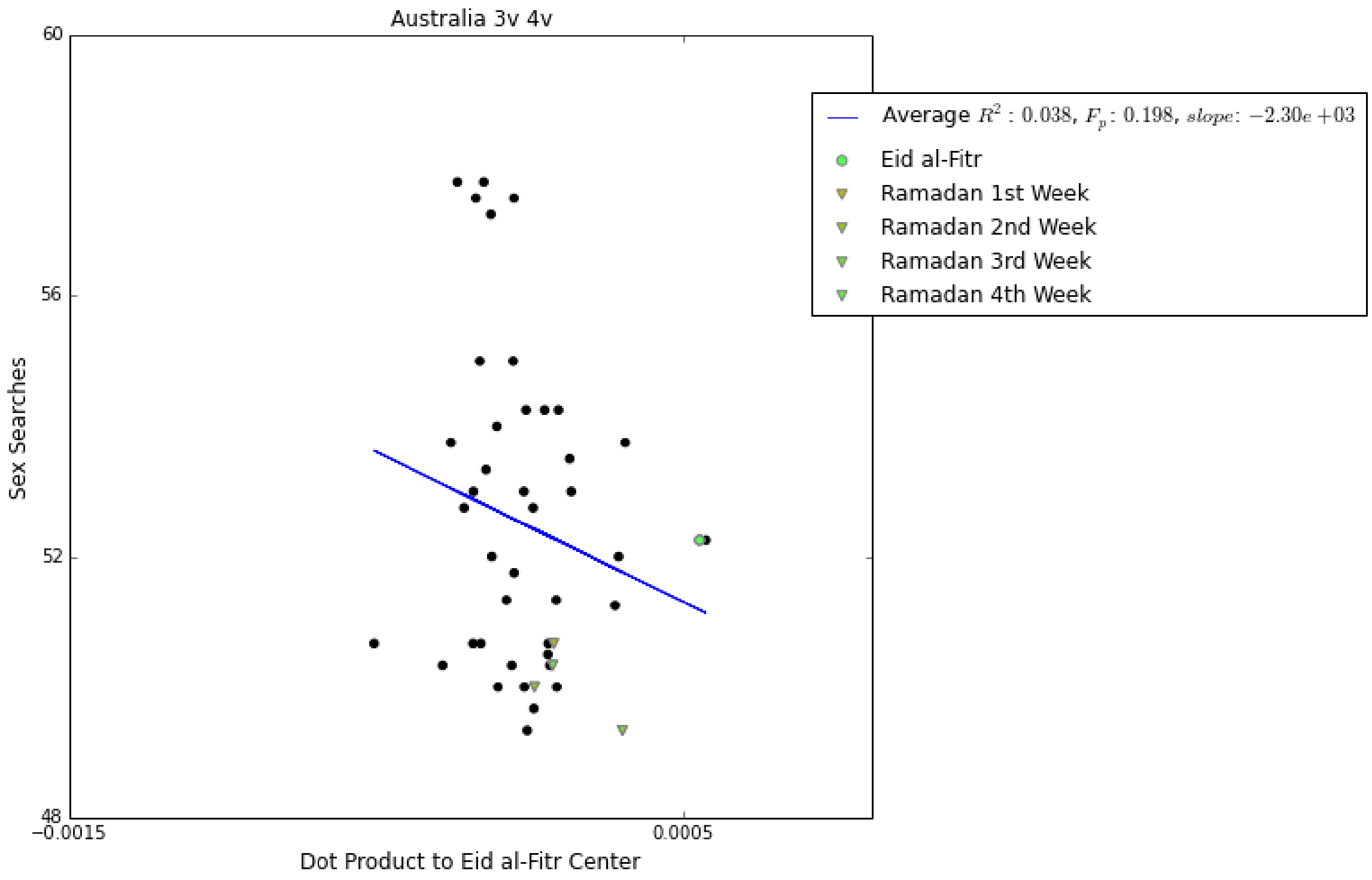

## *Brazil Christmas*

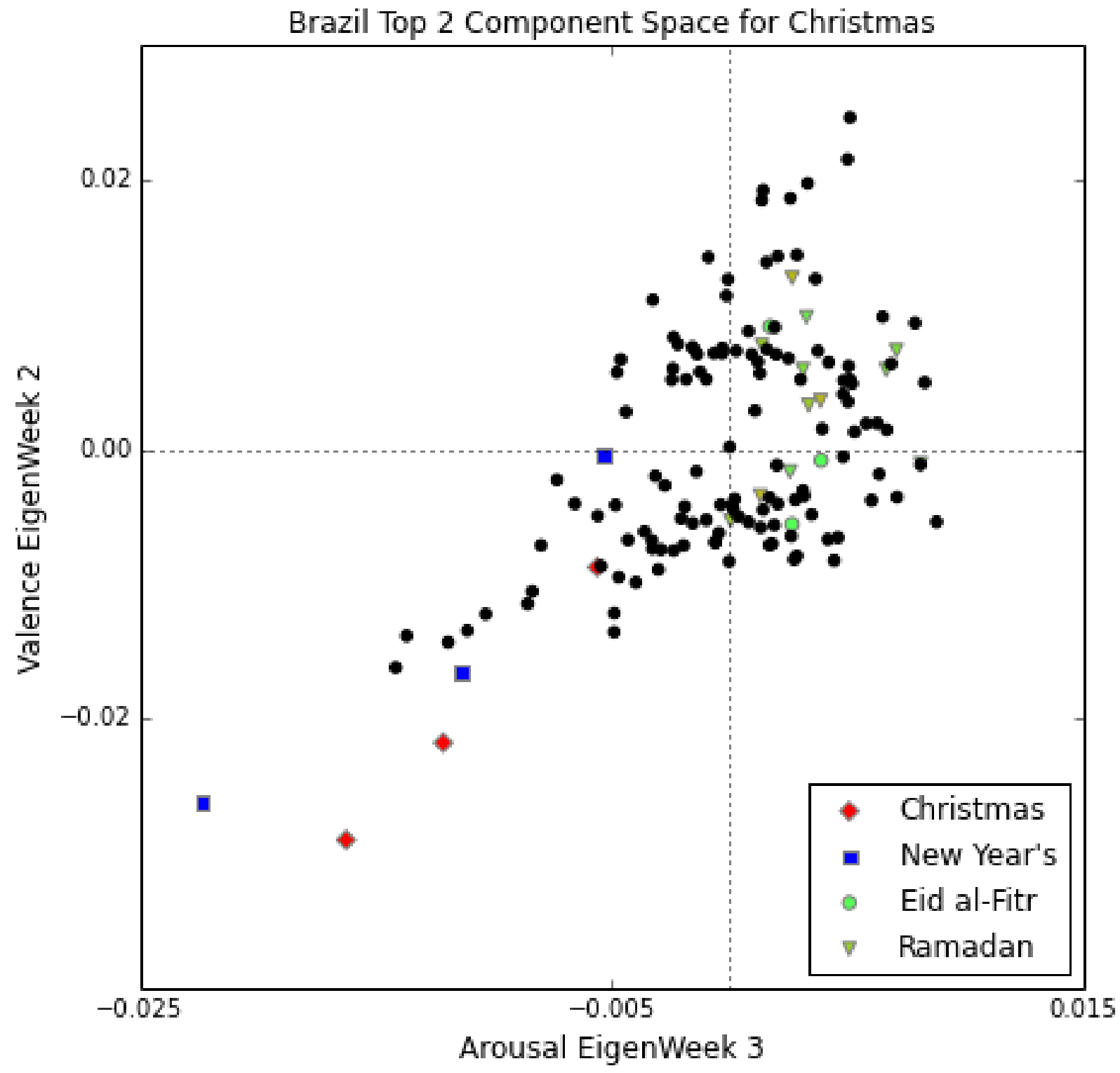

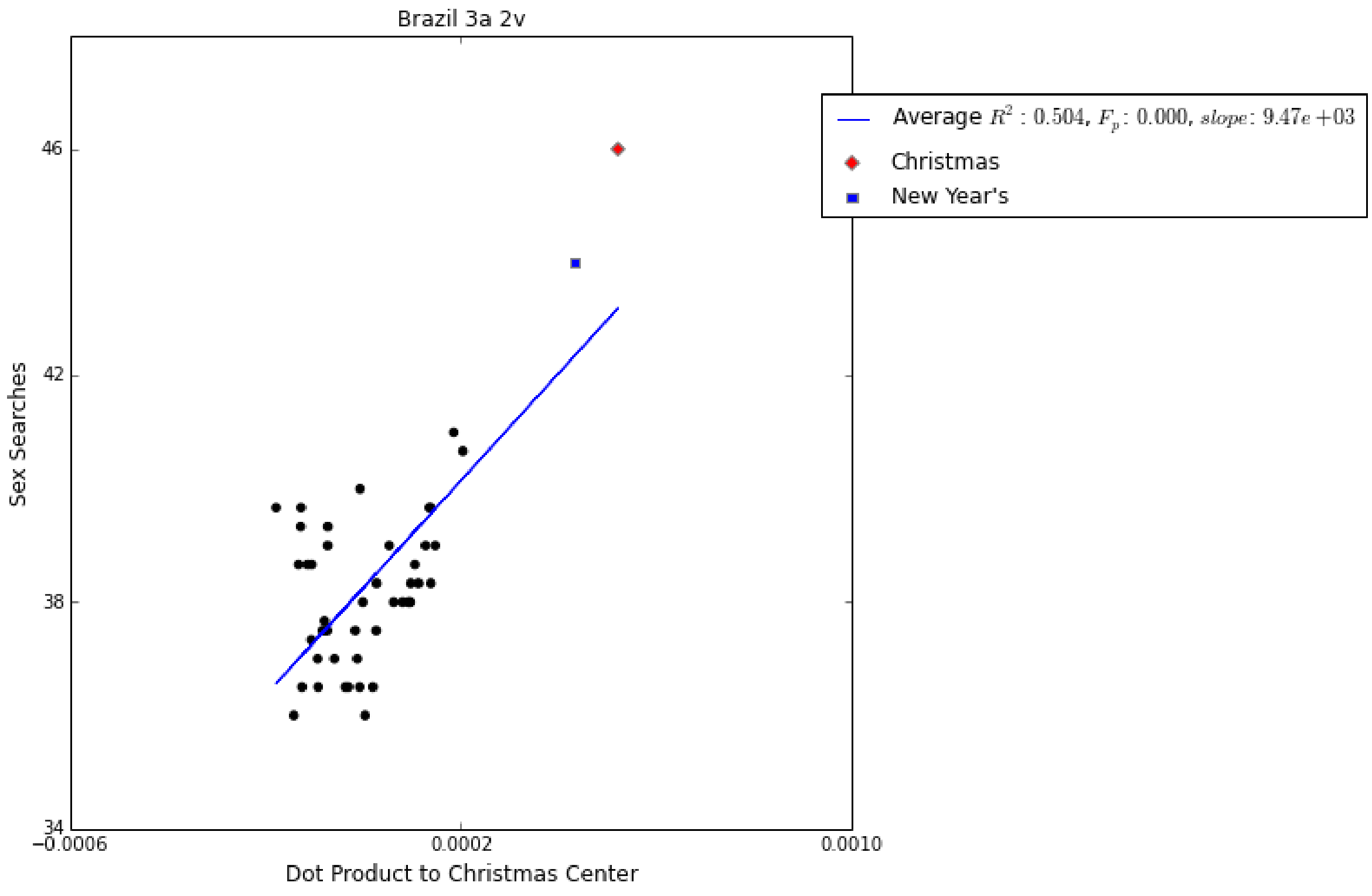

## *Brazil Eid-al-Fitr*

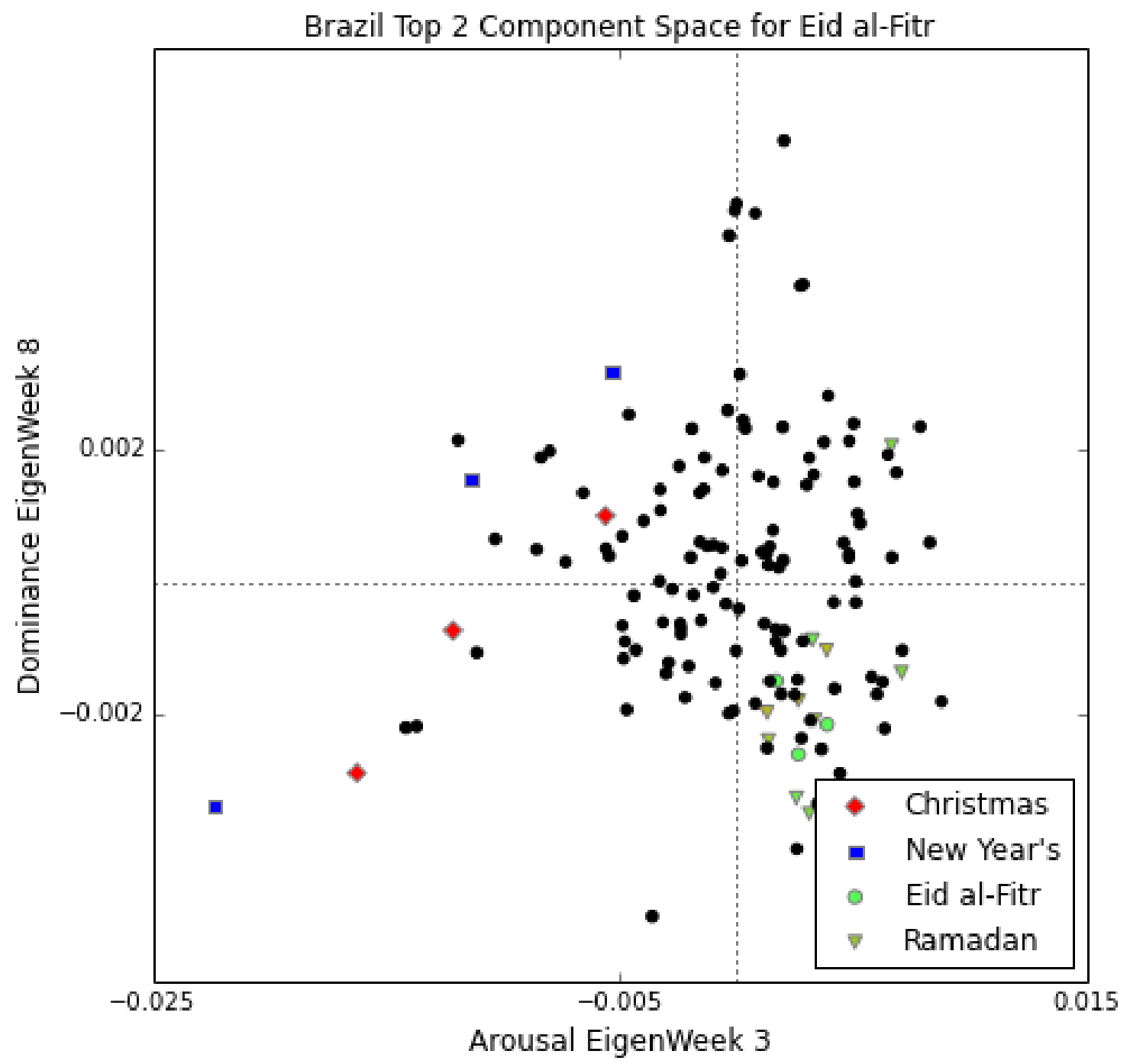

Brazil Top 2 Component Space for Eid al-Fitr
Dominance EigenWeek 8
0.002
−0.002
−0.025
−0.005
0.015
Arousal EigenWeek 3
Christmas
New Year's
Eid al-Fitr
Ramadan

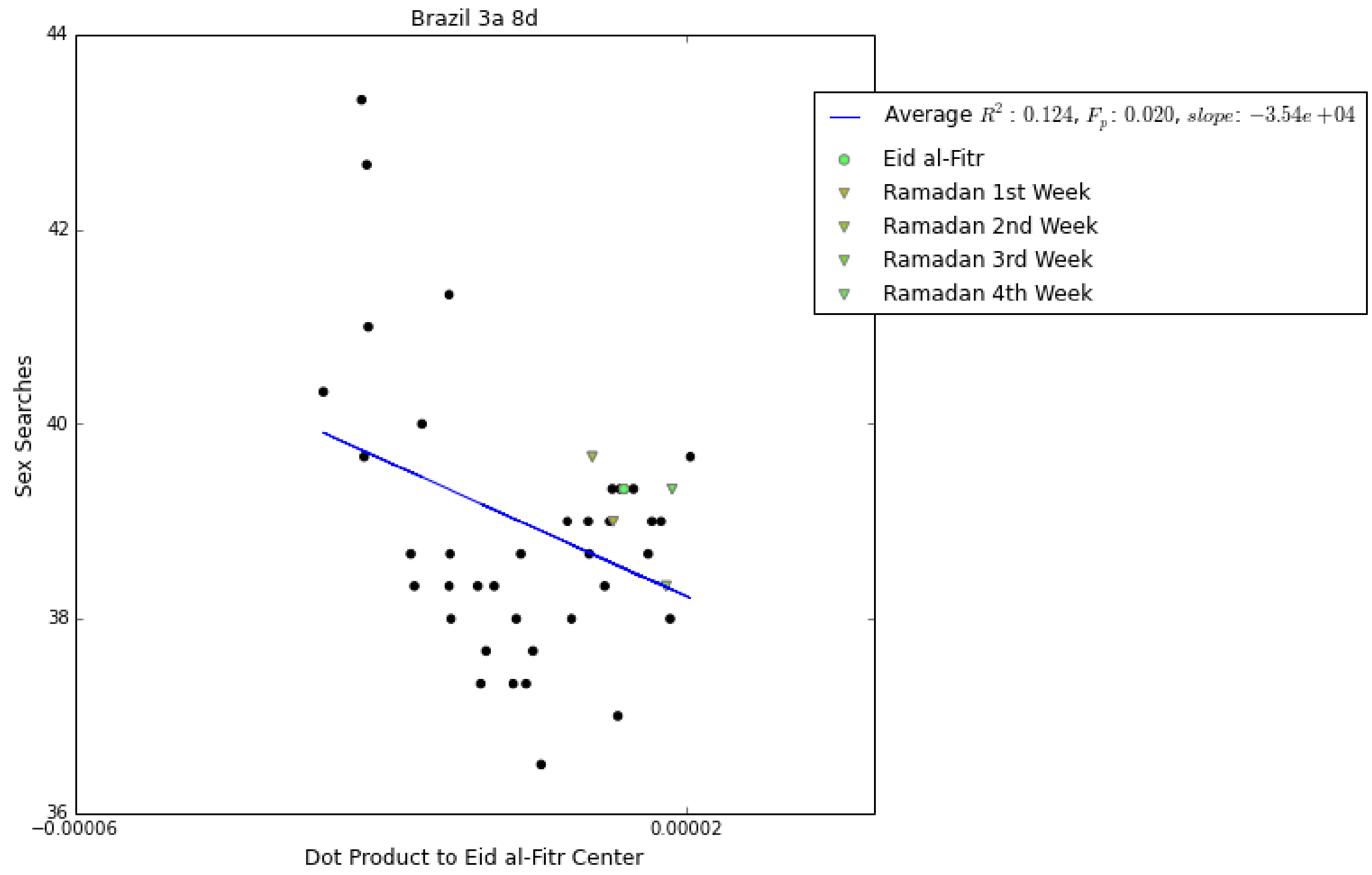

Brazil 3a 8d
Sex Searches
44
42
40
38
36
−0.00006
0.00002
Dot Product to Eid al-Fitr Center
Average $R^2$ : 0.124, $F_p$: 0.020, $slope$: $-3.54e+04$
Eid al-Fitr
Ramadan 1st Week
Ramadan 2nd Week
Ramadan 3rd Week
Ramadan 4th Week

## *Argentina Christmas*

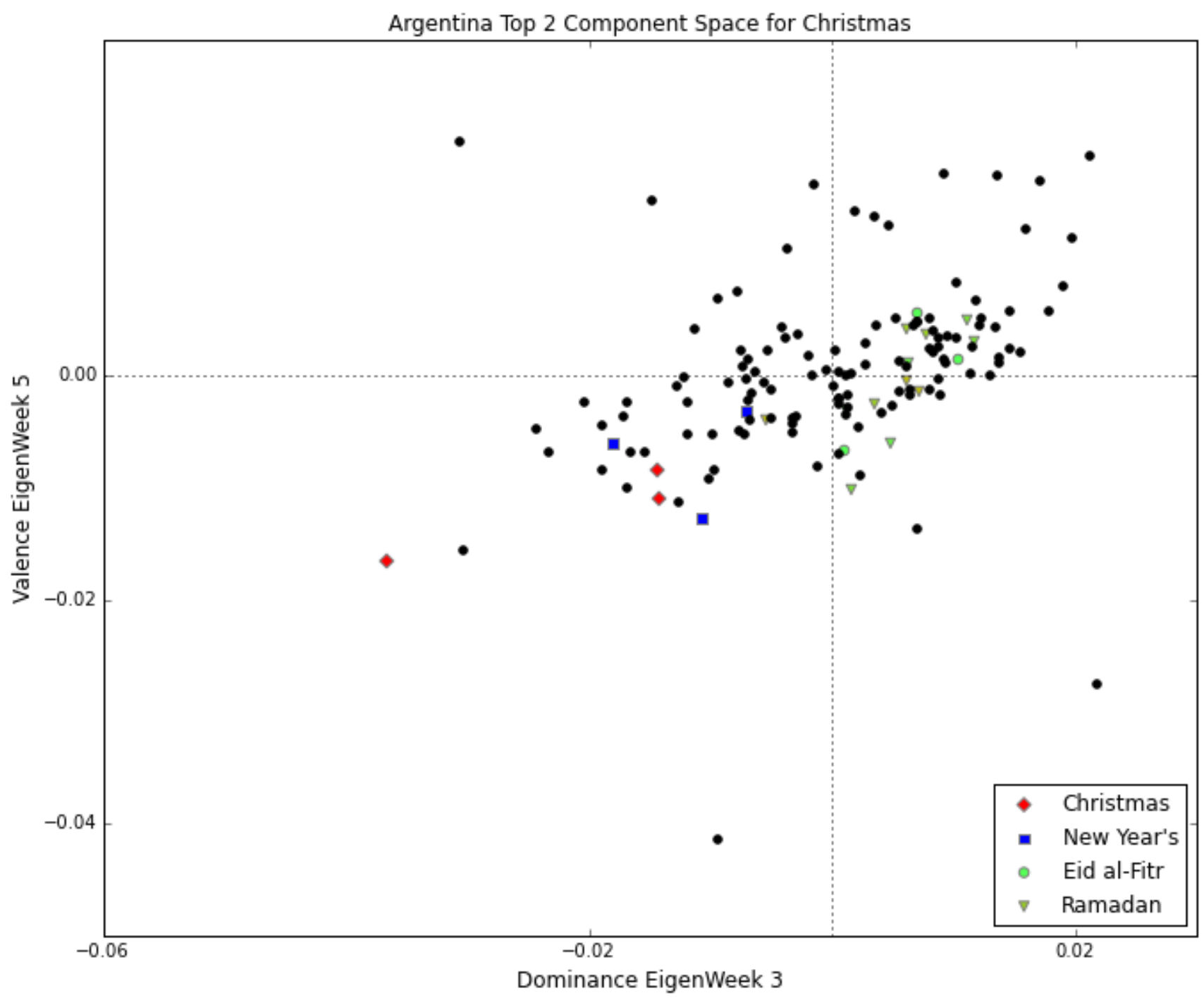

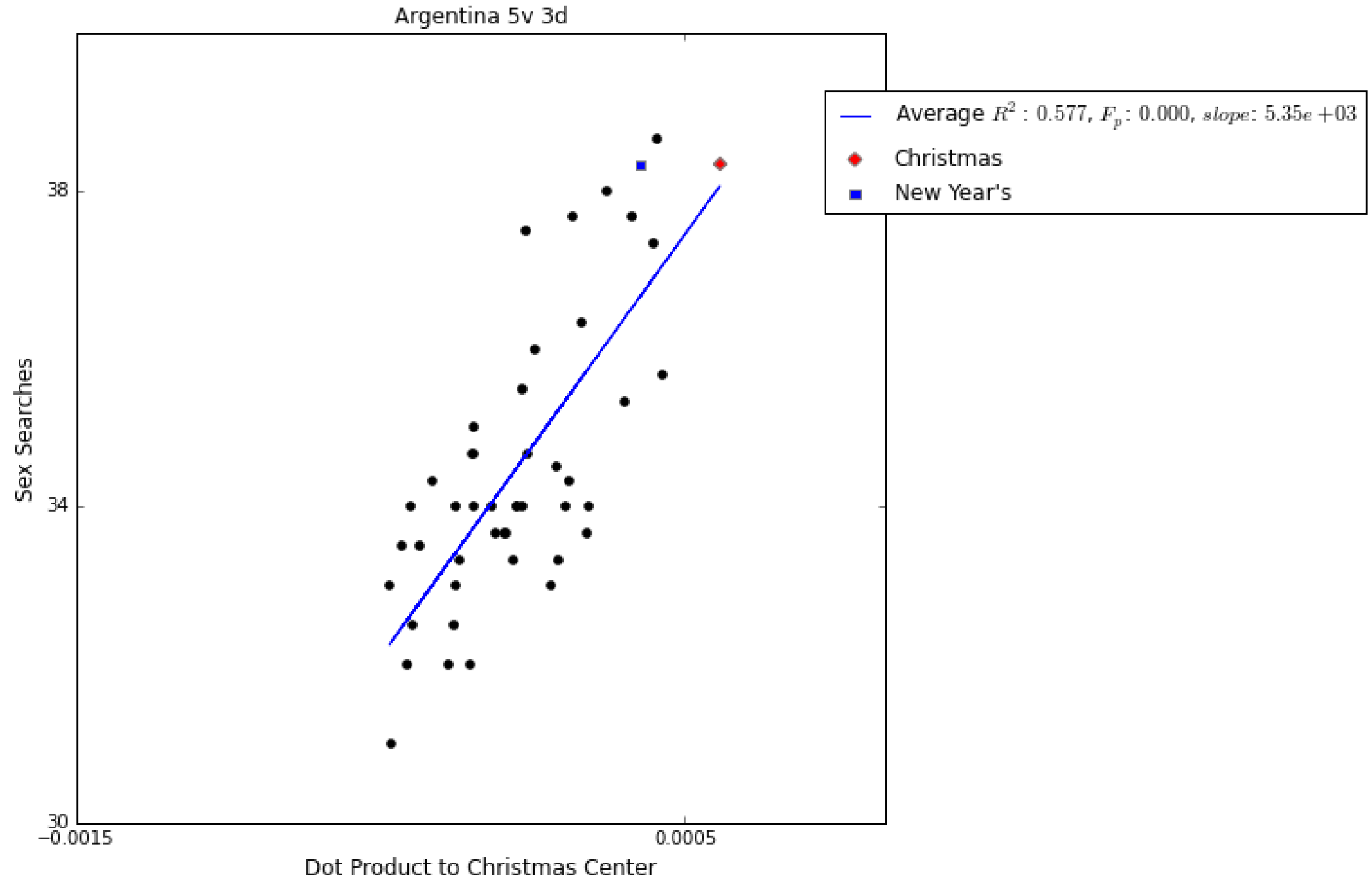

## *Argentina Eid-al-Fitr*

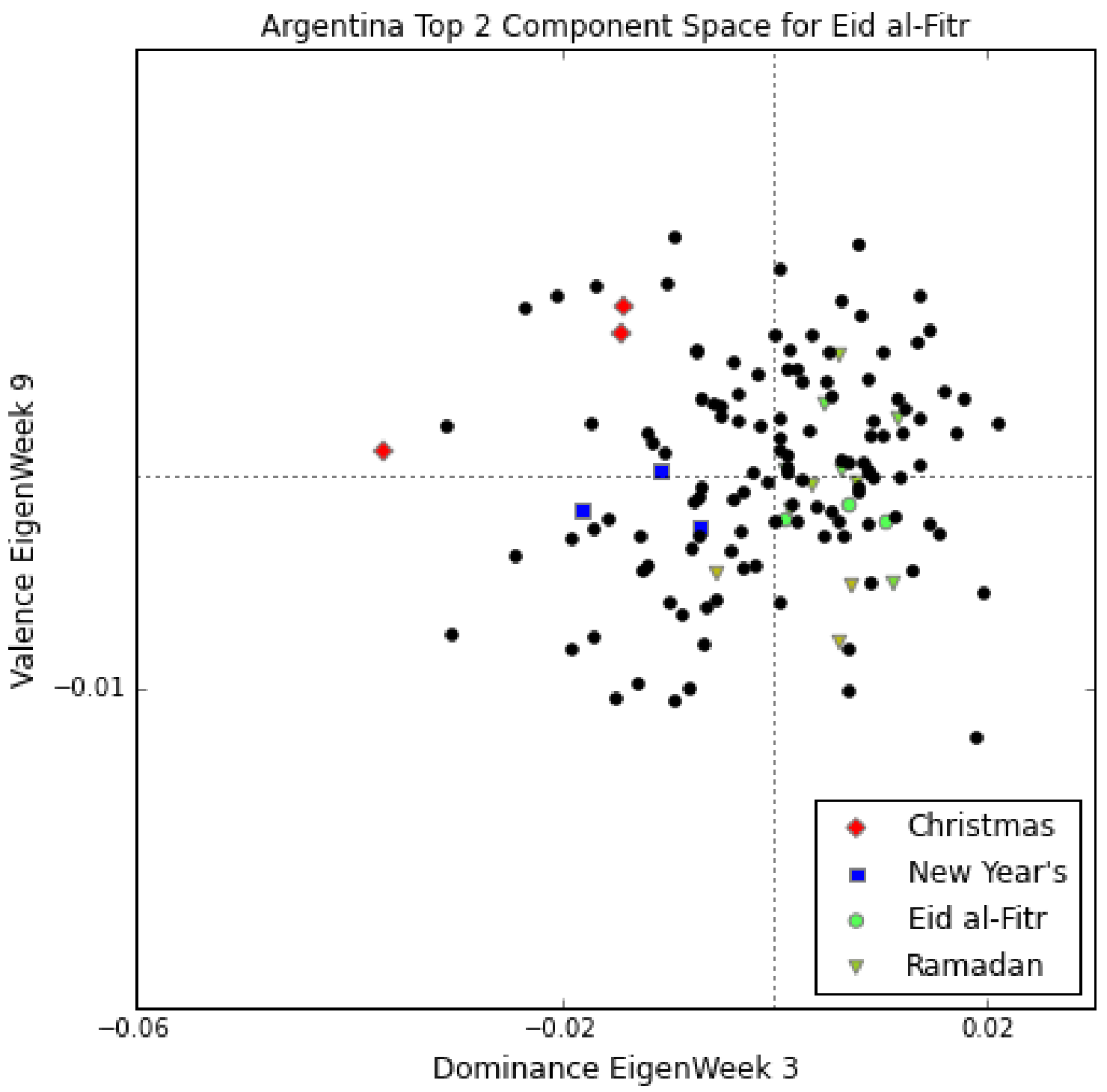

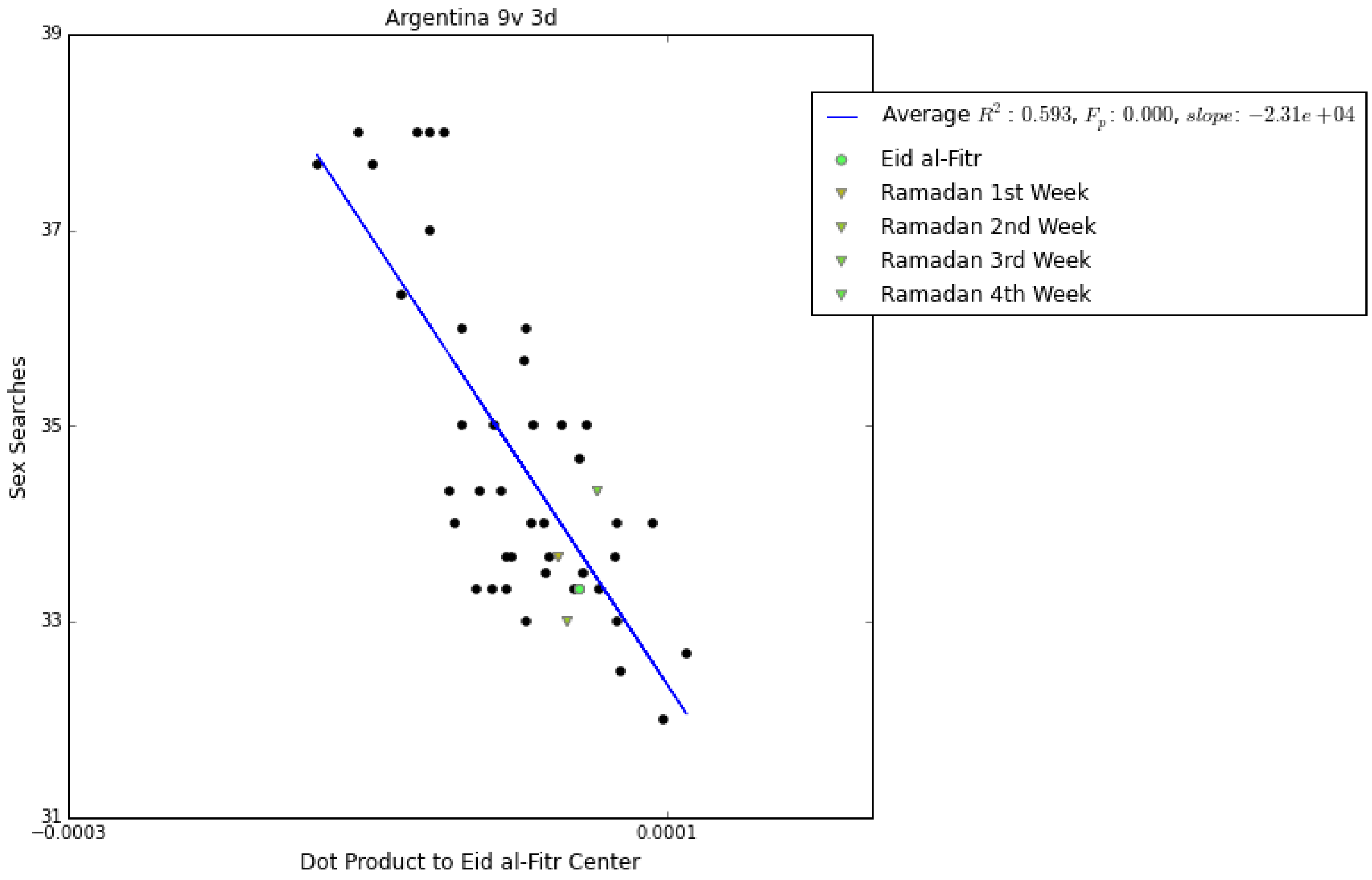

## *Chile Christmas*

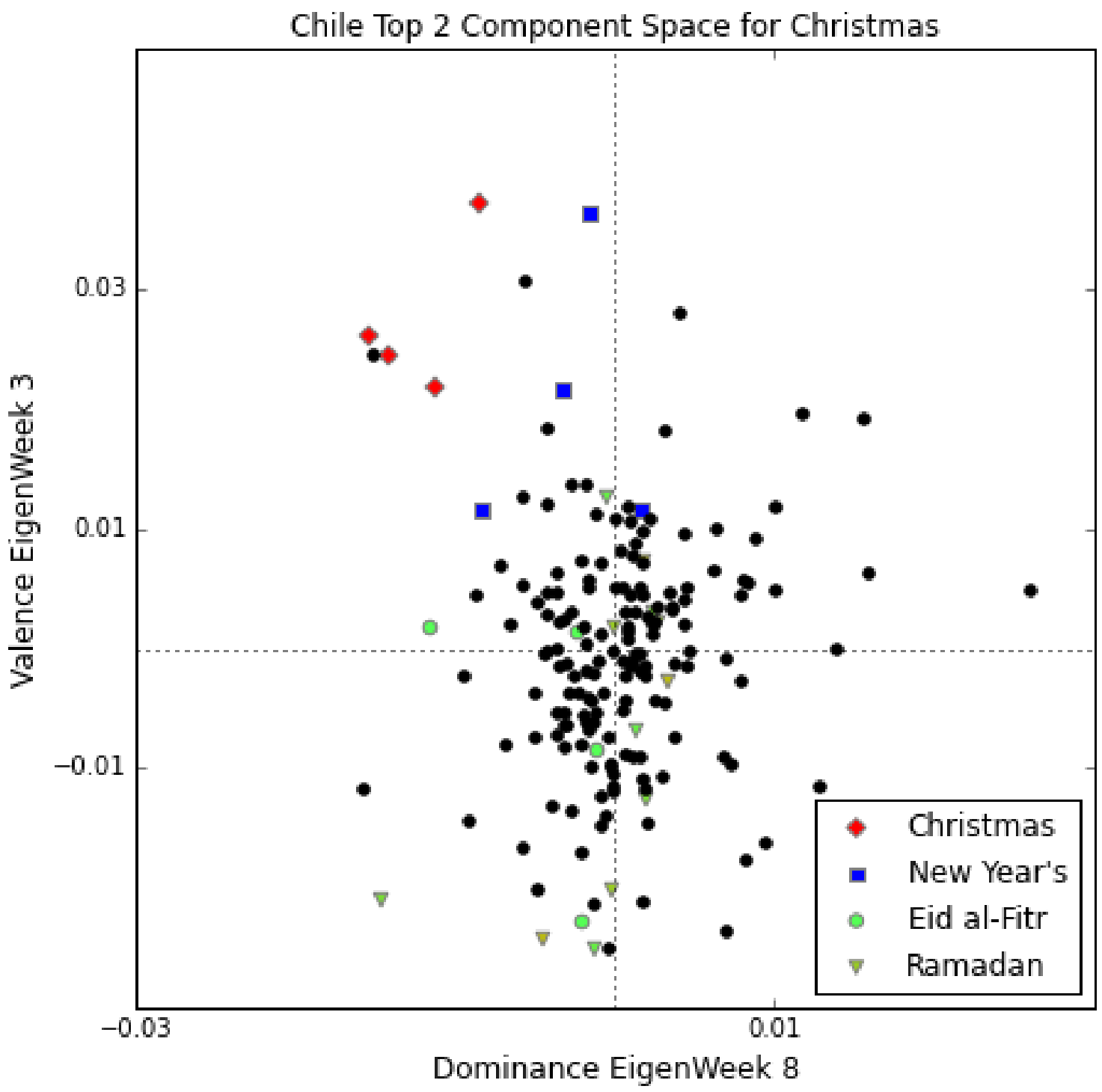

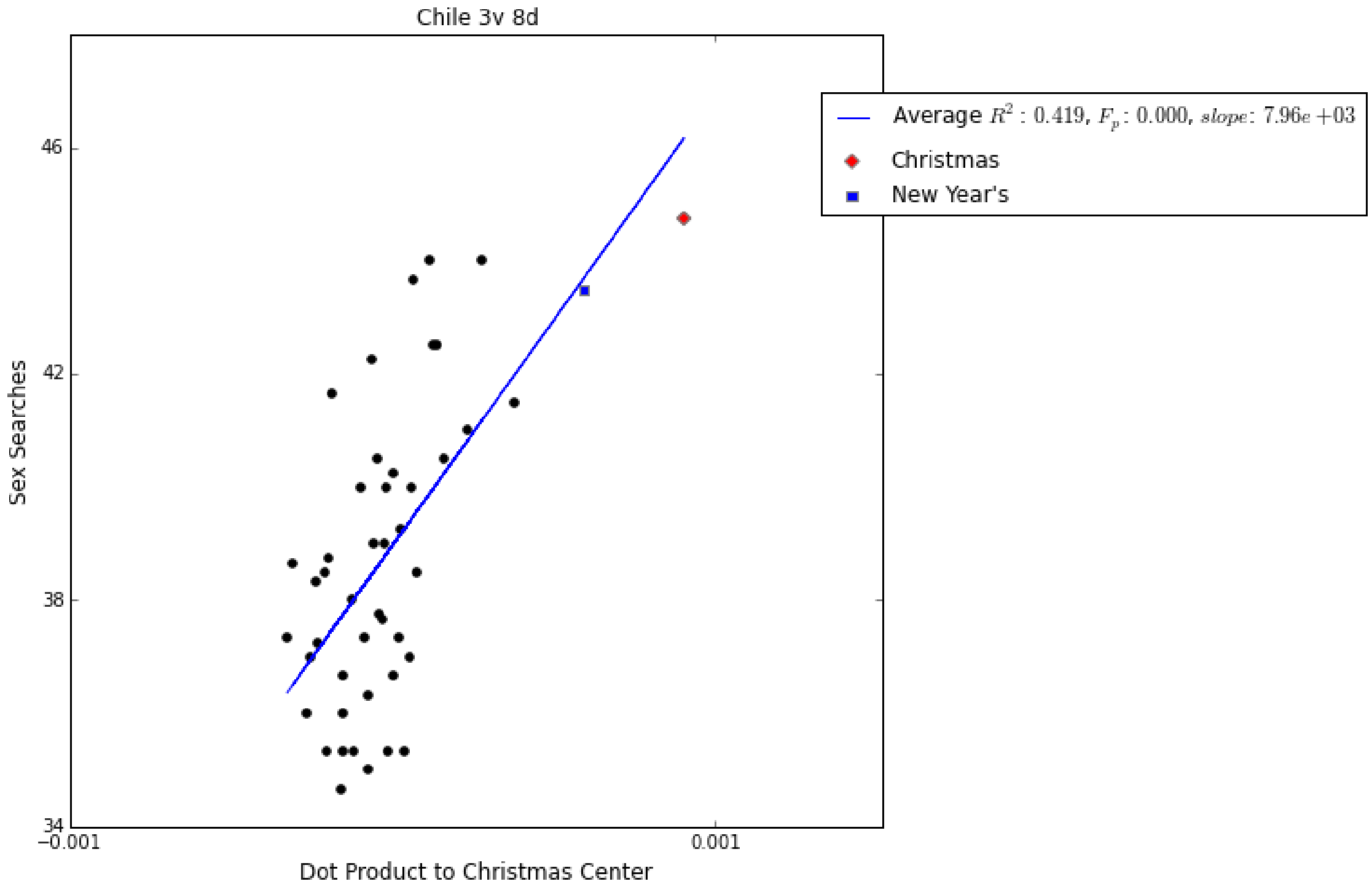

## *Chile Eid-al-Fitr*

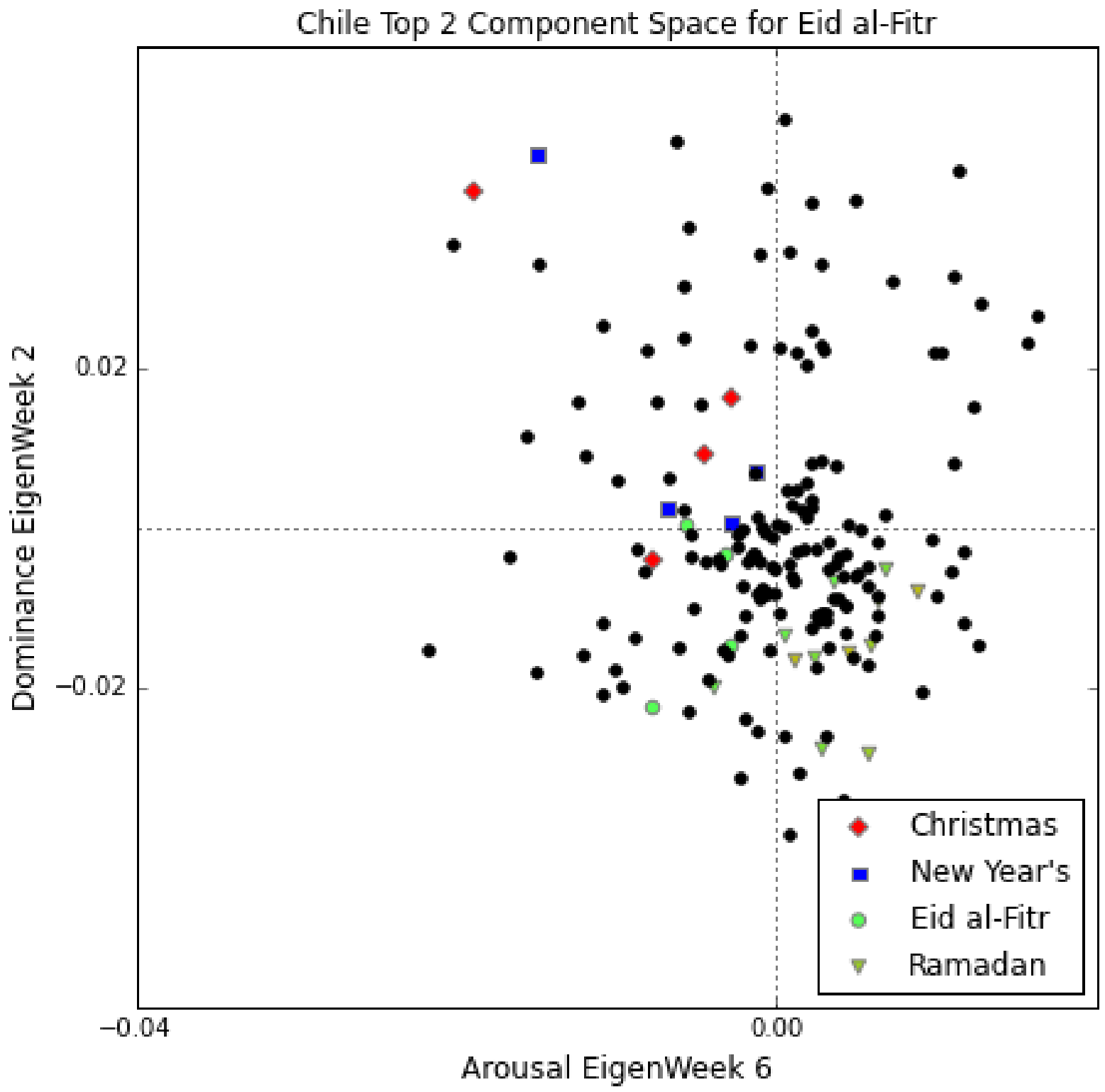

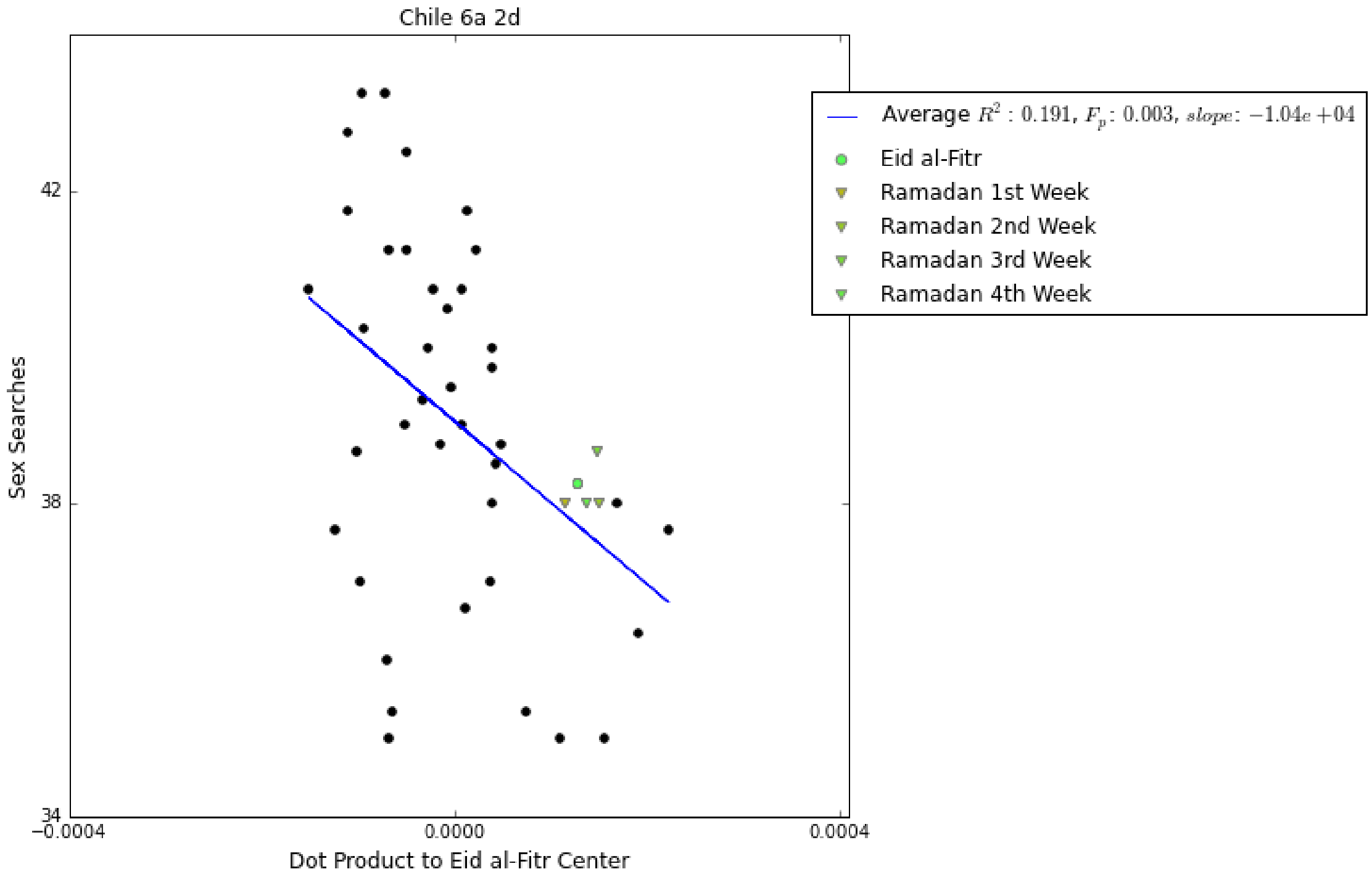

## *Indonesia Christmas*

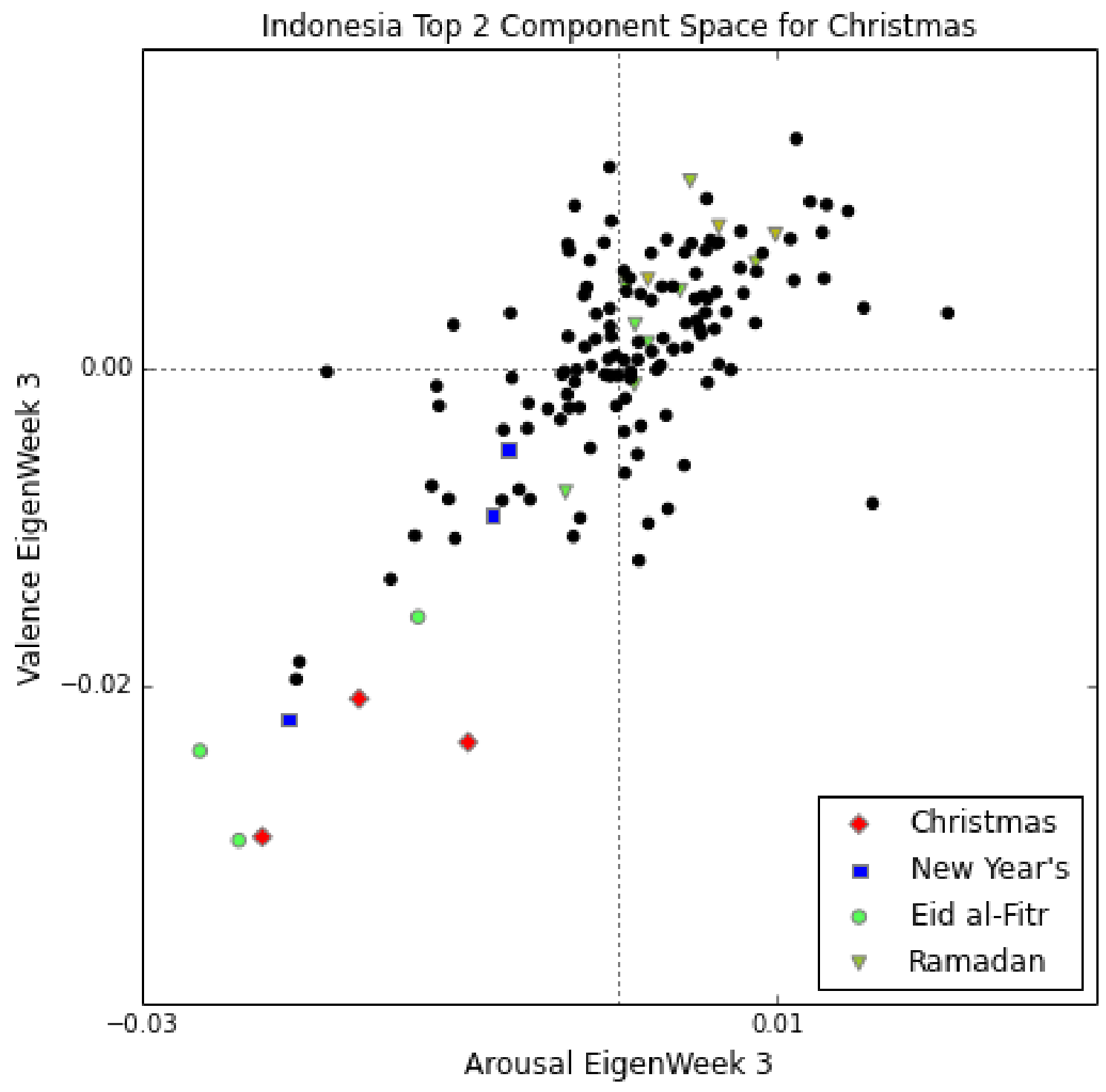

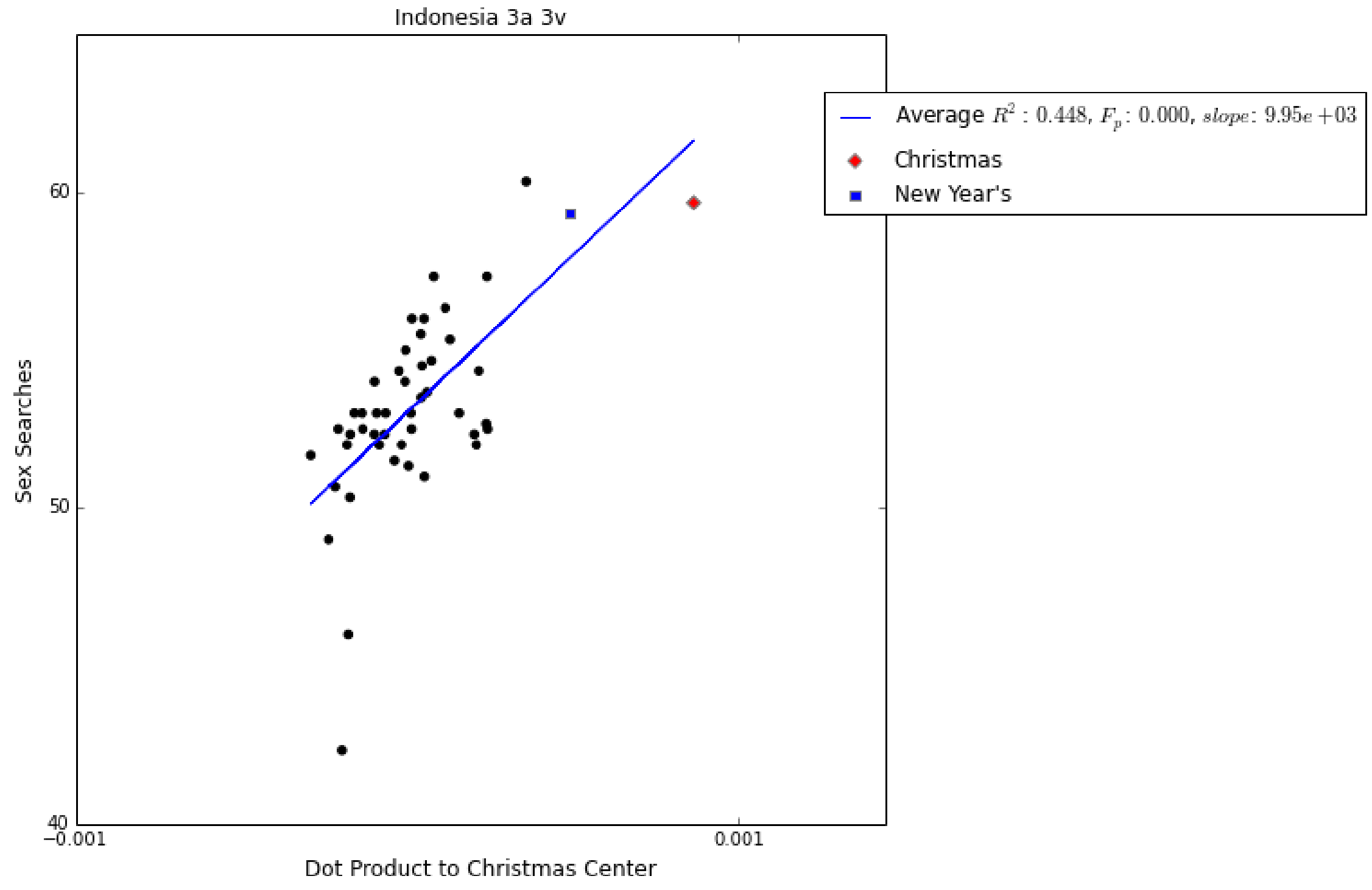

## *Indonesia Eid-al-Fitr*

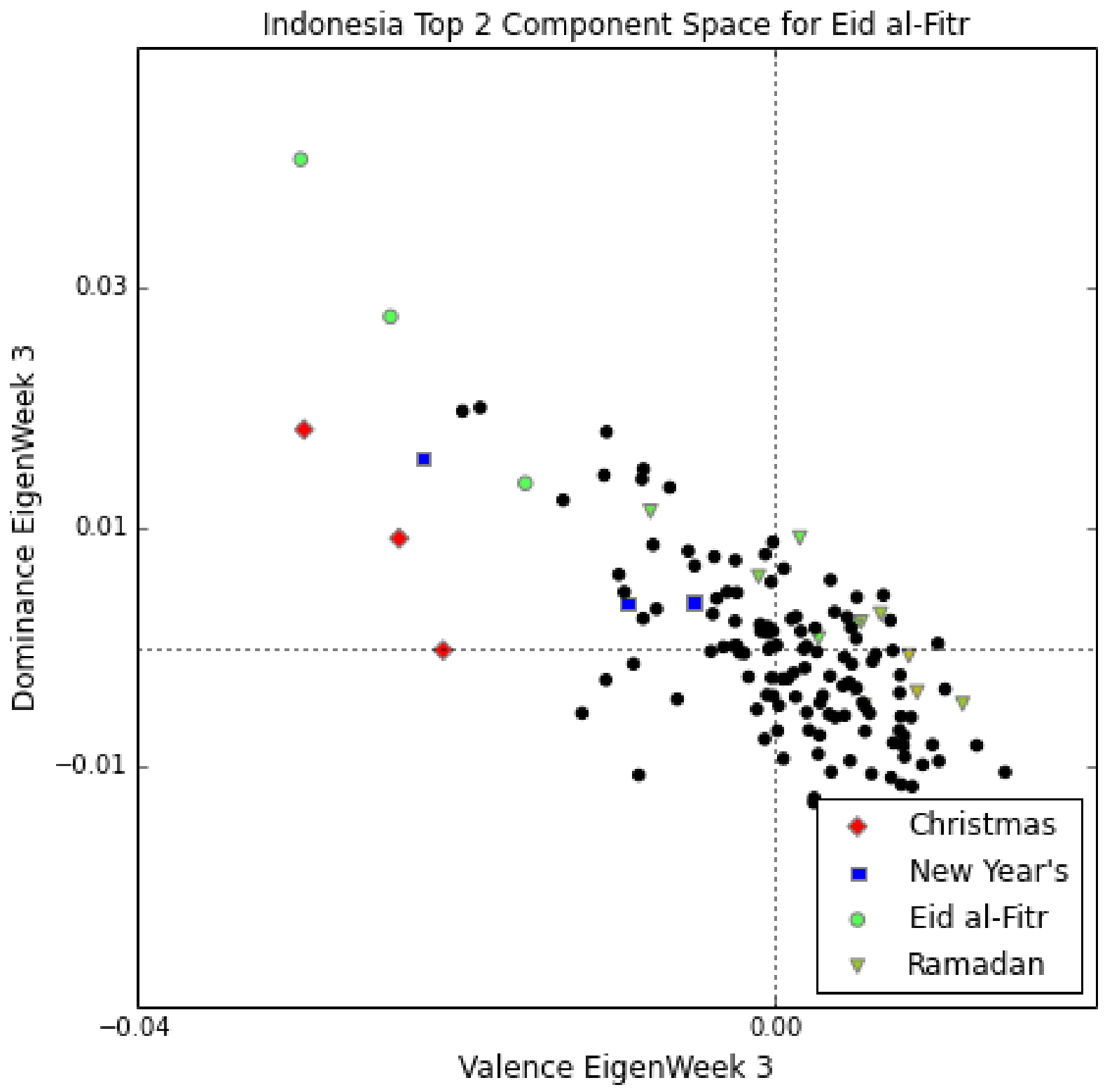

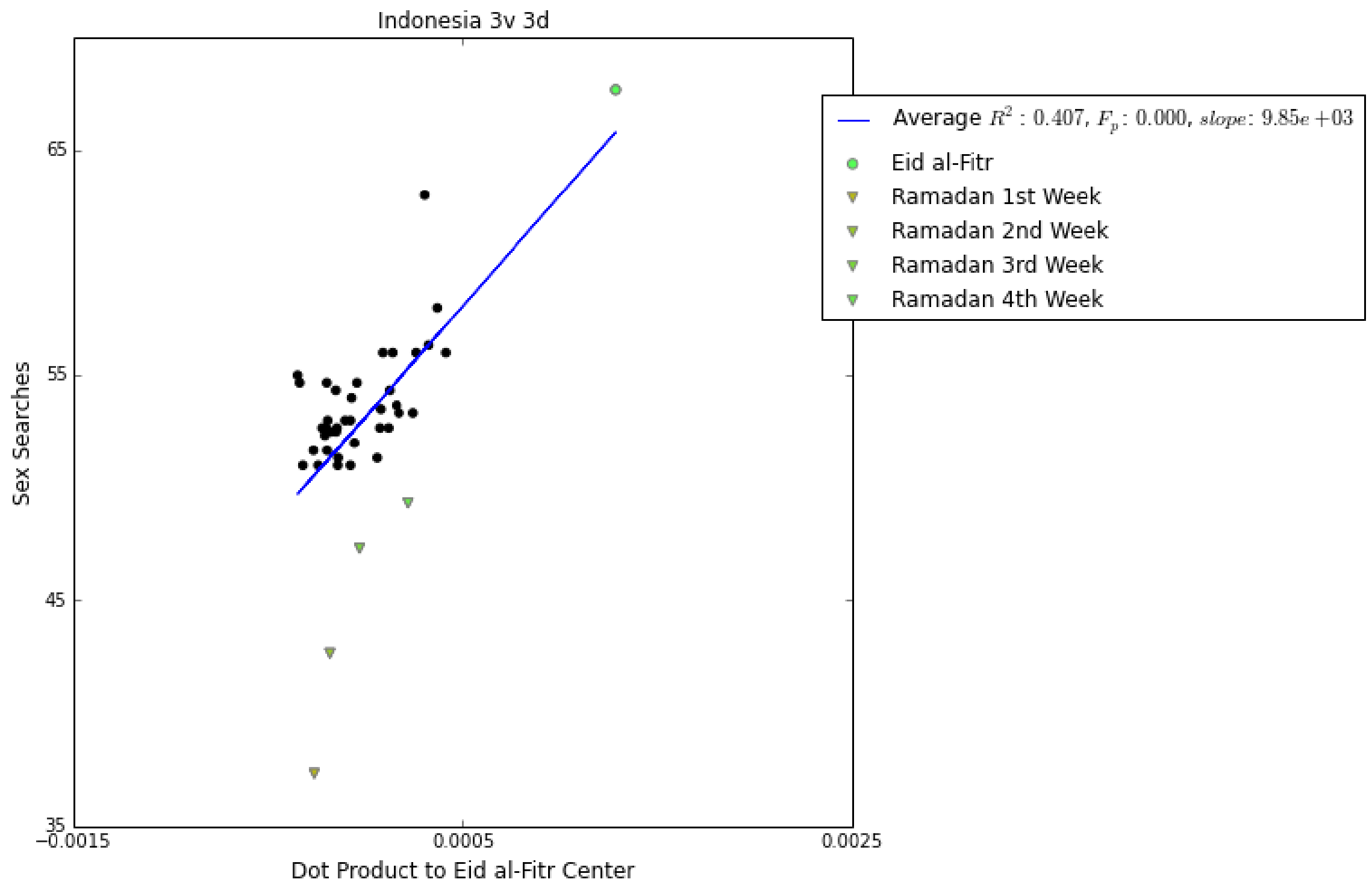

## *Turkey Christmas*

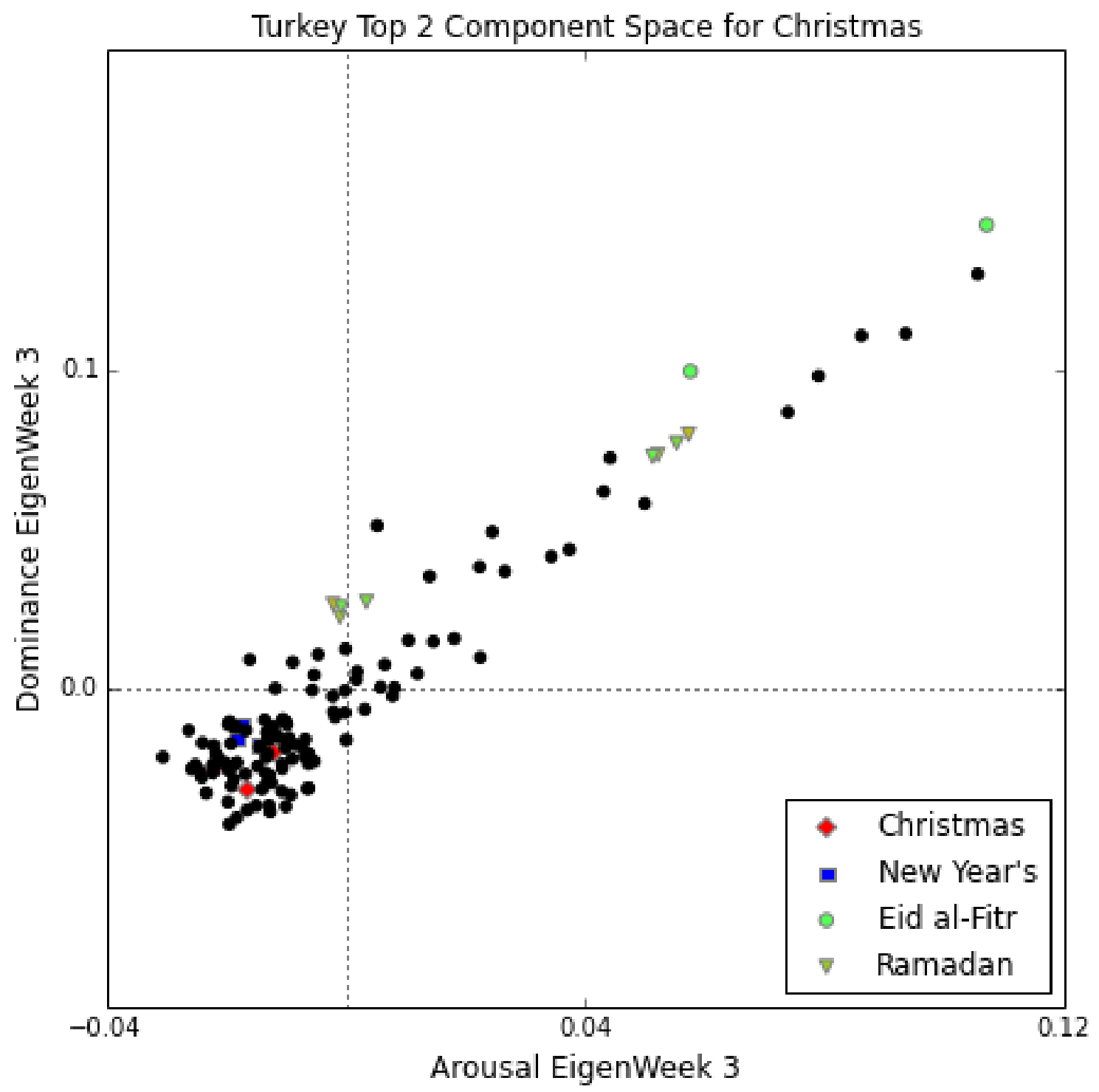

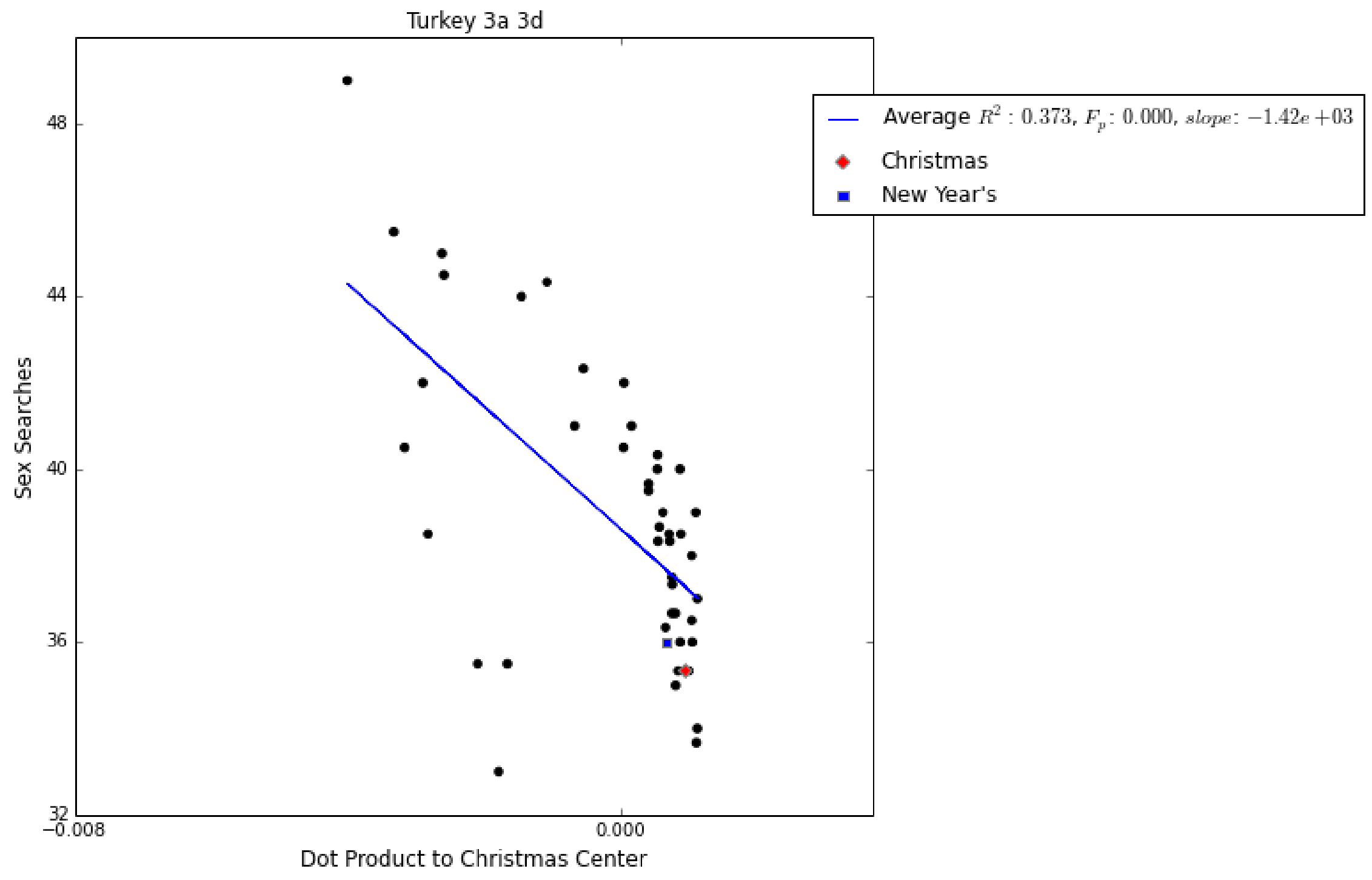

## *Turkey Eid-al-Fitr*

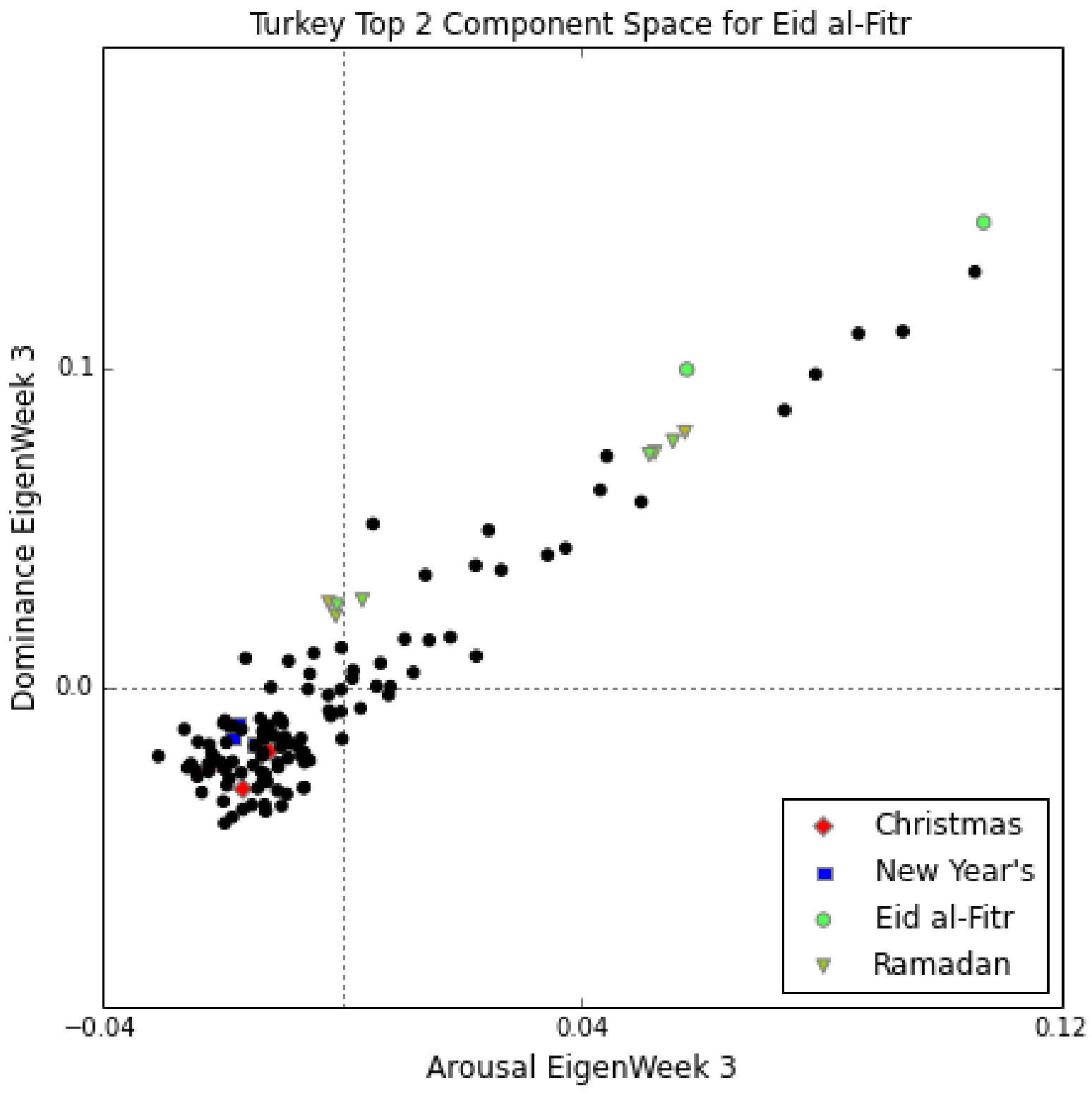

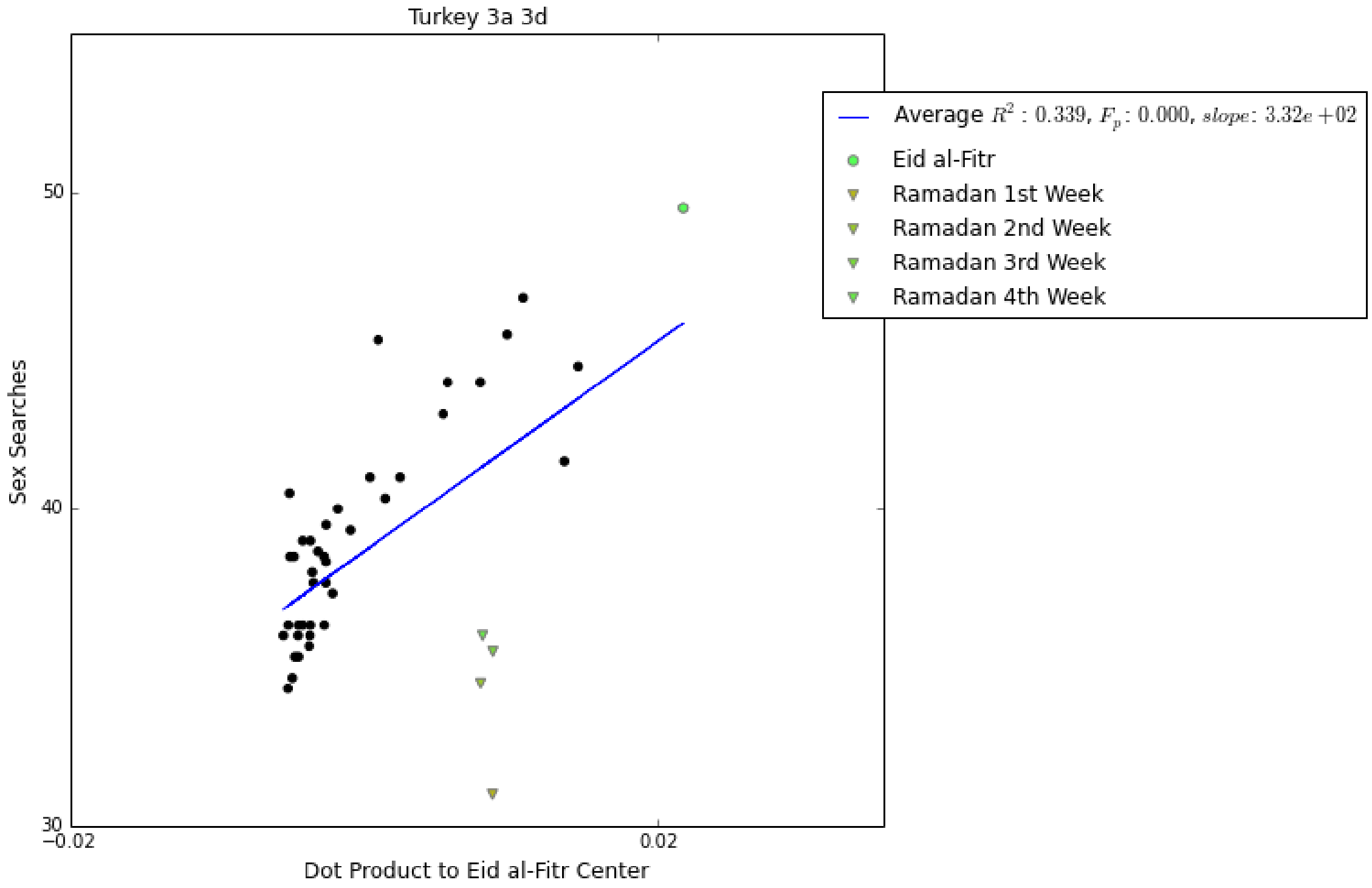

## *Supplementary Tables*

**Table S1.** Searches for "sex" in select countries. Search queries for "sex" are issued in select countries, representing sexual interest in different cultures, hemispheres, and languages. Google Trends ™ allows the retrieval of search volume time series for multiple search terms. We downloaded GT data for 2 search queries: (1) for the term "sex" and (2) for its translation in the local language as detailed in Supplementary Methods. Table S1 shows the 25 countries and languages that retrieved a sufficiently significant search volume in the local language to support our analysis. From left to right, columns show the: "Countries" for which the analysis was performed; "Search term" in GT; the "Top 5 words associated with the search term", provided and ranked by Google Trends; the "Search Volume Ratio", calculated as the number of searches for "sex" divided by the number of searches for the corresponding translation; and the "Correlation between the two time series ("sex" and the translated word).
The English word "sex" is either more searched for than the corresponding word in the local language (blue to red in the 4th column) or there is a strong correlation between the search terms (red in the 5th column). This is consistent with the fact that the top 5 broad searches most associated with "sex" returned by GT refer to interest in sexual content and pornography in every country (3rd column) and that sexual materials and pornography are widely available in English. The two exceptions are Russia and Israel and neither of these countries is relevant to our analysis.

| Country | Search Term | Top 5 Words associated with search term | Search volume ratio | Correlation |
|---|---|---|---|---|
| Argentina | sex | sex free, sex videos, porn sex, porn, video sex | | |
| | sexo | videos, **sexo** videos, sexo gratis, videos de sexo, porno | 0.27 | 0.81 |
| Australia | sex | free sex, porn sex, porn, sex stories, sex videos | | |
| Brazil | sex | videos sex, sexo, videos, sex video, sex shop | | |
| | sexo | videos sexo, videos, videos de sexo, sexo video, sexo gratis | 0.17 | 0.71 |
| Bulgaria | sex | porno, sex porno, sex free, sex video, sex bg | | |
| | секс | **секс** порно (sex porn), порно (porn), секс игри (sex game), секс клипове (sex videos), клипове (videos) | 5.11 | -0.08 |
| Chile | sex | sex free, sex videos, video sex, sex porn, sexo | | |
| | sexo | videos, **sexo** videos, sexo gratis, videos de sexo, porno | 0.23 | 0.81 |
| Egypt | sex | sex free, sex arab, sex tube, sex movies, hot sex | | |
| | جنس | سكس (sex), سكس جنس (sex sex), افلام جنس (sex movies), افلام (films), قصص جنس (sex stories) | 8.04 | 0.83 |
| | سكس | سكس سكس (sex sex), افلام سكس (sex videos), افلام (films), صور (picture), صور سكس (photo) | 1.15 | -0.56 |
| France | sex | sex video, sex free, free, porn sex, porn | | |
| | sexe | video **sexe,** video, sexe gratuit, sexe amateur, amateur | 0.98 | 0.82 |
| Germany | sex | sex free, video sex, sex videos, porn, sex porn | | |
| Greece | sex | sex free, video sex, sex videos, porn, sex porn | | |
| | σεξ | **σεξ** βιντεο (sex video), βιντεο (video), σεξ πορνο (sex porn), πορνο (porn), ιστοριες σεξ (sex stories) | 15.60 | 0.43 |
| India | sex | indian sex, sex videos, free sex, sex stories, hot sex | | |
| Indonesia | sex | cerita sex, cerita, video sex, video, foto sex | | |
| | seks | cerita (story), cerita **seks,** video seks, video, foto seks | 7.18 | 0.20 |
| Israel | sex | sex free, , sex video, sex porn, porn | | |
| | סקס | סקס (sex), סרטי סקס (sex videos), סקס חינם (free sex), , , סרטי סקס חינם (free sex movies) | 0.72 | 0.24 |
| Italy | sex | video, sex video, free sex, porno, porno sex | | |
| | sesso | video **sesso,** video, porno, sesso porno, sesso gratis | 1.22 | 0.54 |
| Japan | sex | sex xxx, xxx, sex, sex, sex | | |
| | セック | セックス動画 (sex video), セックス画像 (sex image), エロ (hello), 無料セックス (free sex) | 1.70 | -0.30 |
| New Zealand | sex | sex free, porn, sex porn, sex stories, sex videos | | |
| Portugal | sex | free sex, videos sex, porn sex, porn, sex video | | |
| | sexo | videos sexo, videos, videos de sexo, sexo gratis, sexo filmes | 0.52 | 0.85 |
| Russia | sex | sex video, sex free, porno, porno sex, porn sex | | |
| | секс | порно секс (porno sex), порно (porn), видео секс (sex video), фото секс (photo sex), онлайн секс (online sex) | 0.61 | -0.48 |
| South Africa | sex | porn sex, porn, free sex, sex videos, sex pics | | |
| Spain | sex | sex free, videos, videos sex, sex porn, porn | | |
| | sexo | sexo gratis, videos, videos sexo, videos de sexo, porno | 0.46 | 0.68 |
| Sweden | sex | free sex, porn, sex porn, sex video, sex tube | | |
| Tunisia | sex | sex sex, porno sex, video sex, video, porno | | |
| | جنس | **جنس** عربي (arabic sex), افلام جنس, (sex movies), افلام (films), سكس (sex), قصص جنس (sex stories) | 32.74 | 0.62 |
| | سكس | **سكس** سكس (sex), سكس عربي (sex arabic), افلام (films), افلام سكس (sex videos), قصص سكس (sex stories) | 19.32 | 0.71 |
| Turkey | sex | porno sex, porno, sex izle, sex hikayeleri, porn | | |
| | seks | porno **seks,** porno, seks hikayeleri (sex stories), seks izle (watch sex), sex | 3.29 | 0.92 |
| UK | sex | free sex, sex porn, porn, sex videos, sex tape | | |
| USA | sex | free sex, sex videos, sex porn, porn, video sex | | |
| Vietnam | sex | phim (movies), phim sex (porn movies), truyen sex (manga sex), truyen (manga), anh sex (he sex) | | |
| Worldwide | sex | sex free, free, sex videos, sex porn, porn | | |

**Table S2.** Countries analyzed and categorized according to religion and geographical location (hemisphere)

The 1st column shows the international country code, the 2nd columns shows the Country name; the third column (Week) shows the first week for which we could find stable GT™ data. A country was considered "culturally Christian" when at least half of its population identified as Christian (Catholic, Protestant, Orthodox, or other) according to [13]. A country was considered "culturally Muslim" when at least half of its population identified as Muslim according to [14]. A country was labeled as "Other" when the majority of its population didn't identify as either Christian or Muslim. The 4th column, "Country Set" shows how each country was categorized and the 5th and 6th columns show the percentage of the population that identify as Christian or Muslim, respectively. The 7th and 8th columns show the continent and the Hemisphere to which each country belongs, according to Wikipedia.

| Code | Country Name | First Week | Country Set | % Christian | % Muslim | Continent | Hemisphere |
|---|---|---|---|---|---|---|---|
| AE | United Arab Emirates | 04-01-2004 | Muslim | 2.6 (2.6;) | 76 | Asia | North |
| AF | Afghanistan | 12-11-2006 | Muslim | 0.02 (;) | 99.8 | Asia | North |
| AL | Albania | 06-11-2005 | Muslim | 17 (7;10) | 82.1 | Europe | North |
| AR | Argentina | 04-01-2004 | Christian | 90 (77;13) | 2.5 | South America | South |
| AT | Austria | 04-01-2004 | Christian | 68.4 (62.4;6) | 5.7 | Europe | North |
| AU | Australia | 04-01-2004 | Christian | 63 (25.8;37) | 1.9 | Oceania | South |
| AW | Aruba | 04-06-2006 | Christian | 88 (80.8;7.8) | 0 | North America | South |
| BA | Bosnia and Herzegovina | 04-01-2004 | Christian | 52 (15;37) | 41.6 | Europe | North |
| BD | Bangladesh | 04-01-2004 | Muslim | 0.3 (0.3;) | 90.4 | Asia | North |
| BE | Belgium | 04-01-2004 | Christian | 55.4 (57;7) | 6 | Europe | North |
| BG | Bulgaria | 04-01-2004 | Christian | 84 (1;83) | 13.4 | Europe | North |
| BH | Bahrain | 04-01-2004 | Muslim | 9 (;9) | 81.2 | Asia | North |
| BN | Brunei | 08-01-2006 | Muslim | 11 (;) | 51.9 | Asia | North |
| BO | Bolivia | 04-01-2004 | Christian | 89 (76;13) | 2.5 | South America | South |
| BR | Brazil | 04-01-2004 | Christian | 90.2 (63;27) | 0.1 | South America | South |
| BS | Bahamas | 05-06-2005 | Christian | 81 (13.5;67.6) | 0 | Central America | North |
| BY | Belarus | 01-01-2006 | Christian | 55.4 (7.1;48.3) | 0.2 | Europe | North |
| CA | Canada | 04-01-2004 | Christian | 67.3 (38.7;29) | 2.8 | North America | North |
| CH | Switzerland | 04-01-2004 | Christian | 71 (38;33) | 5.7 | Europe | North |
| CL | Chile | 04-01-2004 | Christian | 87.2 (67;20) | 0 | South America | South |
| CM | Cameroon | 26-08-2007 | Christian | 65 (38.4;26.3) | 18 | Africa | North |
| CN | China | 04-01-2004 | Other | 5 (1;4) | 1.8 | Asia | North |
| CO | Colombia | 04-01-2004 | Christian | 90 (75;15) | 0 | South America | North |
| CR | Costa Rica | 04-01-2004 | Christian | 83 (69;14) | 0 | Central America | North |
| CY | Cyprus | 04-01-2004 | Christian | 79.3 (4.3;75) | 22.7 | Europe | North |
| CZ | Czech Republic | 04-01-2004 | Other | 11.2 (10.4;0.8) | 0 | Europe | North |
| DE | Germany | 04-01-2004 | Christian | 62 (30;32) | 5 | Europe | North |
| DJ | Djibouti | 06-01-2008 | Muslim | 6 (1;5) | 97 | Africa | North |
| DK | Denmark | 04-01-2004 | Christian | 81 (1;80) | 4.1 | Europe | North |

| Code | Country Name | First Week | Country Set | % Christian | % Muslim | Continent | Hemisphere |
|---|---|---|---|---|---|---|---|
| DO | Dominican Republic | 04-01-2004 | Christian | 95 (95;) | | North America | North |
| DZ | Algeria | 04-01-2004 | Muslim | 2 (1;1) | 98.2 | Africa | North |
| EC | Ecuador | 04-01-2004 | Christian | 94 (74;20) | 0 | South America | South |
| EE | Estonia | 04-01-2004 | Other | 23.9 (0;23) | 0.1 | Europe | North |
| EG | Egypt | 04-01-2004 | Muslim | 18 (0;18) | 94.7 | Africa | North |
| ES | Spain | 04-01-2004 | Christian | 73 (71;2) | 2.3 | Europe | North |
| ET | Ethiopia | 04-01-2004 | Christian | 63.4 (0;63.4) | 33.8 | Africa | North |
| FI | Finland | 04-01-2004 | Christian | 81.6 (0;81) | 0.8 | Europe | North |
| FJ | Fiji | 03-09-2006 | Christian | 64.4 (8.9;55.5) | 6.3 | Oceania | South |
| FR | France | 04-01-2004 | Christian | 65 (63;2) | 7.5 | Europe | North |
| GE | Georgia | 01-05-2005 | Christian | 88.6 (0.9;87.7) | 10.5 | Europe | North |
| GH | Ghana | 16-10-2005 | Christian | 68.8 (13.1;55.5) | 16.1 | Africa | North |
| GP | Guadalupe | 09-03-2008 | Christian | 96 (95;1) | | North America | North |
| GR | Greece | 04-01-2004 | Christian | 97 (0;97) | 4.7 | Europe | North |
| GT | Guatemala | 04-01-2004 | Christian | 87 (47;40) | 0 | Central America | North |
| GU | Guam | 17-12-2006 | Christian | 85 (;) | 0.1 | Oceania | South |
| HN | Honduras | 04-09-2005 | Christian | 87.6 (47;40) | 0.1 | Central America | North |
| HR | Croatia | 04-01-2004 | Christian | 90(70;20) | 1.3 | Europe | North |
| HU | Hungary | 04-01-2004 | Christian | 82.7 (70.1;11.6) | 0.3 | Europe | North |
| ID | Indonesia | 04-01-2004 | Muslim | 10(3;7) | 88.1 | Asia | South |
| IE | Ireland | 04-01-2004 | Christian | 94.1 (82;12) | 0.9 | Europe | North |
| IL | Israel | 04-01-2004 | Other | 3.5(;3.5) | 17.7 | Asia | North |
| IN | India | 04-01-2004 | Other | 2.6 (1.6;1) | 14.6 | Asia | North |
| IQ | Iraq | 12-12-2004 | Muslim | 3(;3) | 98.9 | Asia | North |
| IR | Iran | 04-01-2004 | Muslim | 0.4(;) | 99.7 | Asia | North |
| IS | Iceland | 04-01-2004 | Christian | 95 (2.5;92.5) | 0.1 | Europe | North |
| IT | Italy | 04-01-2004 | Christian | 85.1 (85;0) | 2.6 | Europe | North |
| JM | Jamaica | 04-01-2004 | Christian | 65.3 (2;63.3) | 0 | Central America | North |
| JO | Jordan | 04-01-2004 | Muslim | 6 (;) | 98.8 | Asia | North |
| JP | Japan | 04-01-2004 | Other | 2 (1;1) | 0.1 | Asia | North |
| KE | Kenya | 04-01-2004 | Christian | 85.1 (23.4;61.7) | 7 | Africa | North |
| KH | Cambodia | 05-12-2004 | Other | 1 (0.15;0.85) | 1.6 | Asia | North |
| KR | South Korea | 04-01-2004 | Other | (;) | 0.2 | Asia | North |

| Code | Country Name | First Week | Country Set | % Christian | % Muslim | Continent | Hemisphere |
|---|---|---|---|---|---|---|---|
| KW | Kuwait | 04-01-2004 | Muslim | 15 (3.2;12.8) | 86.4 | Asia | North |
| KZ | Kazakhstan | 01-10-2006 | Muslim | 51 (0.16;50) | 56.4 | Europe | North |
| LA | Laos | 15-04-2007 | Other | 2.2 (1;1) | 0 | Asia | North |
| LB | Lebanon | 04-01-2004 | Muslim | 41 (26;15) | 59.7 | Asia | North |
| LK | Sri Lanka | 04-01-2004 | Other | 7.5 (6.1;1.4) | 8.5 | Asia | North |
| LT | Lithuania | 04-01-2004 | Christian | 84.9 (77.2;7.6) | 0.1 | Europe | North |
| LU | Luxemburg | 04-01-2004 | Christian | 71 (69;2) | 2.3 | Europe | North |
| LV | Latvia | 04-01-2004 | Christian | 57 (25;32.2) | 0.1 | Europe | North |
| MA | Morocco | 04-01-2004 | Muslim | 2.1 (0.1;2) | 99.9 | Africa | North |
| MD | Moldova | 02-10-2005 | Christian | 97.53 (0;93) | 0.4 | Europe | North |
| ME | Montenegro | 13-11-2005 | Christian | 78.8 (3.4;72.07) | 18.5 | Europe | North |
| MK | Macedonia | 04-01-2004 | Christian | 65.1 (0.3;64.8) | 34.9 | Europe | North |
| MM | Myanmar | 04-12-2005 | Other | 7.9 (1;6.9) | 3.8 | Asia | North |
| MN | Mongolia | 14-08-2005 | Other | 2.1 (;) | 4.4 | Asia | North |
| MT | Malta | 04-01-2004 | Christian | 97 (;) | 0.3 | Europe | North |
| MU | Mauritius | 10-07-2005 | Other | 32.2 (-;-) | 16.6 | Africa | South |
| MV | Maldives | 04-01-2004 | Muslim | 41 (26;15) | 98.4 | Asia | North |
| MX | Mexico | 04-01-2004 | Christian | 92 (;) | 0.1 | North America | North |
| MY | Malaysia | 04-01-2004 | Muslim | 12.1 (;) | 61.4 | Asia | North |
| MZ | Mozambique | 24-02-2008 | Christian | 56.1 (28.4;27.7) | 22.8 | Africa | South |
| NA | Namibia | 27-06-2010 | Christian | 90 (13.7;76.3) | 0.4 | Africa | South |
| NG | Nigeria | 04-01-2004 | Christian | 50.01 (14;36) | 47.9 | Africa | North |
| NI | Nicaragua | 16-08-2009 | Christian | 89.6 (58.8;30.8) | 0 | Central America | North |
| NL | Netherlands | 04-01-2004 | Other | 44 (24;20) | 5.5 | Europe | North |
| NO | Norway | 04-01-2004 | Christian | 86.2 (3;83.5) | 3 | Europe | North |
| NP | Nepal | 04-01-2004 | Other | 0.9 (0.1;0.8) | 4.2 | Asia | North |
| NZ | New Zealand | 04-01-2004 | Christian | 55.6 (28.7;24.9) | 0.9 | South America | South |
| OM | Oman | 04-01-2004 | Muslim | 2.5 (2.1;0.4) | 87.7 | Asia | North |
| PA | Panama | 15-02-2004 | Christian | 92 (80;12) | 0.7 | Central America | North |
| PE | Peru | 04-01-2004 | Christian | 96 (81;15) | 0 | South America | South |
| PH | Philippines | 04-01-2004 | Christian | 93 (80;13) | 5.1 | Asia | North |
| PK | Pakistan | 04-01-2004 | Muslim | 1.6 (0.8;0.8) | 96.4 | Asia | North |
| PL | Poland | 04-01-2004 | Christian | 94.3 (86.3;8) | 0.1 | Europe | North |
| PR | Puerto Rico | 04-01-2004 | Christian | 97 (50;47) | 0 | North America | North |

| Code | Country Name | First Week | Country Set | % Christian | % Muslim | Continent | Hemisphere |
|---|---|---|---|---|---|---|---|
| PS | Palestine | 04-01-2004 | Muslim | (;) | 97.5 | Asia | North |
| PT | Portugal | 04-01-2004 | Christian | 95.7 (81;14.7) | 0.6 | Europe | North |
| PY | Paraguay | 12-02-2006 | Christian | 96 (88;7.9) | 0 | South America | South |
| QA | Qatar | 04-01-2004 | Muslim | 13.8 (;) | 77.5 | Asia | North |
| RO | Romania | 04-01-2004 | Christian | 99.5 (5.7;93.8) | 0.3 | Europe | North |
| RS | Serbia | 04-01-2004 | Christian | 93.5 (4.97;79.4) | 3.7 | Europe | North |
| RU | Russia | 04-01-2004 | Christian | 60 (0;60) | 11.7 | Europe | North |
| SA | Saudi Arabia | 04-01-2004 | Muslim | 5.5 (3.5;2) | 97.1 | Asia | North |
| SD | Sudan | 11-01-2004 | Muslim | 2 (;) | 71.4 | Africa | North |
| SE | Sweden | 04-01-2004 | Christian | 67.2 (2;65) | 4.9 | Europe | North |
| SG | Singapore | 04-01-2004 | Other | 18 (5.7;12) | 14.9 | Asia | North |
| SI | Slovenia | 04-01-2004 | Christian | 79.2 (57;22.2) | 2.4 | Europe | North |
| SK | Slovakia | 04-01-2004 | Christian | 86.5 (75.2;11.3) | 0.1 | Europe | North |
| SV | El Salvador | 04-01-2004 | Christian | 81.9 (52.6;29.3) | 0 | Central America | North |
| SY | Syria | 04-01-2004 | Muslim | 10 (0;10) | 92.8 | Asia | North |
| TH | Thailand | 04-01-2004 | Other | 0.7 (0.4;0.3) | 5.8 | Asia | North |
| TN | Tunisia | 04-01-2004 | Muslim | 0.2 (;0.2) | 99.8 | Africa | North |
| TR | Turkey | 04-01-2004 | Muslim | 0.2 (;) | 98.6 | Europe | North |
| TT | Trinidad and Tobago | 04-01-2004 | Christian | 57.6 (21.5;33.4) | 5.8 | Central America | North |
| TW | Taiwan | 04-01-2004 | Other | 3.9 (2.6;1.3) | 0.1 | Asia | North |
| TZ | Tanzania | 04-01-2004 | Christian | 62 (;) | 29.9 | Africa | South |
| UA | Ukraine | 04-01-2004 | Christian | 83.8 (5.9;76.7) | 0.9 | Europe | North |
| UG | Uganda | 08-01-2006 | Christian | 88.6 (41.9;46.7) | 12 | Africa | North |
| UK | United Kingdom | 04-01-2004 | Christian | 59.3 (8.9;50) | 4.6 | Europe | North |
| US | United States of America | 04-01-2004 | Christian | 73 (22;51) | 0.8 | North America | North |
| UY | Uruguay | 04-01-2004 | Christian | 58.4 (47;11) | 0 | South America | South |
| UZ | Uzbekistan | 17-10-2004 | Muslim | 2.6 (2.6;) | 96.5 | Asia | North |
| VE | Venezuela | 04-01-2004 | Christian | 87 (79;8) | 9.3 | South America | North |
| VN | Vietnam | 04-01-2004 | Other | 8 (7;1) | 0.2 | Asia | North |
| YE | Yemen | 04-01-2004 | Muslim | 0.0013 (0.0013;) | 99 | Asia | North |
| ZA | South Africa | 04-01-2004 | Christian | 80 (5;75) | 1.5 | Africa | South |
| ZM | Zambia | 06-05-2007 | Christian | 97.6 (25;72) | 0.4 | Africa | South |
| ZW | Zimbabwe | 05-03-2006 | Christian | 85 (7;77) | 0.9 | Africa | South |

**Table S3**. Correlation Table for the averaged time series of all countries grouped either by hemisphere (Northern or Southern ) or by religion (Muslim or Christian ).
Table S3a shows $R^2$ and Table S3b shows the corresponding p-values.

**Table S3a**.

| | Northern | Southern | Christian | Muslim |
|---|---|---|---|---|
| Northern | 1 | | | |
| Southern | 0.536811 | 1 | | |
| Christian | 0.890322 | 0.627146 | 1 | |
| Muslim | 0.415906 | 0.309619 | 0.192213 | 1 |

**Table S3b.**

| | Northern | Southern | Christian | Muslim |
|---|---|---|---|---|
| Northern | 1 | | | |
| Southern | 5.89E-90 | 1 | | |
| Christian | 1.3E-254 | 9.1E-115 | 1 | |
| Muslim | 2.07E-63 | 2.98E-44 | 3.27E-26 | 1 |

**Table S4.** The three major Muslim holidays, in regard to the Gregorian calendar, for the period under analysis.

| *Beginning of Ramadan* | *Eid-al-Fitr* | *Eid al-Adha* |
|---|---|---|
| 15 Oct 2004 | 14 Nov 2004 | 21 Jan 2005 |
| 4 Oct 2005 | 3 Nov 2005 | 10 Jan 2006 |
| 24 Sep 2006 | 23 Oct 2006 | 31 Dec 2006 |
| 13 Sep 2007 | 13 Oct 2007 | 20 Dec 2007 |
| 1 Sep 2008 | 1 Oct 2008 | 8 Dec 2008 |
| 22 Aug 2009 | 20 Sep 2009 | 27 Nov 2009 |
| 11 Aug 2010 | 10 Sep 2010 | 16 Nov 2010 |
| 1 Aug 2011 | 30 Aug 2011 | 6 Nov 2011 |
| 20 Jul 2012 | 19 Aug 2012 | 26 Oct 2012 |
| 9 Jul 2013 | 8 Aug 2013 | 15 Oct 2013 |

**Table S5 -** Starting day of the "Christian Calendar", starting day of the weeks that included December 25th – Christmas (always on week 26), the last week of each centered year and the discarded exception weeks after centering.

| *1* | *26* | *week 52* | *exception week* |
|---|---|---|---|
| - | | 6/20/2004 | - |
| 6/27/2004 | 12/19/2004 | 6/19/2005 | 6/26/2005 |
| 7/26/2005 | 12/25/2005 | 6/25/2006 | |
| 7/2/2006 | 12/24/2006 | 6/24/2007 | - |
| 7/1/2007 | 12/23/2007 | 6/22/2008 | - |
| 6/29/2008 | 12/21/2008 | 6/21/2009 | - |
| 6/28/2009 | 12/20/2009 | 6/20/2010 | - |
| 6/27/2010 | 12/19/2010 | 6/19/2011 | 26 June 2011 |
| 7/3/2011 | 12/25/2011 | 6/24/2012 | - |
| 7/1/2012 | 12/23/2012 | 6/23/2013 | - |
| 6/30/2013 | 12/22/2013 | - | - |

**Table S6.** Weeks that included Eid-al-Fitr and the discarded exception weeks after centering.

| *1* | *25* | *week 50* | *exception week* |
|---|---|---|---|
| - | | 5/23/2004 | - |
| 5/30/2004 | 11/14/2004 | 5/8/2005 | - |
| 5/15/2005 | 10/30/2005 | 4/23/2006 | 4/30/2006 |
| 5/7/2006 | 10/22/2006 | 4/15/2007 | - |
| 4/22/2007 | 10/7/2007 | 3/30/2008 | 4/6/2008 |
| 4/13/2008 | 9/28/2008 | 3/22/2009 | 3/29/2009 |
| 4/5/2009 | 9/20/2009 | 3/14/2010 | - |
| 3/21/2010 | 9/5/2010 | 2/27/2011 | 3/6/2011 |
| 3/13/2011 | 8/28/2011 | 2/19/2012 | 2/26/2012 |
| 3/4/2012 | 8/19/2012 | 2/10/2013 | - |
| 2/17/2013 | 8/4/2013 | 1/26/2014 | - |
| 2/2/2014 | - | - | - |

**Table S7.** Z-scores on the corresponding centered week for all countries in the dataset, calculated from the each country's average for each week, as detailed in the Methods. When z>1 for both the Christmas and Eid-al-Fitr centered calendars, classification was based on the higher score (bold).

| | Country | Country Set | Hemisphere | Christmas | Eid-al-Fitr | June Solstice | Dec Solstice |
|---|---|---|---|---|---|---|---|
| AE | United Arab Emirates | Muslim | North | 1.877 | **3.023** | 0.179 | 1.313 |
| AF | Afghanistan | Muslim | North | 0.654 | 0.443 | 0.587 | 0.889 |
| AL | Albania | Muslim | North | 0.372 | 1.417 | 0.399 | 0.491 |
| AR | Argentina | Christian | South | 2.190 | -2.066 | 0.395 | 1.146 |
| AT | Austria | Christian | North | 3.598 | -0.089 | -0.724 | 1.879 |
| AU | Australia | Christian | South | 3.598 | -0.089 | -0.724 | 1.879 |
| AW | Aruba | Christian | South | **1.970** | 1.960 | -0.570 | 1.502 |
| BA | Bosnia and Herzegovina | Christian | North | -0.312 | 0.883 | 0.658 | -0.477 |
| BD | Bangladesh | Muslim | North | 1.544 | **2.576** | 0.701 | 1.062 |
| BE | Belgium | Christian | North | 1.713 | 0.315 | 0.770 | 0.350 |
| BG | Bulgaria | Christian | North | 0.843 | 0.476 | 1.169 | -0.443 |
| BH | Bahrain | Muslim | North | 1.151 | 2.492 | 1.128 | 1.879 |
| BN | Brunei | Muslim | North | 1.183 | 2.075 | 0.912 | 2.005 |
| BO | Bolivia | Christian | South | 3.028 | 0.831 | 0.074 | 1.159 |
| BR | Brazil | Christian | South | 3.658 | -0.580 | 0.231 | 1.921 |
| BS | Bahamas | Christian | North | 0.185 | -0.069 | 0.298 | 0.069 |
| BY | Belarus | Christian | North | 0.403 | 0.217 | 0.106 | -0.534 |
| CA | Canada | Christian | North | 2.397 | 0.327 | 1.159 | 0.868 |
| CH | Switzerland | Christian | North | 4.012 | -0.374 | 0.553 | 0.984 |
| CL | Chile | Christian | South | 1.966 | -2.006 | -0.634 | 1.232 |
| CM | Cameroon | Christian | North | 1.410 | 0.926 | 1.021 | 0.650 |
| CN | China | Other | North | -0.650 | -0.349 | 0.300 | -1.083 |
| CO | Colombia | Christian | North | 2.641 | -1.164 | 0.596 | 1.995 |
| CR | Costa Rica | Christian | North | 3.671 | -0.728 | -0.038 | 2.110 |
| CY | Cyprus | Christian | North | 2.274 | -0.376 | 0.057 | 0.390 |
| CZ | Czech Republic | Other | North | 2.718 | -0.166 | 0.952 | 1.020 |
| DE | Germany | Christian | North | 3.800 | 0.043 | 0.759 | 0.974 |
| DJ | Djibouti | Muslim | North | -0.506 | 1.507 | 0.692 | -0.071 |
| DK | Denmark | Christian | North | 2.842 | -0.558 | 0.602 | 0.844 |
| DO | Dominican Republic | Christian | North | 2.379 | -0.861 | 0.649 | 1.240 |
| DZ | Algeria | Muslim | North | 0.503 | 0.872 | 1.611 | 0.153 |
| EC | Ecuador | Christian | South | 3.203 | -0.521 | 0.513 | 2.062 |
| EE | Estonia | Other | North | 1.302 | -0.344 | 1.598 | 0.541 |
| EG | Egypt | Muslim | North | 1.056 | **2.278** | -0.302 | 0.841 |
| ES | Spain | Christian | North | 1.587 | -0.063 | 0.391 | 0.056 |
| ET | Ethiopia | Christian | North | -0.967 | -0.585 | -0.164 | 0.013 |
| FI | Finland | Christian | North | 2.260 | -0.858 | 1.690 | 0.854 |

| | Country | Country Set | Hemisphere | Christmas | Eid-al-Fitr | June Solstice | Dec Solstice |
|---|---|---|---|---|---|---|---|
| FJ | Fiji | Christian | South | 3.087 | -0.002 | -0.437 | 1.683 |
| FR | France | Christian | North | 2.239 | -0.050 | 0.600 | 1.242 |
| GE | Georgia | Christian | North | -0.033 | 0.158 | 0.674 | -0.964 |
| GH | Ghana | Christian | North | 3.869 | -0.417 | 0.389 | 1.850 |
| GP | Guadalupe | Christian | North | 1.550 | -0.059 | 1.814 | 1.751 |
| GR | Greece | Christian | North | 1.241 | -0.158 | -0.156 | 0.056 |
| GT | Guatemala | Christian | North | 3.170 | -1.062 | 0.561 | 2.496 |
| GU | Guam | Christian | South | 0.028 | 1.379 | 1.080 | -0.229 |
| HN | Honduras | Christian | North | 2.903 | -0.713 | 0.279 | 2.062 |
| HR | Croatia | Christian | North | 0.953 | 0.236 | 1.712 | -0.234 |
| HU | Hungary | Christian | North | 1.244 | 0.588 | 0.928 | 0.114 |
| ID | Indonesia | Muslim | South | 2.792 | **3.584** | -0.415 | 1.337 |
| IE | Ireland | Christian | North | 3.498 | 0.072 | 0.477 | 1.052 |
| IL | Israel | Other | North | -1.235 | 0.085 | 1.261 | -1.446 |
| IN | India | Other | North | 1.850 | 1.315 | -0.363 | 0.756 |
| IQ | Iraq | Muslim | North | -0.833 | 0.514 | -0.704 | -0.066 |
| IR | Iran | Muslim | North | -0.597 | 0.497 | 0.714 | -1.260 |
| IS | Iceland | Christian | North | 1.913 | -0.698 | 0.824 | 1.064 |
| IT | Italy | Christian | North | 1.811 | 0.107 | 0.056 | 0.266 |
| JM | Jamaica | Christian | North | 1.255 | -0.357 | 1.799 | 1.190 |
| JO | Jordan | Muslim | North | -0.169 | 2.317 | 1.463 | -0.334 |
| JP | Japan | Other | North | 1.067 | 0.257 | 0.468 | -0.734 |
| KE | Kenya | Christian | North | **4.217** | 1.686 | -0.604 | 3.297 |
| KH | Cambodia | Other | North | 1.064 | 0.988 | -0.475 | -0.242 |
| KR | South Korea | Other | North | 0.994 | -1.400 | 1.172 | -0.305 |
| KW | Kuwait | Muslim | North | 1.730 | **2.384** | 0.145 | 1.855 |
| KZ | Kazakhstan | Muslim | North | 0.151 | -0.458 | 1.537 | -0.248 |
| LA | Laos | Other | North | 1.559 | 0.670 | 0.273 | 0.290 |
| LB | Lebanon | Muslim | North | 1.389 | **2.497** | 0.843 | 0.205 |
| LK | Sri Lanka | Other | North | 2.505 | 0.443 | -0.390 | 0.970 |
| LT | Lithuania | Christian | North | 0.942 | 0.594 | 1.249 | -0.277 |
| LU | Luxemburg | Christian | North | 4.643 | -0.968 | 0.611 | 1.418 |
| LV | Latvia | Christian | North | 1.087 | -0.082 | 2.154 | -0.139 |
| MA | Morocco | Muslim | North | 0.148 | 0.484 | 1.173 | -0.669 |
| MD | Moldova | Christian | North | 0.648 | -0.115 | 0.626 | -0.154 |
| ME | Montenegro | Christian | North | 0.004 | -0.514 | 0.145 | 0.773 |
| MK | Macedonia | Christian | North | -0.789 | -0.233 | 0.786 | -0.920 |
| MM | Myanmar | Other | North | 1.753 | 1.324 | -1.771 | 1.998 |
| MN | Mongolia | Other | North | 0.087 | -0.694 | 0.785 | -0.143 |
| MT | Malta | Christian | North | 1.547 | -0.059 | 1.718 | 1.145 |
| MU | Mauritius | Other | South | 2.627 | -0.528 | -0.212 | 1.745 |
| MV | Maldives | Muslim | North | -0.475 | 0.704 | -0.215 | 0.133 |

| | Country | Country Set | Hemisphere | Christmas | Eid-al-Fitr | June Solstice | Dec Solstice |
|---|---|---|---|---|---|---|---|
| MX | Mexico | Christian | North | 3.092 | -1.378 | 0.739 | 1.967 |
| MY | Malaysia | Muslim | North | 1.838 | **3.709** | 0.174 | 0.602 |
| MZ | Mozambique | Christian | South | 2.243 | -0.531 | -0.048 | 1.702 |
| NA | Namibia | Christian | South | 3.757 | -1.345 | 0.064 | 2.812 |
| NG | Nigeria | Christian | North | **4.650** | 1.208 | -0.227 | 3.060 |
| NI | Nicaragua | Christian | North | 1.199 | -0.917 | -0.321 | 2.106 |
| NL | Netherlands | Other | North | 1.692 | 0.031 | 0.891 | 0.197 |
| NO | Norway | Christian | North | 3.694 | -1.155 | 0.932 | 2.015 |
| NP | Nepal | Other | North | 1.095 | **1.588** | -0.454 | 0.281 |
| NZ | New Zealand | Christian | South | 3.230 | -0.254 | -0.495 | 1.660 |
| OM | Oman | Muslim | North | 0.873 | 1.943 | 0.611 | 1.054 |
| PA | Panama | Christian | North | 1.955 | 0.914 | 0.009 | 1.456 |
| PE | Peru | Christian | South | 2.317 | -2.338 | -0.130 | 1.514 |
| PH | Philippines | Christian | North | 2.444 | 0.981 | -1.614 | 1.819 |
| PK | Pakistan | Muslim | North | **2.282** | 2.126 | -0.124 | 1.787 |
| PL | Poland | Christian | North | 1.414 | 0.083 | 1.341 | 0.215 |
| PR | Puerto Rico | Christian | North | 2.606 | -1.690 | 1.211 | 2.274 |
| PS | Palestine | Muslim | North | 1.152 | **1.609** | 0.458 | 0.215 |
| PT | Portugal | Christian | North | 2.226 | -0.074 | 0.699 | 0.859 |
| PY | Paraguay | Christian | South | 1.952 | -2.259 | -1.242 | 1.278 |
| QA | Qatar | Muslim | North | 1.783 | **2.986** | -1.061 | 0.835 |
| RO | Romania | Christian | North | 1.458 | 0.401 | 0.960 | -0.073 |
| RS | Serbia | Christian | North | -0.163 | 0.474 | 1.130 | -0.390 |
| RU | Russia | Christian | North | 0.042 | -0.455 | 1.443 | -0.371 |
| SA | Saudi Arabia | Muslim | North | 0.271 | 2.698 | -0.037 | 0.330 |
| SD | Sudan | Muslim | North | 0.460 | 1.662 | 0.602 | 0.682 |
| SE | Sweden | Christian | North | 1.764 | -0.609 | 1.547 | 0.383 |
| SG | Singapore | Other | North | **2.238** | 1.525 | 1.339 | 1.140 |
| SI | Slovenia | Christian | North | 0.742 | -0.170 | 1.275 | 0.018 |
| SK | Slovakia | Christian | North | 2.172 | 0.125 | 0.913 | 0.123 |
| SV | El Salvador | Christian | North | 3.076 | 0.144 | -0.263 | 1.603 |
| SY | Syria | Muslim | North | 0.136 | 2.361 | 0.845 | 0.101 |
| TH | Thailand | Other | North | 0.658 | -0.094 | -0.761 | -0.361 |
| TN | Tunisia | Muslim | North | 0.083 | 2.042 | 0.523 | 1.618 |
| TR | Turkey | Muslim | North | -1.084 | 2.988 | 1.447 | -1.123 |
| TT | Trinidad Tobago | Christian | North | 3.526 | 1.158 | -0.016 | 1.704 |
| TW | Taiwan | Other | North | 1.458 | -0.249 | 0.185 | 0.382 |
| TZ | Tanzania | Christian | South | 2.475 | -0.365 | 1.200 | 1.710 |
| UA | Ukraine | Christian | North | 0.497 | 0.158 | 0.270 | -0.051 |
| UG | Uganda | Christian | North | 3.703 | 0.921 | -1.054 | 2.327 |
| UK | United Kingdom | Christian | North | 3.982 | 0.208 | -0.086 | 1.559 |

| | Country | Country Set | Hemisphere | Christmas | Eid-al-Fitr | June Solstice | Dec Solstice |
|---|---|---|---|---|---|---|---|
| US | United States of America | Christian | North | 3.100 | -0.306 | 1.009 | 1.137 |
| UY | Uruguay | Christian | South | 2.140 | -0.462 | -1.259 | 0.879 |
| UZ | Uzbekistan | Muslim | North | -0.590 | 2.098 | 1.472 | -0.960 |
| VE | Venezuela | Christian | North | 3.768 | -0.982 | -0.292 | 2.287 |
| VN | Vietnam | Other | North | -0.033 | 1.300 | 0.436 | -0.380 |
| YE | Yemen | Muslim | North | -0.367 | 1.963 | 0.325 | -0.181 |
| ZA | South Africa | Christian | South | 3.815 | 0.048 | -0.108 | 2.375 |
| ZM | Zambia | Christian | South | 1.804 | 0.915 | -0.098 | 2.308 |
| ZW | Zimbabwe | Christian | South | 3.783 | -0.146 | 1.001 | 2.569 |

**Table S8A**. Correlation between the Z-scores' time series for all countries in the data set. Calendars were centered around each of the events and the z-cores calculated, as detailed in the Methods. The high correlation between the Z-score variation around Christmas and around the December Solstice is due to the fact that Christmas often falls on the same week or very close to the December Solstice.

| | *Christmas* | *Eid-al-Fitr* | *June Solstice* | *December Solstice* |
|---|---|---|---|---|
| Christmas | 1.00 | | | |
| Eid-al-Fitr | -0.28 | 1.00 | | |
| June Solstice | -0.29 | -0.06 | 1.00 | |
| December Solstice | 0.80 | -0.15 | -0.36 | 1.00 |

**Table S8B**. Percentage of countries that were originally classified as Christian, Muslim, or as being located in one of the hemispheres (rows) that showed increased sex-searches (z-scores>1) during Christmas, Eid-al-Fitr or the Solstices (columns).

| | | **Increased sex-searches around:** | | | |
|---|---|---|---|---|---|
| | | Christmas | Eid-al-Fitr | June Sltc | Dec Sltc |
| Identified as | Christian | 80% | 6% | 25% | 56% |
| | Muslim | 40% | 77% | 23% | 30% |
| | Southern Hemisphere | 95% | 14% | 14% | 90% |
| | Northern Hemisphere | 64% | 28% | 26% | 36% |

**Table S9.** Monthly birth data available for countries from Supplementary Table 2. First column, countries that belong to the ”Other” country set are marked with a blue background, countries that belong to the ”Muslim” country set with a green background, and countries belonging to the ”Christian” country set with a white background. At the bottom of the table are the only four countries from the Southern Hemisphere for which we could find birth data, and all four were classified as Christian. Dark shaded area coincides with the period for which we have GT data and these were the years used in all birth plots.

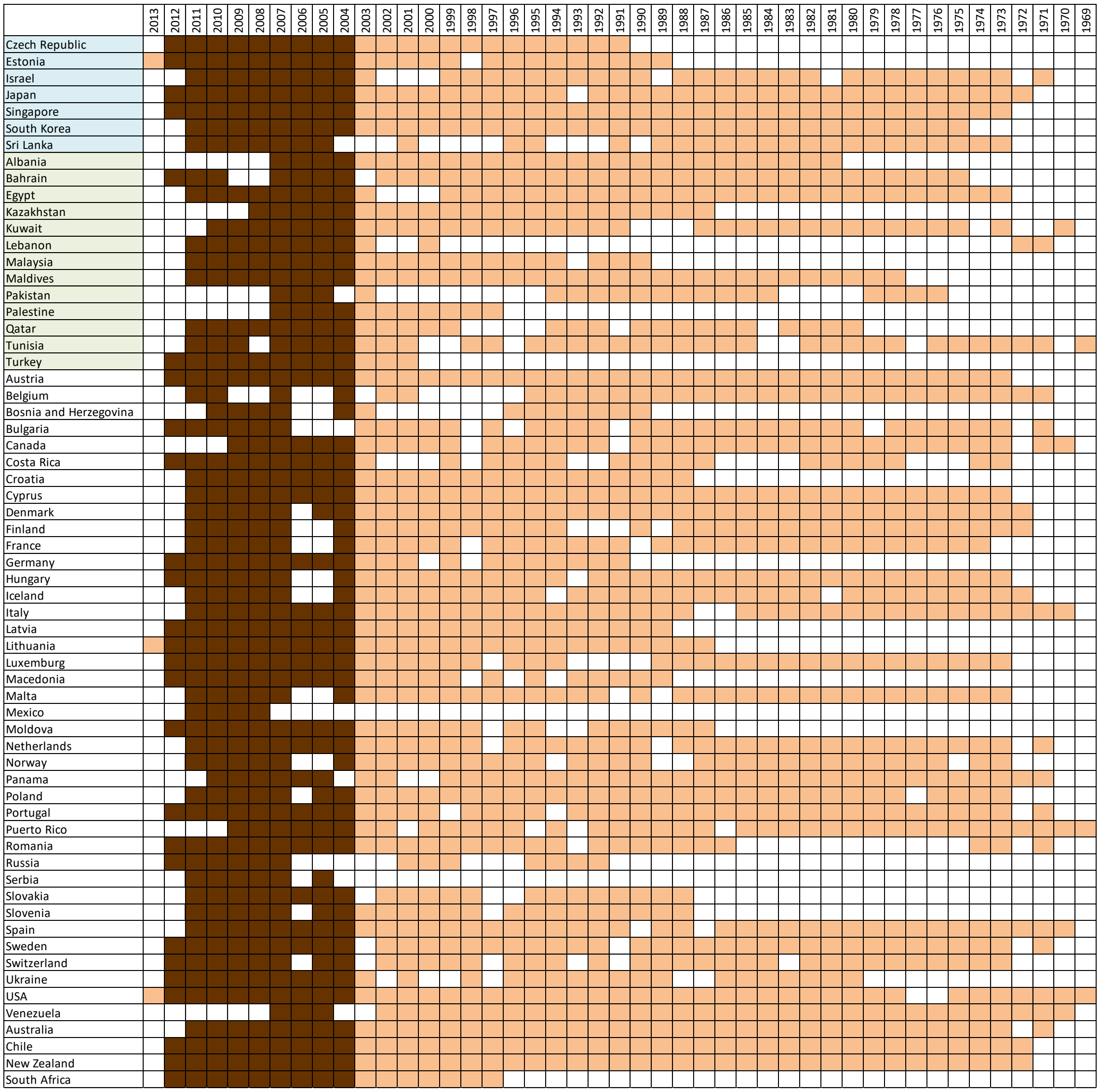

**Table S10.** Multiple linear regression statistics with all three ANEW dimensions, using weekly ANEW means as independent variables and sex search volume as dependent variable. A) Regression over all years of data. B) Regression over an average year centered on Christmas (USA, Australia, Brazil, Argentina, Chile) and Eid-al-Fitr (Indonesia and Turkey) – Independent variables are: [mean ANEW values averaged across years – the holiday center] (i.e, Christmas is 0,0,0), dependent variable is the number of sex-searches averaged across years of data. $R^2$ columns indicate the coefficient of determination for the regression, $F_p$ columns indicate the p-value for the F-statistic of the overall model, B columns indicate the coefficients for the independent variables in the regression. t-test p columns indicate the individual t-test p values for the independent variables. Bold values denote significance at α=0.05, italicized values denote Bonferroni corrected significance over countries per variable choice α=0.05/7= 0.00714.

**A**

| *Country* | $R^2$ | Valence B | Dominance B | Arousal B | $F_p$ | Valence t-test p | Dominance t-test p | Arousal t-test p |
|---|---|---|---|---|---|---|---|---|
| *USA* | 0.399 | 197.69 | -379.75 | -0.36 | ***1.18E-20*** | ***4.76E-18*** | ***1.06E-12*** | 0.972 |
| *Australia* | 0.274 | 55.77 | -92.18 | -25.05 | ***2.91E-12*** | ***1.10E-07*** | ***4.22E-06*** | ***9.79E-06*** |
| *Brazil* | 0.401 | 12.47 | 37.78 | 90.74 | ***1.19E-15*** | 0.416 | 0.423 | ***4.79E-08*** |
| *Argentina* | 0.388 | 39.22 | -36.67 | -8.79 | ***1.40E-14*** | ***2.59E-09*** | ***1.91E-03*** | 0.0786 |
| *Chile* | 0.240 | 4.93 | 26.63 | -28.93 | ***1.68E-10*** | 0.602 | 0.280 | ***6.36E-10*** |
| *Indonesia* | 0.187 | 72.13 | -127.96 | -12.95 | ***1.87E-06*** | ***1.24E-07*** | ***5.28E-04*** | 0.366 |
| *Turkey* | 0.135 | 6.66 | -1.72 | 16.11 | ***4.22E-04*** | 0.128 | 0.893 | ***1.83E-04*** |

**B**

| *Country* | $R^2$ | Valence B | Dominance B | Arousal B | $F_p$ | Valence t-test p | Dominance t-test p | Arousal t-test p |
|---|---|---|---|---|---|---|---|---|
| *USA* | 0.426 | 193.002 | -427.958 | 96.678 | ***6.20E-06*** | ***2.94E-07*** | ***2.77E-05*** | 0.0632 |
| *Australia* | 0.566 | 95.519 | -128.290 | 19.318 | ***8.40E-09*** | ***4.67E-08*** | ***1.44E-03*** | 0.225 |
| *Brazil* | 0.488 | 90.086 | -148.254 | 35.561 | ***4.15E-07*** | ***3.06E-05*** | 0.0340 | 0.116 |
| *Argentina* | 0.530 | 57.468 | -65.493 | -2.228 | ***5.51E-08*** | ***3.61E-07*** | 0.0145 | 0.871 |
| *Chile* | 0.697 | 70.497 | -81.955 | 12.632 | ***1.73E-12*** | ***8.21E-08*** | 0.0123 | 0.0606 |
| *Indonesia* | 0.267 | 144.696 | -272.516 | -53.604 | ***2.34E-03*** | **8.40E-03** | 0.0213 | 0.271 |
| *Turkey* | 0.260 | 7.835 | -81.301 | 41.880 | ***2.94E-03*** | 0.503 | 0.0103 | **0.0220** |

**Table S11.** Linear regression statistics for individual ANEW dimensions, using weekly ANEW means as independent variables and sex search volume as dependent variable. A: Regression over all years of data. B: Regression over an average year centered on Christmas (USA, Australia, Brazil, Argentina, Chile) and Eid-al-Fitr (Indonesia and Turkey) – Independent variables are: [mean ANEW value averaged across years – the holiday center] (i.e, Christmas is 0), dependent variable is the number of sex-searches averaged across years of data. Independent variables from top to bottom: Valence, Dominance, and Arousal. $R^2$ columns indicate the coefficient of determination for the regression, $F_p$ columns indicate the p-value for the F-statistic of the overall model, B columns indicate the coefficients for the independent variables in the regression. Bold values denote significance at α=0.05, italicized values denote Bonferroni corrected significance over countries per variable choice α=0.05/7 = 0.00714.

**A**

| ***Country*** | **Valence $R^2$** | **Valence $F_p$** | **Valence B** |
|---|---|---|---|
| *USA* | 0.057 | ***8.99E-04*** | 54.80 |
| *Australia* | 0.065 | ***5.34E-04*** | -10.74 |
| *Brazil* | 0.004 | 0.434 | 6.57 |
| *Argentina* | 0.255 | ***1.78E-10*** | 13.25 |
| *Chile* | 0.019 | 0.0680 | 8.52 |
| *Indonesia* | 0.091 | ***2.29E-04*** | 27.35 |
| *Turkey* | 0.008 | 0.3.07 | 3.16 |

| ***Country*** | **Dominance $R^2$** | **Dominance $F_p$** | **Dominance B** |
|---|---|---|---|
| *USA* | 0.052 | ***1.46E-03*** | -101.86 |
| *Australia* | 0.120 | ***1.84E-06*** | -30.11 |
| *Brazil* | 0.141 | ***3.21E-06*** | 115.42 |
| *Argentina* | 0.143 | ***3.67E-06*** | 15.86 |
| *Chile* | 0.001 | 0.711 | 4.19 |
| *Indonesia* | 0.007 | 0.302 | 21.45 |
| *Turkey* | 0.031 | **4.84E-02** | 18.09 |

| ***Country*** | **Arousal $R^2$** | **Arousal $F_p$** | **Arousal B** |
|---|---|---|---|
| *USA* | 0.102 | ***6.20E-06*** | -41.47 |
| *Australia* | 0.148 | ***9.78E-08*** | -15.07 |
| *Brazil* | 0.347 | ***6.28E-15*** | 90.83 |
| *Argentina* | 0.026 | 0.0573 | 6.66 |
| *Chile* | 0.186 | ***1.48E-09*** | -23.98 |
| *Indonesia* | 0.007 | 0.323 | 11.16 |
| *Turkey* | 0.105 | ***1.95E-04*** | 13.97 |

**B**

| *Country* | Valence $R^2$ | Valence $F_p$ | Valence B |
|---|---|---|---|
| *USA* | 0.166 | ***2.76E-03*** | 80.924 |
| *Australia* | 0.459 | ***3.39E-08*** | 56.364 |
| *Brazil* | 0.437 | ***9.41E-08*** | 51.052 |
| *Argentina* | 0.418 | ***2.25E-07*** | 31.673 |
| *Chile* | 0.652 | ***4.82E-13*** | 43.522 |
| *Indonesia* | 0.008 | 0.541 | 19.959 |
| *Turkey* | 0.043 | 0.150 | 8.871 |

| *Country* | Dominance $R^2$ | Dominance $F_p$ | Dominance B |
|---|---|---|---|
| *USA* | 0.002 | 0.778 | -20.481 |
| *Australia* | 0.167 | ***2.66E-03*** | 74.668 |
| *Brazil* | 0.214 | ***5.48E-04*** | 121.891 |
| *Argentina* | 0.138 | **6.68E-03** | 39.978 |
| *Chile* | 0.426 | ***1.55E-07*** | 97.791 |
| *Indonesia* | 0.049 | 0.123 | -94.162 |
| *Turkey* | 0.005 | 0.612 | -11.387 |

| *Country* | Arousal $R^2$ | Arousal $F_p$ | Arousal B |
|---|---|---|---|
| *USA* | 0.000 | 0.945 | -3.900 |
| *Australia* | 0.165 | ***2.85E-03*** | 42.919 |
| *Brazil* | 0.010 | 0.490 | -14.894 |
| *Argentina* | 0.006 | 0.598 | 7.270 |
| *Chile* | 0.000 | 0.948 | -0.640 |
| *Indonesia* | 0.146 | ***6.16E-03*** | -112.497 |
| *Turkey* | 0.125 | 0.0119 | 28.478 |

**Table S12 –** Ordinary least squares linear regression statistics for sex-searches v.s proximity in eigenmood to Christmas. The components selected were the two components (eigenbins) that most distinguish the holiday week from other weeks (see Methods S11). In the Components column, v stands for valence, d for dominance, and a for arousal. $R^2$ is the coefficient of determination, $F_p$ is the p-value of the overall F-test for the regression, and the Slope is the slope of regressions. $\rho$ is the Pearson's correlation coefficient between proximity and sex searches, $\rho_D$ is the Brownian distance correlation coefficient, and $DCov_p$ is the p-value for the Brownian distance covariance calculated from a permutation test of the data. Bold denotes significance at α=0.05, italicized values denote Bonferroni corrected significance over countries per variable choice α=0.05/7 = 0.00714, underlined denote Bonferroni corrected significance over all table possibilities α=0.05/21 = 0.00238.

**Christmas**

| *Country* | Components | $R^2$ | Fp | Slope | ρ | ρD | $DCov_p$ |
|---|---|---|---|---|---|---|---|
| *USA* | v4, v5 | 0.38 | ***5.08E-06*** | 6.50E+04 | 0.616 | 0.559 | 0.001 |
| *Australia* | d5, d8 | 0.392 | ***2.52E-06*** | 2.44E+04 | 0.626 | 0.576 | 0.001 |
| *Brazil* | a3, v2 | 0.504 | ***3.35E-08*** | 9.47E+03 | 0.71 | 0.624 | 0.001 |
| *Argentina* | v5, d3 | 0.577 | ***6.11E-10*** | 5.35E+03 | 0.759 | 0.712 | 0.001 |
| *Chile* | v3, d8 | 0.419 | ***1.16E-06*** | 7.96E+03 | 0.647 | 0.646 | 0.001 |
| *Indonesia* | a3, v3 | 0.448 | ***2.66E-07*** | 9.95E+03 | 0.67 | 0.657 | 0.001 |
| *Turkey* | a3, d3 | 0.373 | ***6.46E-06*** | -1.42E+03 | -0.611 | 0.618 | 0.001 |

**Eid-al-Fitr without Ramadan**

| *Country* | Components | $R^2$ | Fp | Slope | ρ | ρD | $DCov_p$ |
|---|---|---|---|---|---|---|---|
| *USA* | a6, v3 | 0.065 | 0.107 | 1.57E+05 | 0.256 | 0.328 | 0.118 |
| *Australia* | v3, v4 | 0.02 | 0.381 | -1.62E+03 | -0.141 | 0.317 | 0.154 |
| *Brazil* | a3, d8 | 0.147 | **0.0147** | -4.07E+04 | -0.383 | 0.539 | 0.001 |
| *Argentina* | v9, d3 | 0.598 | ***3.08E-09*** | -2.32E+04 | -0.773 | 0.735 | 0.001 |
| *Chile* | a6, d2 | 0.189 | ***5.00E-03*** | -1.15E+04 | -0.435 | 0.461 | 0.005 |
| *Indonesia* | v3, d3 | 0.637 | ***6.87E-10*** | 8.70E+03 | 0.798 | 0.712 | 0.001 |
| *Turkey* | a3, d3 | 0.737 | ***6.94E-13*** | 4.81E+02 | 0.859 | 0.858 | 0.001 |

**Eid-al-Fitr**

| *Country* | Components | $R^2$ | Fp | Slope | ρ | ρD | $DCov_p$ |
|---|---|---|---|---|---|---|---|
| *USA* | a6, v3 | 0.077 | 0.0645 | 1.75E+05 | 0.278 | 0.343 | 0.061 |
| *Australia* | v3, v4 | 0.038 | 0.198 | -2.30E+03 | -0.196 | 0.333 | 0.085 |
| *Brazil* | a3, d8 | 0.124 | **0.0204** | -3.54E+04 | -0.353 | 0.516 | 0.001 |
| *Argentina* | v9, d3 | 0.593 | ***6.23E-10*** | -2.31E+04 | -0.77 | 0.73 | 0.001 |
| *Chile* | a6, d2 | 0.191 | ***3.03E-03*** | -1.04E+04 | -0.437 | 0.489 | 0.001 |
| *Indonesia* | v3, d3 | 0.407 | ***3.19E-06*** | 9.85E+03 | 0.638 | 0.621 | 0.001 |
| *Turkey* | a3, d3 | 0.339 | ***3.42E-05*** | 3.32E+02 | 0.582 | 0.634 | 0.001 |

**Table S13**– List of words and expressions removed from the Twitter/ANEW analysis.

“merry christmas”
“merry xmas”
“happy christmas”
“happy xmas”
“happy new year”
“happy newyear”
“happy thanksgiving”
“happy ramadan”
“happy easter”
“happy holidays”
“happy hanukkah”
“happy hanukah”
“happy ramadan”
“happy eid”
“happy halloween”
“happy valentines day”
“happy valentine’s day”
“feliz natal”
“feliz ano”
“feliz pascoa”
“pascoa feliz”
“feliz thanksgiving”
“feliz navidad”
“feliz ano nuevo”
“feliz ano novo”
“feliz ramadan”
“feliz año”
“feliz páscoa”
“páscoa feliz ”
“feliz año nuevo”
“happy anzac day”
“feliz anzac day”
“happy adelaide cup”
“feliz adelaide cup”
“happy all saints day”
“feliz all saints day”
“happy all souls day”
“feliz all souls day”
“happy andalucia day”
“feliz andalucia day”
“happy arafat day”
“feliz arafat day”
“happy armistice day”
“feliz armistice day”
“happy army day”
“feliz army day”
“happy asahna bucha day”
“feliz asahna bucha day”
“happy ascension day”
“feliz ascension day”
“happy ash monday”
“feliz ash monday”
“happy ash wednesday”
“feliz ash wednesday”
“happy ashura”
“feliz ashura”
“happy assumption day”
“feliz assumption day”
“happy asturias”
“feliz asturias”
“happy auckland province”
“feliz auckland province”
“happy august bank holiday”
“feliz august bank holiday”
“happy august holiday”
“feliz august holiday”
“happy australia day”
“feliz australia day”
“happy australia day holiday”
“feliz australia day holiday”
“happy autumnal equinox day”
“feliz autumnal equinox day”
“happy awal muharram”
“feliz awal muharram”
“happy balearic islands”
“feliz balearic islands”
“happy bank holiday”
“feliz bank holiday”
“happy bastille day”
“feliz bastille day”
“happy battle of the boyne”
“feliz battle of the boyne”
“happy benito juarezs birthday”
“feliz benito juarezs birthday”
“happy berchtolds day”
“feliz berchtolds day”
“happy bettagsmontag”
“feliz bettagsmontag”
“happy bhogi”
“feliz bhogi”
“happy bicentennial of the constituent assembly of 1813”
“feliz bicentennial of the constituent assembly of 1813”
“happy birthday of muhammad iqbal”
“feliz birthday of muhammad iqbal”
“happy birthday of prophet muhammad”
“feliz birthday of prophet muhammad”
“happy birthday of quaid-e-azam muhammad ali jinnah”
“feliz birthday of quaid-e-azam muhammad ali jinnah”
“happy birthday of spb yang di pertuan agong”
“feliz birthday of spb yang di pertuan agong”

"happy birthday of the sultan of selangor"
"feliz birthday of the sultan of selangor"
"happy boxing day"
"feliz boxing day"
"happy bridge public"
"feliz bridge public"
"happy buddha purnima"
"feliz buddha purnima"
"happy buddhas birthday"
"feliz buddhas birthday"
"happy canada day"
"feliz canada day"
"happy canary islands"
"feliz canary islands"
"happy canberra day"
"feliz canberra day"
"happy canterbury"
"feliz canterbury"
"happy carnival"
"feliz carnival"
"happy castile-la mancha"
"feliz castile-la mancha"
"happy catalonia"
"feliz catalonia"
"happy celebration of the golden spurs"
"feliz celebration of the golden spurs"
"happy ceuta"
"feliz ceuta"
"happy chanukah"
"feliz chanukah"
"happy chatham islands"
"feliz chatham islands"
"happy childrens day"
"feliz childrens day"
"happy chinese new year"
"feliz chinese new year"
"happy chinese new year eve"
"feliz chinese new year eve"
"happy ching ming"
"feliz ching ming"
"happy christmas day"
"feliz christmas day"
"happy christmas eve"
"feliz christmas eve"
"happy christmas eve day"
"feliz christmas eve day"
"happy christmas"
"feliz christmas"
"happy chulalongkorn day"
"feliz chulalongkorn day"
"happy chung yeung festival"
"feliz chung yeung festival"
"happy cinco de mayo"
"feliz cinco de mayo"
"happy civic day"
"feliz civic day"
"happy columbus day"
"feliz columbus day"
"happy coming of age day"
"feliz coming of age day"
"happy community day"
"feliz community day"
"happy community festival of madrid"
"feliz community festival of madrid"
"happy constitution day"
"feliz constitution day"
"happy constitution memorial day"
"feliz constitution memorial day"
"happy corpus christi"
"feliz corpus christi"
"happy culture day"
"feliz culture day"
"happy day after christmas"
"feliz day after christmas"
"happy day after new years day"
"feliz day after new years day"
"happy day of atonement"
"feliz day of atonement"
"happy day of good will"
"feliz day of good will"
"happy day of national sovereignty"
"feliz day of national sovereignty"
"happy day of reconciliation"
"feliz day of reconciliation"
"happy day of reformation"
"feliz day of reformation"
"happy day of unity"
"feliz day of unity"
"happy day of respect for cultural diversity"
"feliz day of respect for cultural diversity"
"happy day of the battle of salta"
"feliz day of the battle of salta"
"happy day of the constitution of the slovak republic"
"feliz day of the constitution of the slovak republic"
"happy day of the dead"
"feliz day of the dead"
"happy day of the establishment of the slovak republic"
"feliz day of the establishment of the slovak republic"
"happy day of the german-speaking community of belgium"
"feliz day of the german-speaking community of belgium"

"happy day of the virgin of guadalupe"
"feliz day of the virgin of guadalupe"
"happy day of victory over fascism"
"feliz day of victory over fascism"
"happy declaration of independence"
"feliz declaration of independence"
"happy deepavali"
"feliz deepavali"
"happy deewali"
"feliz deewali"
"happy defence of the motherland"
"feliz defence of the motherland"
"happy discovery day"
"feliz discovery day"
"happy double ninth day"
"feliz double ninth day"
"happy dragon boat festival"
"feliz dragon boat festival"
"happy dussehra"
"feliz dussehra"
"happy early may bank holiday"
"feliz early may bank holiday"
"happy easter"
"feliz easter"
"happy easter monday"
"feliz easter monday"
"happy easter sunday"
"feliz easter sunday"
"happy eid al adha"
"feliz eid al adha"
"happy eid al fitr"
"feliz eid al fitr"
"happy eid milad un-nabi"
"feliz eid milad un-nabi"
"happy eid ul-azha day 1"
"feliz eid ul-azha day 1"
"happy eid ul-azha day 2"
"feliz eid ul-azha day 2"
"happy eid-ul-fitr"
"feliz eid-ul-fitr"
"happy emancipation day"
"feliz emancipation day"
"happy epiphany"
"feliz epiphany"
"happy extremadura"
"feliz extremadura"
"happy family & community day"
"feliz family & community day"
"happy family day"
"feliz family day"
"happy fathers day"
"feliz fathers day"
"happy feast of st ambrose"
"feliz feast of st ambrose"
"happy feast of st anthony"
"feliz feast of st anthony"
"happy feast of st john the baptist"
"feliz feast of st john the baptist"
"happy federal territory day"
"feliz federal territory day"
"happy fiesta de san isidro"
"feliz fiesta de san isidro"
"happy foundation day"
"feliz foundation day"
"happy foundation of the independent czechoslovak state"
"feliz foundation of the independent czechoslovak state"
"happy freedom day"
"feliz freedom day"
"happy french community"
"feliz french community"
"happy ganesh chaturthi"
"feliz ganesh chaturthi"
"happy general prayer day"
"feliz general prayer day"
"happy german unity day"
"feliz german unity day"
"happy good friday"
"feliz good friday"
"happy greenery day"
"feliz greenery day"
"happy groundhog day"
"feliz groundhog day"
"happy guru nanak birthday"
"feliz guru nanak birthday"
"happy guy fawkes night"
"feliz guy fawkes night"
"happy h.m. kings birthday"
"feliz h.m. kings birthday"
"happy h.m. queens birthday"
"feliz h.m. queens birthday"
"happy hangeul day"
"feliz hangeul day"
"happy hari hol almarhum sultan iskandar"
"feliz hari hol almarhum sultan iskandar"
"happy hari raya haji"
"feliz hari raya haji"
"happy hari raya nyepi"
"feliz hari raya nyepi"
"happy hari raya puasa"
"feliz hari raya puasa"
"happy harvest festival"
"feliz harvest festival"
"happy hawkes bay"
"feliz hawkes bay"

"happy health-sports day"
"feliz health-sports day"
"happy heritage day"
"feliz heritage day"
"happy hijri new years day"
"feliz hijri new years day"
"happy hispanic day"
"feliz hispanic day"
"happy holi"
"feliz holi"
"happy holy spirit monday"
"feliz holy spirit monday"
"happy human rights day"
"feliz human rights day"
"happy idul adha"
"feliz idul adha"
"happy idul fitr"
"feliz idul fitr"
"happy idul juha"
"feliz idul juha"
"happy immaculate conception day"
"feliz immaculate conception day"
"happy independence day"
"feliz independence day"
"happy independence day of chile"
"feliz independence day of chile"
"happy independence day"
"feliz independence day"
"happy independence of cartagena"
"feliz independence of cartagena"
"happy isra miraj"
"feliz isra miraj"
"happy israa & miaraj night"
"feliz israa & miaraj night"
"happy jan hus day"
"feliz jan hus day"
"happy janmashtami"
"feliz janmashtami"
"happy june holiday"
"feliz june holiday"
"happy kannada rajyothsava"
"feliz kannada rajyothsava"
"happy kashmir day"
"feliz kashmir day"
"happy kings feast"
"feliz kings feast"
"happy knabenschiessen"
"feliz knabenschiessen"
"happy korean new year"
"feliz korean new year"
"happy la rioja"
"feliz la rioja"
"happy labor day"
"feliz labor day"
"happy labour day"
"feliz labour day"
"happy labour thanksgiving day"
"feliz labour thanksgiving day"
"happy labour day"
"feliz labour day"
"happy lady of aparecida"
"feliz lady of aparecida"
"happy lantern festival"
"feliz lantern festival"
"happy late mid autumn festival"
"feliz late mid autumn festival"
"happy liberation day"
"feliz liberation day"
"happy liberation day czech republic"
"feliz liberation day czech republic"
"happy maha shivratri"
"feliz maha shivratri"
"happy maharashtra day"
"feliz maharashtra day"
"happy mahatma gandhi birthday"
"feliz mahatma gandhi birthday"
"happy mahavir jayanti"
"feliz mahavir jayanti"
"happy makha bucha day"
"feliz makha bucha day"
"happy malaysia day"
"feliz malaysia day"
"happy malvinas day"
"feliz malvinas day"
"happy march 1st movement"
"feliz march 1st movement"
"happy marine day"
"feliz marine day"
"happy marlborough"
"feliz marlborough"
"happy martin luther king day"
"feliz martin luther king day"
"happy maulidur rasul"
"feliz maulidur rasul"
"happy maundy thursday"
"feliz maundy thursday"
"happy may bank holiday"
"feliz may bank holiday"
"happy may day"
"feliz may day"
"happy may day revolution"
"feliz may day revolution"
"happy melbourne cup day"
"feliz melbourne cup day"
"happy memorial day"
"feliz memorial day"

"happy mid autumn festival"
"feliz mid autumn festival"
"happy midsummer day"
"feliz midsummer day"
"happy milad-un-nabi"
"feliz milad-un-nabi"
"happy mothering sunday"
"feliz mothering sunday"
"happy mothers day"
"feliz mothers day"
"happy muharram"
"feliz muharram"
"happy murcia"
"feliz murcia"
"happy national day"
"feliz national day"
"happy national flag day"
"feliz national flag day"
"happy national foundation day"
"feliz national foundation day"
"happy national remembrance day"
"feliz national remembrance day"
"happy national sovereignty and children's day"
"feliz national sovereignty and children's day"
"happy national womens day"
"feliz national womens day"
"happy national holiday"
"feliz national holiday"
"happy navy day"
"feliz navy day"
"happy nelson"
"feliz nelson"
"happy new year"
"feliz new year"
"happy new years day"
"feliz new years day"
"happy new years eve"
"feliz new years eve"
"happy new years"
"feliz new years"
"happy orthodox christmas day"
"feliz orthodox christmas day"
"happy orthodox easter monday"
"feliz orthodox easter monday"
"happy orthodox good friday"
"feliz orthodox good friday"
"happy otago province"
"feliz otago province"
"happy our lady of mount carmel"
"feliz our lady of mount carmel"
"happy our lady of the almudena"
"feliz our lady of the almudena"
"happy pakistan day"
"feliz pakistan day"
"happy pancake tuesday"
"feliz pancake tuesday"
"happy parsi new year"
"feliz parsi new year"
"happy passover"
"feliz passover"
"happy peace memorial day"
"feliz peace memorial day"
"happy pentecost"
"feliz pentecost"
"happy picnic day"
"feliz picnic day"
"happy pongal"
"feliz pongal"
"happy portugal day"
"feliz portugal day"
"happy presidential elections"
"feliz presidential elections"
"happy presidents day"
"feliz presidents day"
"happy public holiday"
"feliz public holiday"
"happy purim"
"feliz purim"
"happy queens birthday"
"feliz queens birthday"
"happy race day"
"feliz race day"
"happy ram navami"
"feliz ram navami"
"happy ramazan feast"
"feliz ramazan feast"
"happy reformation day"
"feliz reformation day"
"happy remembrance day"
"feliz remembrance day"
"happy repentance day"
"feliz repentance day"
"happy republic day"
"feliz republic day"
"happy respect for the aged day"
"feliz respect for the aged day"
"happy restoration day"
"feliz restoration day"
"happy restoration day of the independent czech state"
"feliz restoration day of the independent czech state"
"happy restoration of independence"
"feliz restoration of independence"
"happy revolution day"
"feliz revolution day"

"happy sacred heart"
"feliz sacred heart"
"happy sacrifice feast"
"feliz sacrifice feast"
"happy saint leopold"
"feliz saint leopold"
"happy saint nicholas"
"feliz saint nicholas"
"happy saint peter and saint paul"
"feliz saint peter and saint paul"
"happy saint stephens day"
"feliz saint stephens day"
"happy sechselauten"
"feliz sechselauten"
"happy second day of christmas"
"feliz second day of christmas"
"happy showa day"
"feliz showa day"
"happy simchat torah"
"feliz simchat torah"
"happy slovak national uprising anniversary"
"feliz slovak national uprising anniversary"
"happy songkran festival"
"feliz songkran festival"
"happy south canterbury"
"feliz south canterbury"
"happy southland"
"feliz southland"
"happy special administration region (sar) day"
"feliz special administration region (sar) day"
"happy st andrews day"
"feliz st andrews day"
"happy st cyril and methodius day"
"feliz st cyril and methodius day"
"happy st davids day"
"feliz st davids day"
"happy st georges day"
"feliz st georges day"
"happy st james day"
"feliz st james day"
"happy st josephs day"
"feliz st josephs day"
"happy st martins day"
"feliz st martins day"
"happy st patricks day"
"feliz st patricks day"
"happy st stephens day"
"feliz st stephens day"
"happy st wenceslas day"
"feliz st wenceslas day"
"happy struggle for freedom and democracy day"
"feliz struggle for freedom and democracy day"
"happy sukkot"
"feliz sukkot"
"happy swiss federal fast"
"feliz swiss federal fast"
"happy taranaki"
"feliz taranaki"
"happy thaipusam"
"feliz thaipusam"
"happy thanksgiving"
"feliz thanksgiving"
"happy buddhas birthday"
"feliz buddhas birthday"
"happy emperors birthday"
"feliz emperors birthday"
"happy national holiday of quebec"
"feliz national holiday of quebec"
"happy ochi day"
"feliz ochi day"
"happy patron saint of turin"
"feliz patron saint of turin"
"happy thiruvalluvar day"
"feliz thiruvalluvar day"
"happy tiradentes day"
"feliz tiradentes day"
"happy tomb sweeping festival"
"feliz tomb sweeping festival"
"happy tomb sweeping holiday"
"feliz tomb sweeping holiday"
"happy truth and justice memorial day"
"feliz truth and justice memorial day"
"happy uae national day"
"feliz uae national day"
"happy ugadi"
"feliz ugadi"
"happy urs mubarak of hazrat data gunj bakhsh"
"feliz urs mubarak of hazrat data gunj bakhsh"
"happy v-e day"
"feliz v-e day"
"happy valencia"
"feliz valencia"
"happy vernal equinox day"
"feliz vernal equinox day"
"happy vesak day"
"feliz vesak day"
"happy veterans day"
"feliz veterans day"
"happy victoria day"
"feliz victoria day"
"happy victory day"
"feliz victory day"
"happy visakha bucha day"
"feliz visakha bucha day"
"happy waisak day"

“feliz waisak day”
“happy waitangi day”
“feliz waitangi day”
“happy wellington province”
“feliz wellington province”
“happy wesak day”
“feliz wesak day”
“happy westland”
“feliz westland”
“happy whitmonday”
“feliz whitmonday”
“happy womens day”
“feliz womens day”
“happy youth day”
“feliz youth day”
“happy zumbi dos palmares”
“feliz zumbi dos palmares”
“christmas”
“navidad”
“natal”
“valentine”
“san valentín”
“valentín”
“san valentin”
“valentin”
“valentim